%% file: RMITTAL.tex
\documentclass[usenatbib]{mn2e}
\usepackage{times}
\usepackage{txfonts}
\usepackage{graphicx}
\usepackage{setspace}
\usepackage{natbib}
\usepackage{sidecap}
\usepackage{color}
\usepackage{subfigure}
\usepackage{floatflt}
\usepackage{rotating}
\usepackage{multirow}
\usepackage{aalongtable, lscape, afterpage}
\newcommand{\comment}[1]{}
\newcommand{\rul}{\rule[-2.50mm]{0mm}{2mm}}

\begin{document}

\input{makros}
\input{journals}

\title[Herschel Observations of the Perseus
Cluster]{Herschel\thanks{{\it Herschel} is an ESA space observatory
    with science instruments provided by European-led Principal
    Investigator consortia and with important participation from
    NASA.} observations of extended atomic gas in the core of the
  Perseus cluster}

\author[Rupal Mittal et~al.]
{{Rupal Mittal$^{1}$,
J. B. Raymond Oonk$^{2}$,
Gary J. Ferland$^{3}$,
Alastair C. Edge$^{4}$,
}
\newauthor 
{Christopher P. O'Dea$^{5, 6}$, Stefi A. Baum$^{1, 7}$,  John T. Whelan$^{8}$, Roderick M. Johnstone$^{9}$}
\newauthor
{Francoise Combes$^{10}$, Philippe Salom\'e$^{10}$, Andy C. Fabian$^{9}$, Grant R. Tremblay$^{11}$,
}
\newauthor
{Megan Donahue$^{12}$ and Helen Russell$^{13}$} \\
$^{1}$ Chester F. Carlson Center for Imaging Science, Rochester Institute of Technology, Rochester, NY 14623, USA \\
$^{2}$ Netherlands Institute for Radio Astronomy, Postbus 2, 7990 AA Dwingeloo, The Netherlands \\ 
$^{3}$ Department of Physics, University of Kentucky, Lexington, KY 40506, USA \\
$^{4}$ Institute for Computational Cosmology, Department of Physics, Durham University, Durham, DH1 3LE\\
$^{5}$ Department of Physics, Rochester Institute of Technology, Rochester, NY 14623, USA  \\
$^{6}$ Harvard Smithsonian Center for Astrophysics, 60 Garden St. Cambridge, MA 02138 \\
$^{7}$ Radcliffe Institute for Advanced Study, 10 Garden St. Cambridge, MA 02138 \\
$^{8}$ School of Mathematical Sciences and Center for Computational Relativity \& Gravitation, Rochester Institute of Technology, Rochester, NY 14623, USA  \\
$^{9}$ Institute of Astronomy, Madingley Road, Cambridge, CB3 0HA \\
$^{10}$ Observatoire de Paris, LERMA, CNRS, 61 Av. de l'Observatoire, 75014 Paris, France\\
$^{11}$ European Southern Observatory, Karl-Schwarzschild-Str.~2, 85748 Garching 
bei M\"unchen, Germany\\
$^{12}$ Michigan State University, Physics and Astronomy Dept., East Lansing, MI 48824, USA \\
$^{13}$  Department of Physics \& Astronomy, University of Waterloo, Canada, N2L 3G1
}

\date{}

\maketitle

\begin{abstract}
  We present Herschel observations of the core of the Perseus cluster
  of galaxies. Especially intriguing is the network of filaments that
  surround the brightest cluster galaxy, NGC~1275, previously imaged
  extensively in {\ha} and CO. In this work, we report detections of
  far-infrared~(FIR) lines, in particular, {\cii}~158{\mm},
  {\oi}~63{\mm}, {\nii}~122{\mm}, {\oib}~145{\mm} and {\oiii}~88{\mm},
  with Herschel. All lines are spatially extended, except {\oiii},
  with the {\cii} line emission extending up to 25~kpc from the
  core. {\cii} emission is found to be cospatial with {\ha} and
  CO. Furthermore, {\cii} shows a similar velocity distribution to CO,
  which has been shown in previous studies to display a close
  association with the {\ha} kinematics. The spatial and kinematical
  correlation among {\cii}, {\ha} and CO gives us confidence to model
  the different components of the gas with a common heating model.

  With the help of FIR continuum Herschel measurements, together with
  a suite of coeval radio, submm and infrared data from other
  observatories, we performed a spectral energy distribution fitting
  of NGC~1275 using a model that contains contributions from dust
  emission as well as synchrotron AGN emission. This has allowed us to
  accurately estimate the dust parameters. The data indicate a low
  dust emissivity index, $\beta \approx 1$, a total dust mass close to
  $10^7~\ms$, a cold dust component with temperature $38\pm2$~K and a
  warm dust component with temperature of $116\pm9$~K. The FIR-derived
  star formation rate~(SFR) is $24\pm1~\mpy$, which is in agreement
  with the FUV-derived SFR in the core, determined after applying
  corrections for both Galactic and internal reddening. The total
  infrared luminosity in the range 8{\mm} to 1000{\mm} is inferred to
  be $1.5 \times 10^{11}~\ls$, making NGC~1275 a luminous infrared
  galaxy (LIRG).

  We investigated in detail the source of the Herschel FIR and {\ha}
  emissions emerging from a core region $4~$kpc in radius. Based on
  simulations conducted using the radiative transfer code, {\sc
    cloudy}, a heating model comprising old and young stellar
  populations is sufficient to explain these observations. The optical
  line ratios indicate that there may be a need for a second heating
  component. However, stellar photoionization seems to be the dominant
  mechanism.

  We have also detected {\cii} in three well-studied regions of the
  filaments. Herschel, with its superior sensitivity to FIR emission,
  can detect far colder atomic gas than previous studies.  We find a
  {\oi}/{\cii} ratio about 1 dex smaller than predicted by the
  otherwise functional Ferland~(2009) model. That study considered
  optically thin emission from a small cell of gas and by design did
  not consider the effects of reasonable column densities.  The line
  ratio suggests that the lines are optically thick, as is typical of
  galactic PDRs, and implies that there is a large reservoir of cold
  atomic gas. This was not included in previous inventories of the
  filament mass and may represent a significant component.

\end{abstract}

\newcommand{\pers}{NGC~1275}
\newcommand{\cen}{NGC~4696}

\section{Introduction}
\label{intro}

The Perseus cluster of galaxies is preeminent among the class of
cool-core galaxy clusters (those with gas cooling times shorter than
the Hubble time). This is in large part due to its close proximity
($z=0.01756$), allowing detailed studies to be conducted in varying
astrophysical contexts. It is the X-ray brightest galaxy cluster and
has a strongly peaked surface-brightness profile. The intracluster gas
in the inner few tens of kpc has a very short radiative cooling time
(200~Myr to 300~Myr). In the absence of heating, the expected X-ray
mass deposition rate is several $100~\mpy$. {\it FUSE} observations,
on the other hand, suggest an actual cooling rate of $\sim 30~\mpy$
\citep{Bregman2006} and {\it XMM-Newton} RGS observations suggest an
even lower residual cooling rate of $20~\mpy$ \citep[][ and references
therein]{Fabian2006}.

Perseus is the prototype of cluster radio ``mini-haloes''
\citep[e.g.][]{Pedlar1990,Gitti2002}. The brightest cluster galaxy of
Perseus~(a giant cD galaxy), {\pers}, is host to a powerful radio
source, 3C~84. It has a Seyfert-like spectrum and a bolometric radio
core-luminosity of order $10^{43}$~erg~s$^{-1}$
\cite[e.g][]{Vermeulen1994}. 3C~84 is inferred to undergo episodic
bursts of activity, blowing out jets of plasma which interact with the
intracluster medium. Chandra X-ray images, when compared with radio
emission at different wavelengths
\citep[e.g.][]{Boehringer1993,Fabian2000,Fabian2006,Sanders2005,Sanders2007}
reveal the extent to which the central radio source has caused havoc
in the intracluster-medium~(ICM) in the form of bubbles of various
kinds (inner, outer, ghost), sound waves and shocks. On VLBI/VLBA
(milliarcsec) scales, the radio morphology consists of a bright
southern jet \citep[e.g.][]{Vermeulen1994,Taylor2006} believed to be
pointed towards the observer and a dim northern counter jet pointed
away. The radio source shows a clear double-lobed morphology on
smaller scales ($\sim \pp{80}$) that extends into a more amorphous one
on larger scales ($\sim \pp{170}$).

One of the most intriguing aspects of Perseus is the spectacular
network of ionized ({\ha}) and molecular (CO and H$_2$) gas filaments
well beyond the optical stellar emission of the brightest cluster
galaxy~(BCG). A significant number of BCGs in cool-core~(CC) clusters
show similarly extended filamentary structures, such as NGC~4696
(Centaurus), Abell~1795 and Hydra-A
\citep{Johnstone1987,Heckman1989,Sparks1989,Crawford1999,McDonald2010}. The
source of excitation of the filaments is currently one of the most
pertinent issues in our understanding of CC galaxy clusters. In the
case of {\pers}, the filaments extend as far as $\sim 50~$kpc out from
the cluster-centric AGN, with the mean surface-brightness declining
much slower than the inverse-square law. Photoionization from a
central AGN thus seems unlikely. Similarly, ionizing radiation from
hot stars, such as O and B type, has been ruled out based on anomalous
emission lines in the spectra of the filaments
\citep{Johnstone1988,Johnstone2007,Ferland2008}. Motivated by the
observations of strong molecular hydrogen lines in NGC~4696 and
{\pers}, \cite{Ferland2009} showed that non-radiative heating, such as
collisional heating from ionizing particles, can produce the observed
emission. The importance of collisional excitation by energetic
(ionizing) particles was suggested more than two decades ago by
\cite{Johnstone1988}.  Candidate sources for these particles are
either cosmic rays or the thermal electrons of the X-ray emitting
intracluster medium. Motivated by the spatial correspondence between
the brightest low-energy X-rays and the {\ha} filaments,
\cite{Fabian2011} considered the penetration of cold filaments by the
surrounding hot X-ray gas through reconnection diffusion. More
recently, \cite{Sparks2012} reported a detection with the HST ACS
camera and also the COS spectrograph, of {\civ} line emission
spatially coincident with the {\ha} line emission in M87 in Virgo. The
{\civ} line emission is indicative of gas at $\sim 10^5$~K. They
suggest the origin of this line emission as being due to thermal
conduction, i.e., the transport of energy by hot electrons from the
hot ICM to the cold filament gas.

Various independent studies have provided strong evidence of the
presence of dust in cool-core BCGs. These include dust continuum
observations
\citep[e.g.][]{Edge1999,Chapman2002,Egami2006,ODea2008,Rawle2012} and
HST observations of BCGs with dust absorption features in them
\citep[e.g.][]{McNamara1996,Pinkney1996,Laine2003,Oonk2011}. Furthermore,
observations of H$_2$ and CO molecular gas
\cite[e.g.][]{Donahue2000,Edge2001,Edge2002,Salome2006} suggest that
there are substantial amounts of dust present which provide shielding.
In \cite{Mittal2011b}, we demonstrated the need for an overabundance
of dust (low gas-to-dust mass ratio) and metallicity to explain
measurements in NGC~4696. In most of the cases, the gas-to-dust mass
ratios are consistent with Galactic values
\citep{Sparks1989,Edge2001}, however, there is clearly an
overabundance of dust in the filaments relative to the surrounding hot
gas. The origin of the dust in the filaments is not yet clear. More
recently, \cite{Voit2011} have suggested that the main source of dust
in BCGs of CC clusters may be the stars of the central galaxy
themselves. Thus, a stellar origin is plausible although in the case
of NGC~1275 that seems unlikely given the filaments exist out to a
large radius from the galaxy core. Even though the sputtering
timescale for dust grains is short compared to the gas cooling time,
optical observations have revealed magnificent dust lanes in some
objects, which in many cases correlate spatially with the {\ha}
filaments \citep{Crawford2005,Sparks1989,Donahue1993}.
\cite{Fabian1994} and \cite{Voit1995} have proposed that gas-phase
reactions create dust in the cold clouds that cool out of the ICM,
which can then lead to an increase in dust through accretion onto
existing grains. In contrast, several studies argue
\citep[e.g.][]{Sparks1989,Farage2010} that the ionized gas and dust
filaments originate from the stripping of a dust-rich neighbouring
galaxy which may be in the process of merging with the BCG. The origin
of the molecular and ionized filaments mixed with dust is to date an
open question.

In this paper, we present Herschel observations of the core of the
Perseus cluster. The main goal of this work is to investigate the
source of the various emissions originating from the
filaments. Far-infrared data from Herschel \citep{Pilbratt2010} have
proven to be useful diagnostics of the heating mechanisms that account
for the filaments in cool-core clusters
\citep[e.g.][]{Rawle2012,Mittal2011b,Edge2010a,Edge2010b,Pereira2010}.
This work is part of a Herschel Open Time Key Project (PI: Edge) aimed
at understanding the origin of cold gas and dust in a representative
sample of 11 BCGs. We describe the data used and the analysis in
section~\ref{data}. We present some basic results in
section~\ref{results} and move on to discuss the heating mechanisms
prevailing in the core of {\pers} in section~\ref{core} and those
prevailing in the filaments of {\pers} in section~\ref{disc}. We
finally give our conclusions in section~\ref{conclusions}. We assume
throughout this paper the $\Lambda$CDM concordance Universe, with $H_0
= 71~h_{71}$~km~s$^{-1}$~Mpc$^{-1}$, $\Omega_{\st m} = 0.27$ and
$\Omega_{\Lambda} = 0.73$.  This translates into a physical scale of
$\pp{1}=0.352$~kpc and a luminosity distance of 75.3~Mpc at the
redshift of {\pers} ($z=0.01756$). This distance is consistent with
the independent distance inferred from the 2005 Type Ia supernova
SN2005mz \citep{Hicken2009a}. The right ascension and declination
coordinates in figures are in J2000 equinox. Lastly, {\pers} comprises
two systems -- a low-velocity system (LVS) consisting of gas at
5200~km~s$^{-1}$ associated with the BCG and a high-velocity system
(HVS) consisting of gas at 8200~km~s$^{-1}$ associated with a
foreground galaxy north-west of the BCG. In this work, {\pers} refers
to the LVS system only unless otherwise mentioned. For calculation of
line velocities we assume the velocity of the LVS to be the systemic
velocity.

\begin{table*}
  \centering
  \caption{\small {\herschel} PACS spectroscopy observational log of {\pers} at a redshift of 
    0.01756.  All the lines were observed in the line spectroscopy mode and on the same day: 
    30$^{\st{th}}$~Dec.~2009.}
  \label{obs}
  \begin{tabular}{|c| c| c| c| c c| c c| c| c |}
  \hline
  Line &  Peak Rest $\lambda$ & ObsID & Duration & \multicolumn{2}{c|}{Bandwidth} & \multicolumn{2}{c|}{Spectral FWHM} & Spatial FWHM & Mode \\
         &  ({\mm})                       &            &      (s)      & ({\mm}) & (km~s$^{-1}$)       & ({\mm}) & (km~s$^{-1}$) &  & \\
  \hline\hline
  OI   &  63.184   & 1342189962  & 9600 & 0.266 & 1250       & 0.017    & 79  & $\pp{9.4}$  & 5x5 raster, step size $\pp{23.5}$\\
  CII  &  157.741 & 1342214362  & 8600 & 1.499 & 2820       & 0.126    & 237 & $\pp{11.1}$   & 5x5 raster, step size $\pp{23.5}$   \\
  NII  & 121.90    & 1342214363  & 3440 & 1.717 & 4180       & 0.116    & 280 & $\pp{10.6}$   & pointed   \\   
  OIb & 145.525  & 1342202581  & 3440 & 1.576 & 3215       & 0.123    & 250 & $\pp{9.7}$   & pointed   \\ 
  OIII & 88.356    & 1342214363  & 3680 & 0.495 & 1660       & 0.033    & 110 & $\pp{8.5}$   & pointed   \\   
  SiI   & 68.473    & 1342202581  & 3840 & 0.218 & 945         & 0.014    & 62   & $\pp{8.3}$   & pointed   \\
  \hline
  \end{tabular}
\end{table*}

\section{Data and analysis}
\label{data}

\subsection{Herschel data}
\label{herschel}

We used the PACS spectrometer \citep{Poglitsch2010} to observe the
{\cii} line at 157.74{\mm} and the {\oi} line at 63.18{\mm}, the two
primary coolants of the interstellar-medium~(ISM). The {\cii} and
{\oi} fine-structure lines are very often the brightest emission lines
in galaxy spectra. In addition, we observed {\oib} at 145.52{\mm},
{\si} at 68.470{\mm}, {\nii} at 121.90{\mm} and {\oiii} at
88.36{\mm}. The {\oi} and {\cii} lines were observed in the
raster-mapping mode, consisting of 5 raster lines and 5 points per
line with a line step of $\pp{23.5}$, whereas the rest of the lines
were observed in a single-pointing mode.  The observational parameters
are summarized in Table~\ref{obs}. The line observations were
conducted in the line-spectroscopy mode using the chopping and nodding
technique (using a chopper throw of $\p{6}$) to subtract the telescope
background, the sky background and the dark current.

The PACS photometric observations were made in large-scan mapping mode
at a speed of $\pp{20}$s$^{-1}$ at blue-short~(BS)~(70{\mm}),
blue-long~(BL)~(100{\mm}) and red~(R)~(160{\mm}) wavelengths (PI:
E.~Sturm, ObsIDs: 1342204217, 1342204218, 1342216022 and
1342216023). The scans consisted of 18 scan line legs of $\p{4}$
length and of a cross-scan step of $\pp{15}$. The ``scan'' and
orthogonal ``cross-scan'' observations were individually calibrated
before being combined into a single map of $\p{9} \times \p{9}$.  The
PACS photometer has a resolution of $\pp{5.2}, \pp{7.7}$ and $\pp{12}$
at 70{\mm}, 100{\mm} and 160{\mm} respectively. The PACS photometer
performs observations at BS and BL simultaneously with the R band so
we have two sets of scans in the R band. The SPIRE photometric
observations was also made in the large-scan mapping mode and the data
were recorded simultaneously at 250{\mm}, 350{\mm} and 500{\mm} (PI:
E.~Sturm, ObsID: 1342203614). The SPIRE photometer \citep{Griffin2010}
has a resolution of about $\pp{18}$, $\pp{25}$ and $\pp{36}$ at these
wavelengths, respectively.

The basic calibration of the data (spectral and photometric) was done
using the Herschel Interactive Processing Environment~({\sc HIPE})
\citep{Ott2010} version 7.0~CIB~1931. For the PACS spectral data, the
standard pipeline routines described in the PACS data reduction
guideline~(PDRG) were adopted to process the spectral data from their
raw to a fully-calibrated level. HIPE 7.0.1931 contains PACS
calibration files that provide the response calibration based on
in-orbit measurements. Hence, no ground-to-flight correction factors
had to be applied.  The PACS cubes were rebinned in wavelength using
the Nyquist-Shannon sampling, corresponding to oversample=2 and
upsample=1. The spatial full-width at half maximum~(FWHM) varies from
$\pp{8}$ for the {\si} 68.47{\mm} line to $\pp{11}$ for the {\cii}
157.74{\mm} line. The line fluxes were determined using the method
described in \cite{Mittal2011b}. Briefly, the routine {\sc
  specproject} was used to obtain a final projection of the different
pointings and nods onto the sky plane. These maps can be readily used
to conduct `aperture photometry' and measure fluxes. For PACS
photometry, the data were reduced using the pipeline for the ScanMap
observing mode, particularly designed to detect extended emission ($>
\pp{100}$). The pipeline employs a second-level deglitching algorithm,
which uses the redundancy in a pixel to flag outliers, so that bright
sources are not erroneously flagged as glitches. For SPIRE photometry,
the data were reduced using the pipeline for the LargeScanMap
observing mode and the na\"ive map-maker.

\subsection{{\ha} data}
\label{ha}

The {\ha} flux measurements were made using the continuum-subtracted
data from \cite{Conselice2001}. Based on comparison with our newer HST
data \citep{Fabian2008}, the calibration of the WIYN data appears to
over-estimate the flux by a factor of about three. The source of this
discrepancy is yet not clear and is under thorough investigation
(Johnstone et al. 2012, in prep.). To verify the calibration offset of
the WIYN continuum-subtracted image, we calculated the total flux in
the WIYN {\ha} image in counts~s$^{-1}$ and converted it into
erg~s$^{-1}$~cm$^{-2}$ using the conversion given in
\cite{Conselice2001}. We were only able to derive the total luminosity
quoted in \cite{Conselice2001}, and which is in good agreement with
\cite{Heckman1989}, after lowering the total flux by a factor of
three. Hence we have scaled down the measured WIYN {\ha} fluxes by a
factor of three.

The HST broad band filter F625W admits light from the {\oiopt},
{\niiopt} and {\siiopt} doublets as well as {\ha}. The ratios of these
lines to {\ha} are variable with position in the nebula. Such
spectroscopic data are available for only a small fraction of the
total area covered by the entire nebula. For the WIYN data the
contamination is only from the {\niiopt} doublet. This is the main
reason why we prefer to use the WIYN data.

\cite{Hatch2006} found a radial gradient in the ratio {\niiopt}/{\ha},
which may be due to a spatially varying metallicity or excitation
mechanism \cite[see also][]{Johnstone1988}. They used the multi-object
spectrograph instrument on Gemini and found that the {\niiopt}/{\ha}
ratio varies from 0.5 to 0.85 in the Horseshoe region. Assuming that
the {\niiopt}/{\ha} ratio does not change significantly with azimuthal
angle, we subtracted the contribution of {\niiopt} from the measured
{\ha} flux in the Horseshoe, southwest and Blue Loop knots, using an
average ratio of {\niiopt}/{\ha}=0.65. The Gemini measurement close to
the nucleus indicates a {\niiopt}/{\ha} ratio close to unity,
therefore we halved the measured {\ha} flux in the core
region. Similarly, we also corrected for the {\nii}$\lambda$6548 line,
usually a third in intensity of the {\niiopt} line \citep{Hatch2006}.

Due to the uncertainties in the {\niiopt}/{\ha} ratio and the
calibration of the WIYN data, we caution the reader concerning the
absolute values of the {\ha} fluxes. However, the relative values of
{\ha} flux between spatial positions should be accurate. The {\ha}
amplitude plays a role in section~\ref{core}, where the Herschel and
{\ha} measurements are used as constraints to determine the heating
mechanisms giving rise to the various emissions. Fortunately, there
are enough FIR-derived constraints that the best-fit model parameters
do not rely solely on the {\ha} flux measurement.

\subsection{Dust Extinction}
\label{extinction}

\begin{figure*}
  \begin{minipage}{0.33\textwidth}
    \centering
    \includegraphics[width=\textwidth]{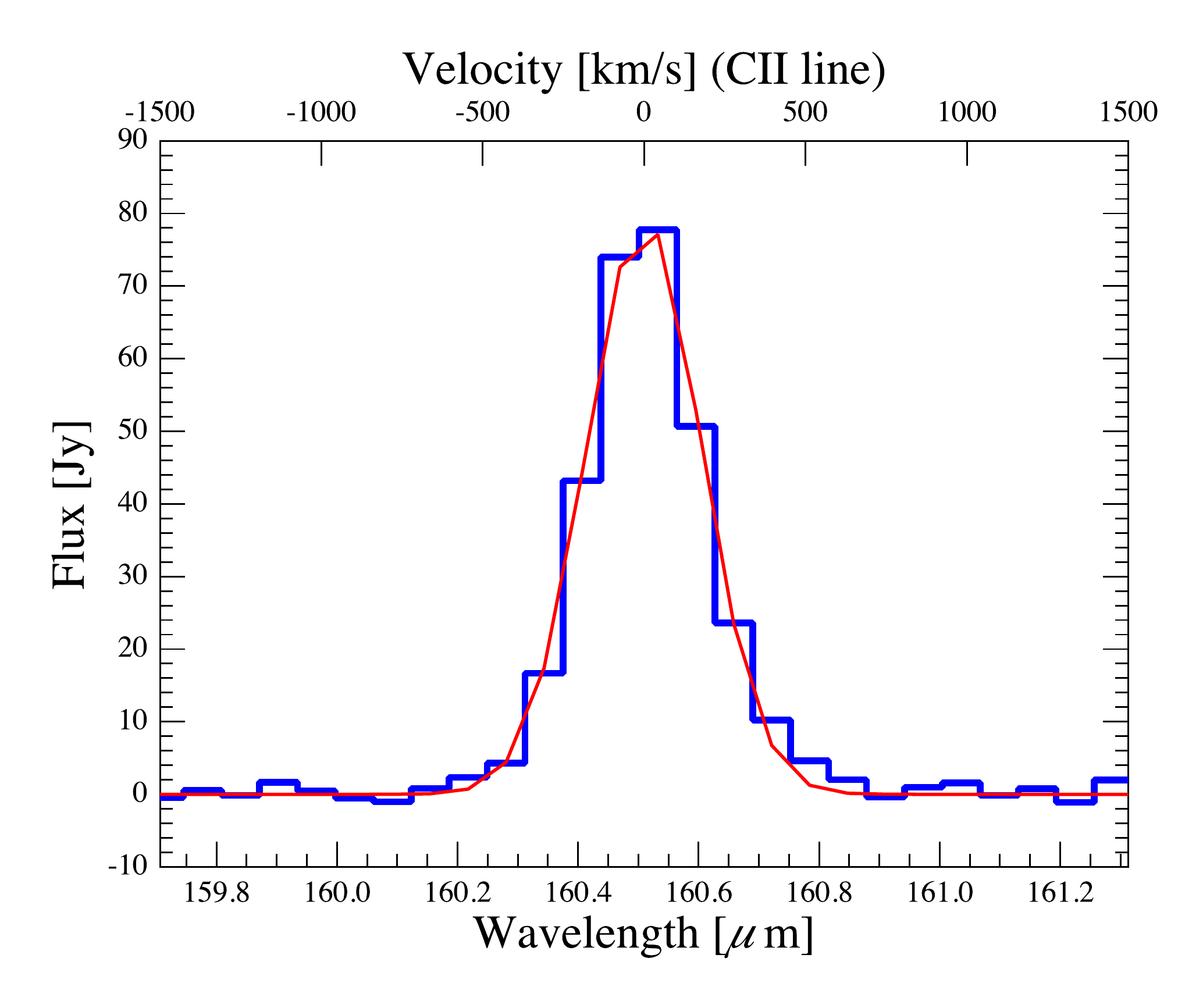}
  \end{minipage}%
  \begin{minipage}{0.33\textwidth}
    \centering
    \includegraphics[width=\textwidth]{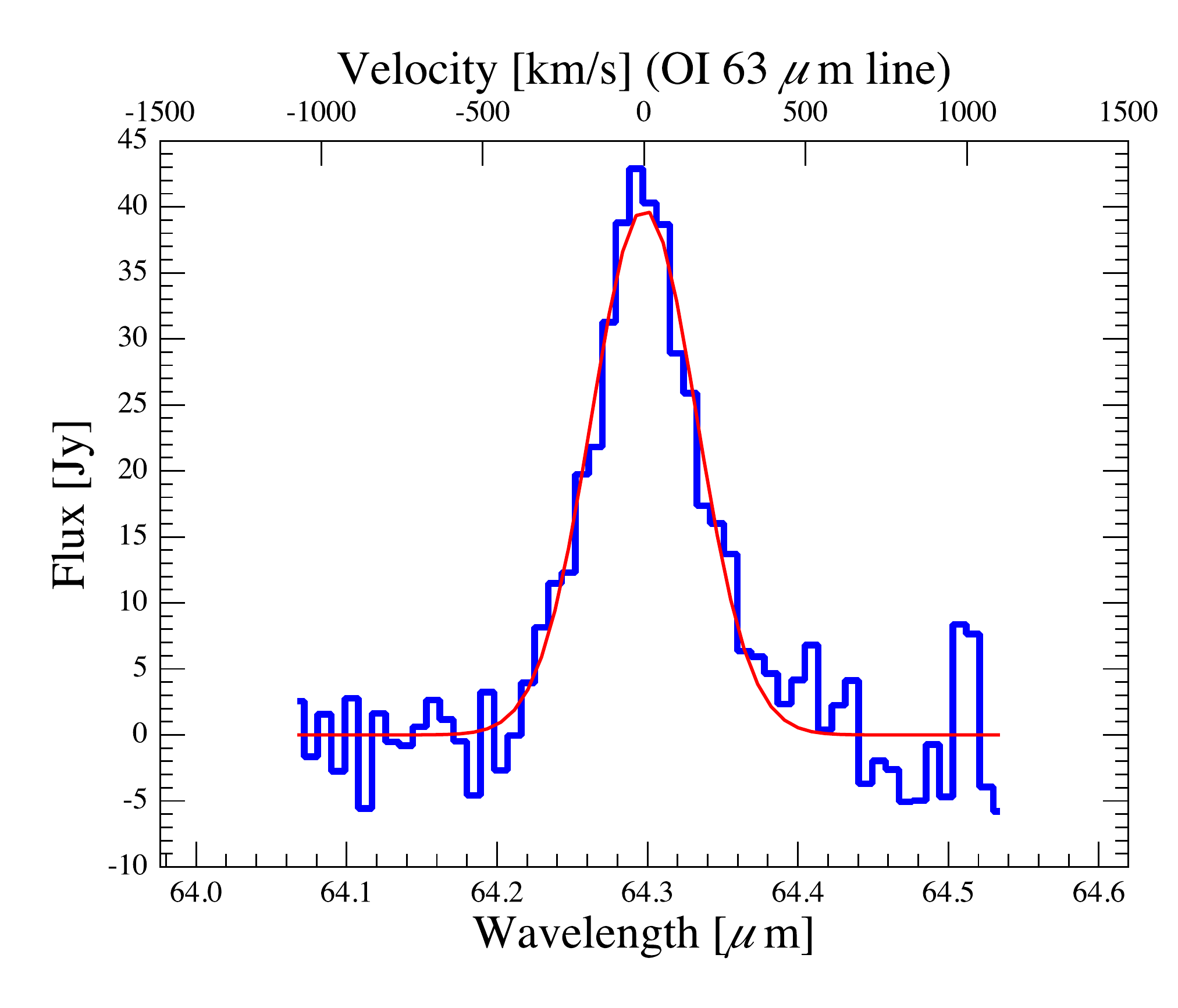}
  \end{minipage}%
  \begin{minipage}{0.33\textwidth}
    \centering
    \includegraphics[width=\textwidth]{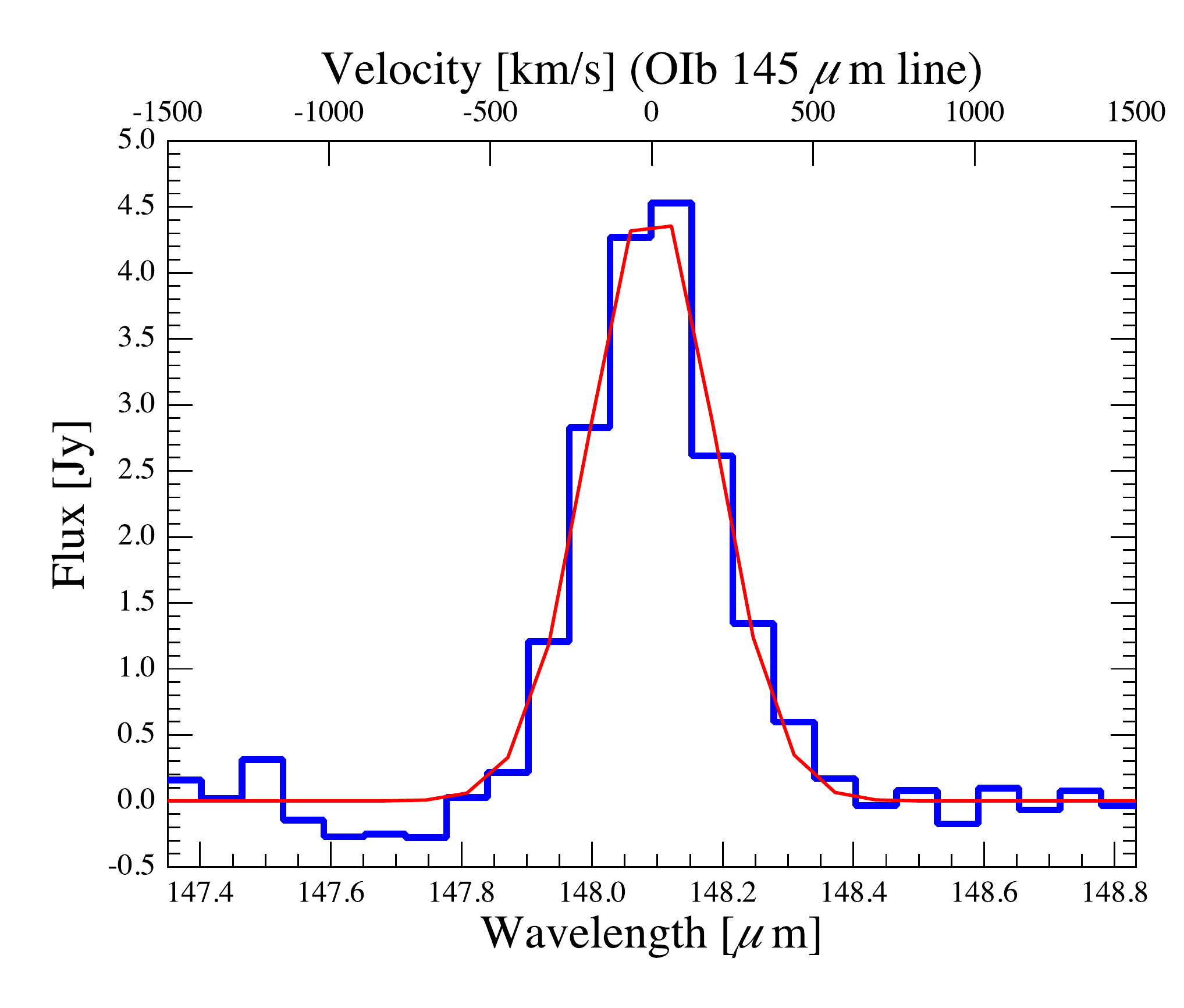}
  \end{minipage}\\
  \begin{minipage}{0.33\textwidth}
    \centering
    \includegraphics[width=\textwidth]{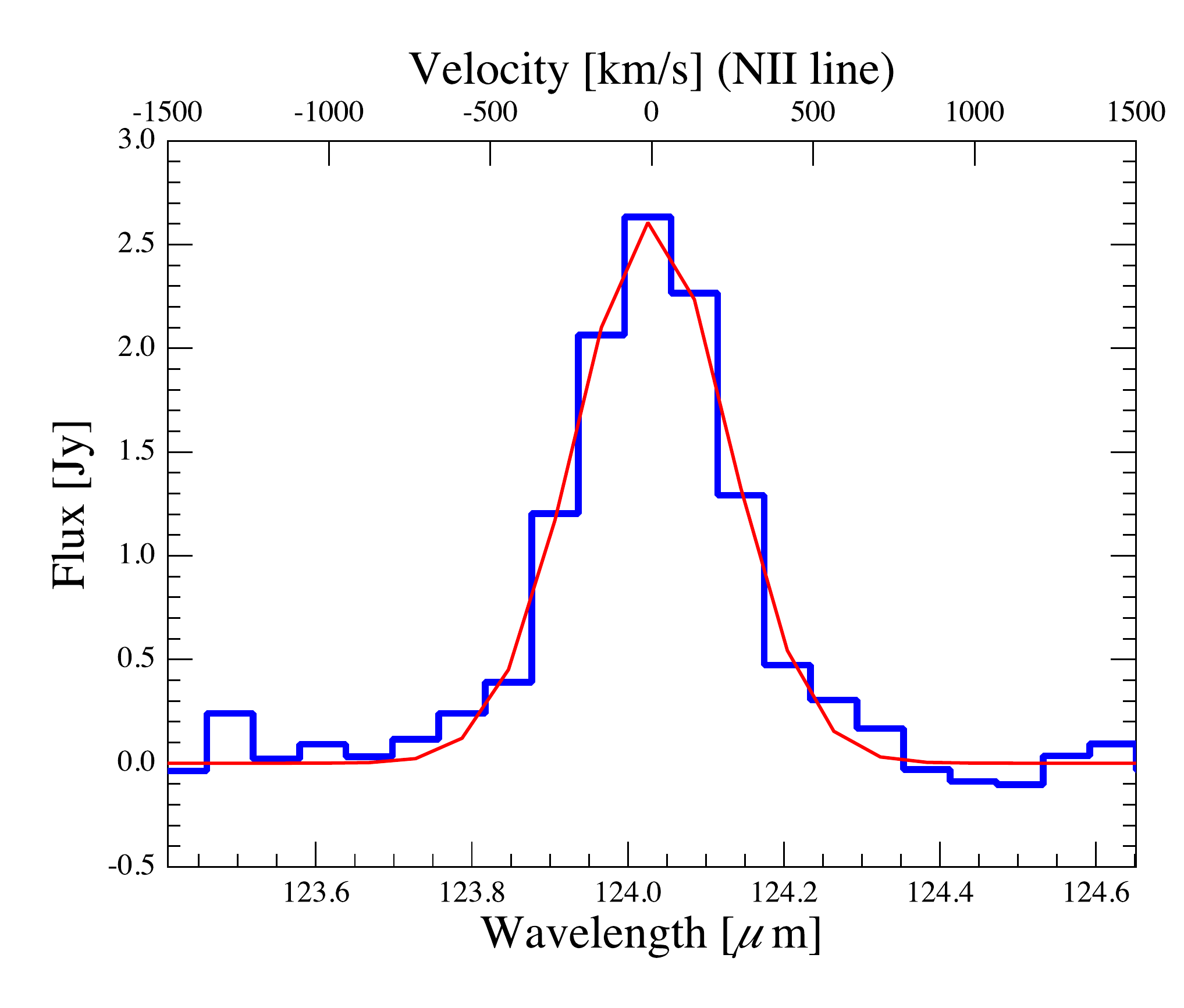}
  \end{minipage}
  \begin{minipage}{0.33\textwidth}
    \centering
    \includegraphics[width=\textwidth]{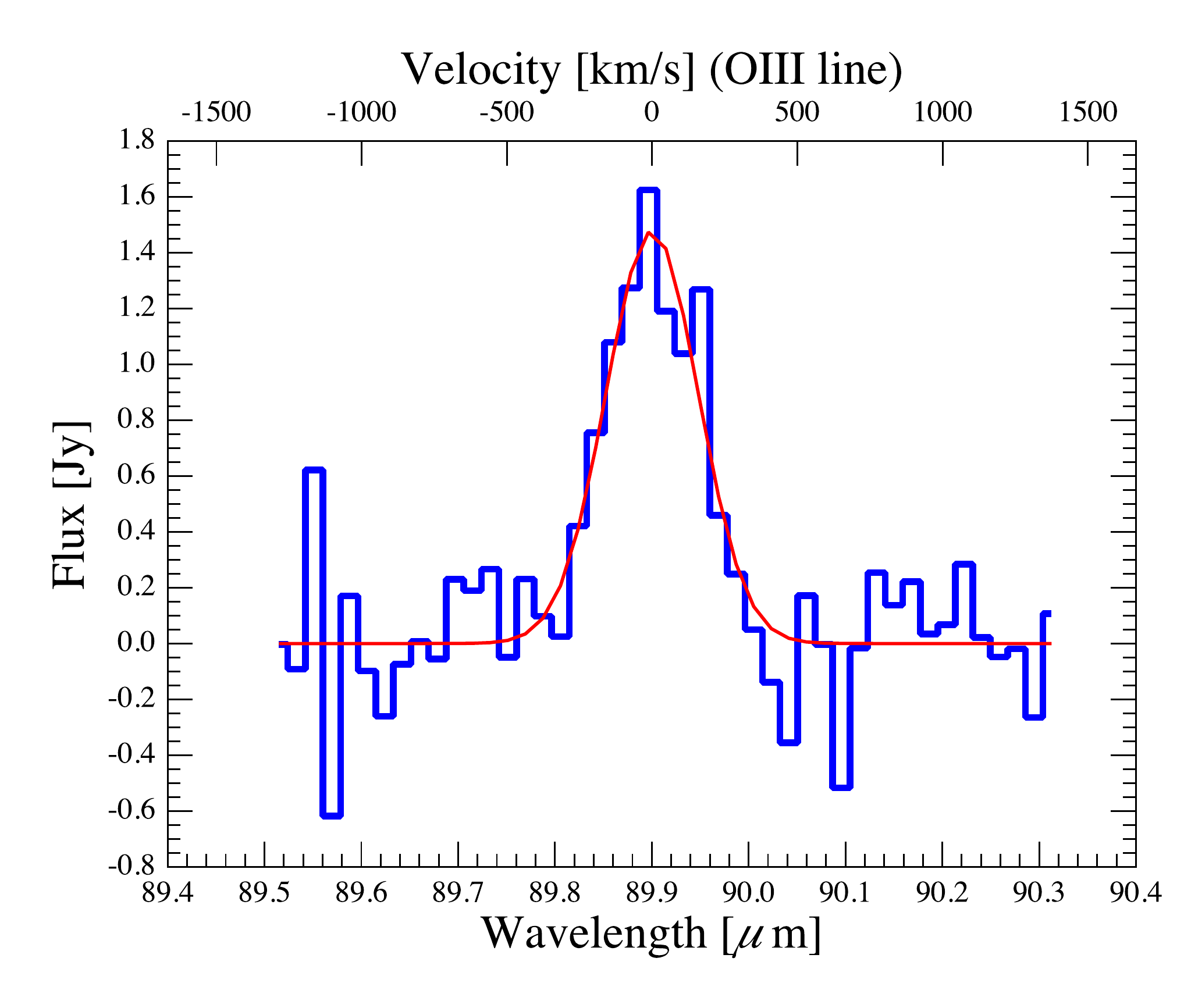}
  \end{minipage}
  \caption{\small The forbidden far-infrared line detections in the
    center of {\pers} made with the {\herschel} PACS instrument. The
    lines are spatially integrated. {\it Top Row:} {\cii} 157.74{\mm}
    (left), {\oi} 63.18{\mm} (middle) and {\oib} 145.52{\mm} (right).
    {\it Bottom Row:} {\nii} 121.90{\mm} (left) and {\oiii} 88.36{\mm}
    (right). }
  \label{persspec}
\end{figure*}

SCUBA observations of {\pers} have been used to infer the presence of
a large amount of dust ($6\times10^7~\ms$) present
\citep{Irwin2001}. The source of dust is not yet clear and
understanding it is one of the goals of this study. However, dust also
presents a hindrance due to the extinction it causes at optical and
higher frequencies. In this work, both the Galactic and internal
extinction corrections, such as for {\ha} (section~\ref{ha}) and
far-ultraviolet measurements (section~\ref{fuv}), were calculated with
the help of the mean extinction laws given in \cite{Cardelli1989}. We
assumed the Galactic extinction law, $R_V=3.1$, and an $E(B-V)$ value
of 0.163 from the NASA/IPAC Extragalactic Database
(NED\footnote{http://nedwww.ipac.caltech.edu}). Internal extinction
was calculated using the observed Balmer decrements, which were
compared to the Case-B value of \ha/\hb = 2.86. For the filaments
located far out from the core, an internal extinction of $E(B-V)=0.38$
was estimated based on the Galactic-extinction corrected Balmer
decrement, \ha/\hb = 4.2, as measured in the Horseshoe knot
\citep{Ferland2009}. For the core, a similar internal extinction was
estimated, $E(B-V) = 0.37$, based on the Galactic-extinction corrected
Balmer decrement, \ha/\hb=4.08, as measured $\pp{18}$ SW of the
nucleus by \cite{Kent1979}. Note that \cite{Kent1979} obtain an
internal reddening of $E(B-V)=0.43$ assuming a Galactic extinction of
$E(B-V)=0.1$. We obtain a lower internal reddening due to the higher
Galactic extinction adopted.

The correction factor for the internal reddening is
model-dependent. If the {\ha} and {\hb} emissions are produced by
another mechanism than case-B recombination, then the Balmer decrement
will be different from the expected value of 2.86. For example, if
particle heating is responsible for the {\ha} and {\hb} emissions,
then the intrinsic Balmer decrement will be higher and consequently
the deduced internal reddening lower.

\section{Results}
\label{results}

\begin{table*}
  \centering
  \caption{\small Integrated line properties. Also given is the 3-$\sigma$ upper-limit for 
    the {\si} line flux. The spatial extents are based on visual inspection.}
  \label{persparam}
  \begin{tabular}{c c c c c c c c c c}
    \hline
    Line &  $\lambda$ ($\mu$m)  & \multicolumn{2}{c}{Offset (km~s$^{-1}$)}    &  \multicolumn{2}{c}{FWHM (km~s$^{-1}$)} & Line Flux &\multicolumn{2}{c}{Spatial Extent (Radius)} \\
              &                                     & z$_{\st {bcg}}$        & z$_{\st{cl}}$         &  Obs.              &  Intrinsic       & ($10^{-18}$ W/m$^2$) &  ($\pp{}$)  &  (kpc) \\   
    \hline\hline
    {\oi}   &  64.298$\pm $0.002    &   39$\pm$9       & -61$\pm$9       & 383$\pm$20 & 375$\pm$20 &  2525.8$\pm$73.7 & $\pp{34}$    & 12 \\ 
    {\cii}  &  160.510$\pm$0.002   &   -0.6$\pm$4    & -100$\pm$4     & 419$\pm$9   & 347$\pm$11 & 2205.3$\pm$26.5  & $\pp{71}$    & 25\\
    {\nii}  & 124.031$\pm$0.004    &   -24$\pm$10   & -124$\pm$10   & 558$\pm$22 & 482$\pm$25 &  125.0$\pm$2.8     & $\pp{28}$    & 10 \\          
    {\oib} & 148.091$\pm$0.003    &   32$\pm$7       & -68$\pm$7       & 458$\pm$17 & 384$\pm$20 &  150.5$\pm$3.1     & $\pp{29}$  & 10 \\       
    {\oiii} &  89.900$\pm$0.005     &   -37$\pm$17   & 137$\pm$17     & 375$\pm$42 & 358$\pm$44 &  65.6$\pm$4.1       & $<\pp{9.4}$ &  $< 3.3$ \\       
    {\si}     & ...  & ...  & ...  & ... & ... &  $< 3.6$ & ... & ... \\                    
    \hline
  \end{tabular}
\end{table*}


\begin{figure*}
  \begin{minipage}{0.49\textwidth}
    \centering
    \includegraphics[width=\textwidth]{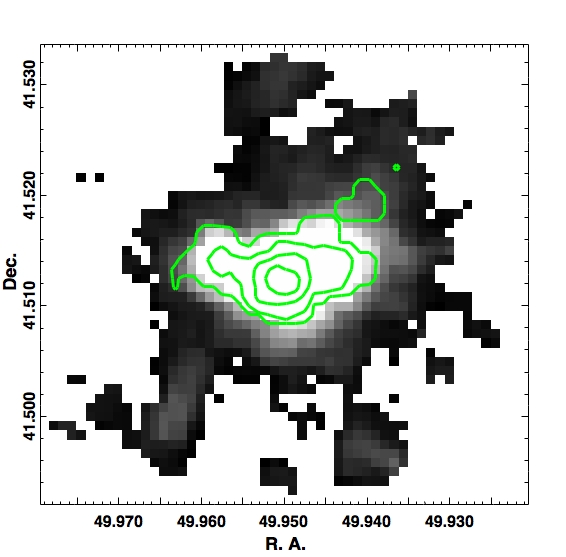}
  \end{minipage}%
  \begin{minipage}{0.49\textwidth}
    \centering
    \includegraphics[width=\textwidth]{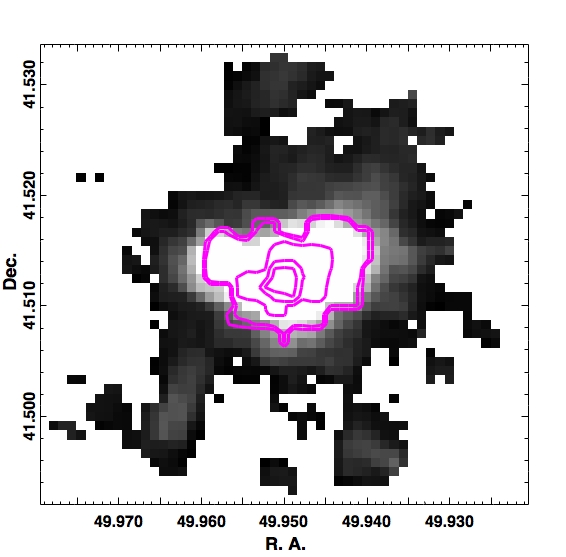}
  \end{minipage}\\
  \begin{minipage}{0.49\textwidth}
    \centering
    \includegraphics[width=\textwidth]{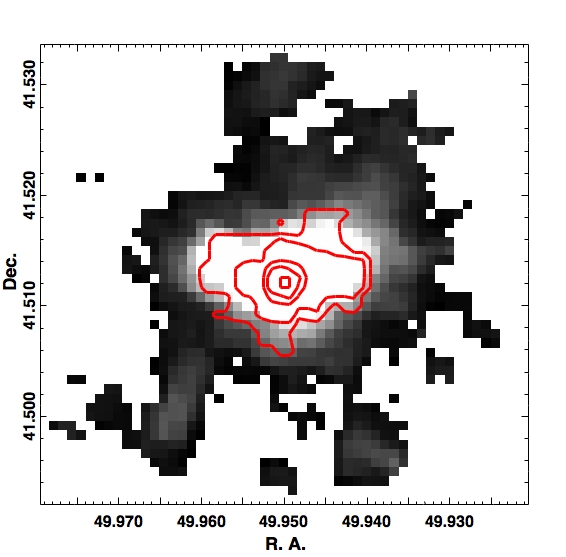}
  \end{minipage}%
  \begin{minipage}{0.49\textwidth}
    \centering
    \includegraphics[width=\textwidth]{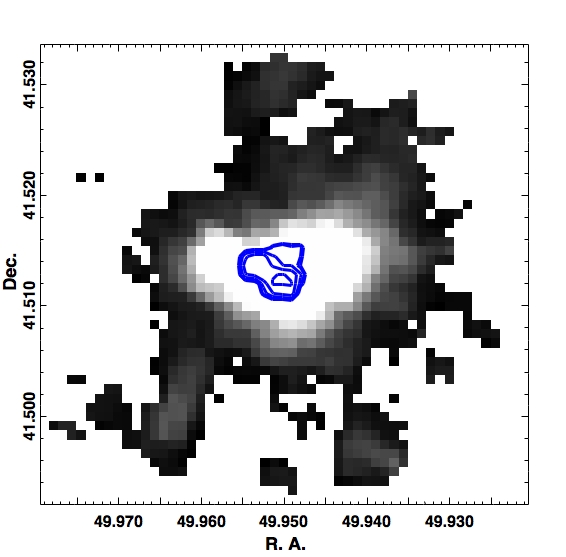}
  \end{minipage}
  \caption{\small A comparison of the Herschel emission
    lines. Overlaid on the {\cii} image (gray scale) are (starting
    from upper left corner and going clockwise) the {\oi} contours
    (green), {\nii} contours (magenta), {\oib} contours (red) and
    {\oiii} contours (blue). All emission lines have a pixel detection
    threshold of SNR$>2$.}
  \label{lines-overlay}
\end{figure*}

Of the six lines observed, we detected all except {\si} at 68.47{\mm}.
Even though only {\cii} and {\oi} observations were designed to detect
extended emission, all the detected lines except {\oiii} are spatially
extended. The integrated line profiles are shown in
Fig.~\ref{persspec} and their relative spatial extensions in
Fig.~\ref{lines-overlay}, where the pixel threshold has been set to
SNR$>$2. The SNR corresponds to the ratio of the line peak to the
standard deviation of the data about the fitted model. Listed in
Table~\ref{persparam} are the integrated line properties. 

\begin{figure*}
    \centering
    \includegraphics[width=1.1\textwidth]{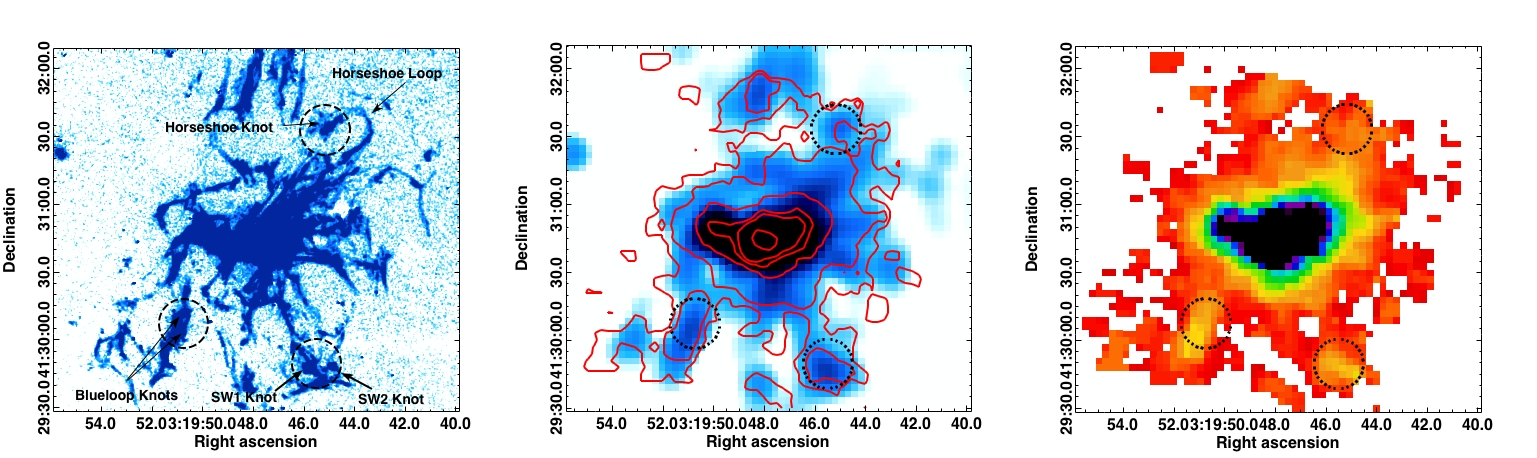}
    \caption{\small {\it: Left} The continuum-subtracted WIYN {\ha}
      line emission with a spatial resolution of $\pp{0.6}$ to
      $\pp{0.8}$ \citep{Conselice2001}. {Middle:} The {\ha} image
      smoothed to the resolution of {\cii}, with the {\cii} line
      contours overlaid in red.  {Right:} The Herschel 
      {\cii} line emission with a spatial resolution of $\pp{11}$
      (this work). The optical {\ha} emission and infrared {\cii}
      emission show a remarkably similar distribution. Marked are the
      Horseshoe knot in the northwest 20~kpc from the BCG, the SW1/SW2
      knots in the southwest 21~kpc from the BCG and the Blue Loop
      knots in the southeast 16~kpc from the BCG. }
  \label{CIIHalpha}
\end{figure*}

The {\cii} line is detected over a spatial region of extent $\pp{140}$
(50~kpc). Shown in the left panel of Fig.~\ref{CIIHalpha} is a
{\ha}+{\niioptii} image displaying the ionized gas filaments taken
with the WIYN~3.5~m telescope \citep{Conselice2001}. Shown in the
right panel of Fig.~\ref{CIIHalpha} is the FIR {\cii} emission with
the pixel detection threshold set to SNR $>1$. The middle panel
displays the {\ha} emission smoothed to match the resolution of the
{\cii} line using the {\sc ciao} tool {\sc`aconvolve'} and the {\sc
  pyraf} tool {\sc `blkavg'}, with the {\cii} line contours overlaid
(in red). The {\cii} emission traces the {\ha} emission very well
despite a much lower resolution. Both {\cii} and {\ha} reveal a
central elongation, about 20~kpc in total extent, with an east-west
alignment. This elongation is clearly visible also in the recent
narrow-band imaging of a ro-vibrational transition line of molecular
hydrogen (H$_2$) \citep{Lim2012}, which shows very good overall
morphological resemblance with {\ha} emission. In \cite{Mittal2011b},
we showed a close spatial and kinematical correspondence between
different emissions, such as {\ha} and {\cii}, in NGC~4696, the BCG of
the Centaurus cluster. {\pers} shows a similarly tight correlation,
both spatial and kinematical, between {\cii}, {\ha} and CO
(section~\ref{kin}), suggesting a common heating process of the gas.

There are three well-studied regions in the {\ha} filaments, seen also
in the {\cii} map, which we describe briefly below: (1) The Horseshoe
is a filament to the north-west of the center of BCG
\citep{Conselice2001}. The filament of gas appears to originate in the
core of the galaxy, rising a projected distance of 26 kpc through the
ICM before looping around towards the center again.  (2) The southwest
knots, at a projected distance of $\sim 23~$kpc from the center of the
BCG, are a part of the southern filaments. 3) The knots in the
southeast filament, also known as the `Blue Loop', are at a projected
distance of $\sim 16~$kpc. The loop is so called due to the blue
optical colour of the knots this region, indicating these are star
forming sites. \cite{Canning2010} studied this region in detail and
found a star formation rate of $\sim20~\mpy$. We refer to the knots as
the `Blue Loop' knots from now on. All three regions have been studied
in detail previously at several wavelengths
\citep{Conselice2001,Salome2006,Salome2008,Canning2010,Salome2011,Hatch2005,Hatch2006,Fabian2003,Fabian2008,Johnstone2007,Ferland2009}.

The three regions are marked in Fig.~\ref{CIIHalpha}.  We extracted
the {\cii} line spectra from them and the core region (see below). The
spectra are shown in Fig.~\ref{filaments}. The Horseshoe region is
weaker in {\cii} emission than the SW and Blue Loop knots. This is
consistent with the study of \cite{Salome2011}, where they find this
region has weaker CO emission in comparison with other
filaments. There is no detection of {\oi} in any of the three regions
of the filaments. The line strengths and ratios in these regions and
the core provide strong constraints for studying the excitation
mechanism(s) in different parts of the BCG. This is followed-up in
detail in section~\ref{disc}.

\cite{Brauher2008} report a flux of
$(2.5\pm0.2)\times10^{-15}$~W~m$^{-2}$ for {\oi} and
$(1.2\pm0.1)\times10^{-15}$~W~m$^{-2}$ for {\cii}, based on
observations with the Long Wavelength Spectrometer~(LWS) on the
Infrared Space Observatory~(ISO). While the {\oi} flux measurement of
$(2.53\pm0.07)\times10^{-15}$~W~m$^{-2}$ obtained in this study
compares well with the ISO measurement, our {\cii} flux of
$(2.21\pm0.03)\times10^{-15}$~W~m$^{-2}$ is higher by a factor of
two. The reason for this discrepancy is that \cite{Brauher2008} obtain
the {\cii} flux based on the assumption that {\pers} is a point
source. The FWHM of the LWS is $\pp{75}$. However, {\cii} emission
clearly extends beyond the FWHM of the LWS ($\sim \pp{140}$ in
diameter). For this reason, we believe that \cite{Brauher2008}
underestimate the line flux in {\cii} by about a factor of two. As a
rough check of this hypothesis, we convolved our {\cii} map with a
gaussian with a FWHM of $\pp{75}$ and obtained a flux of
$0.97\times10^{-15}$~W~m$^{-2}$ within a $\pp{75}$ aperture
diameter. Despite the fact that this test uses a simple gaussian
rather than the true ISO LWS beam profile, the reduction of measured
flux is consistent with the result of \cite{Brauher2008}.

\begin{figure*}
  \begin{minipage}{0.49\textwidth}
    \centering
    \includegraphics[width=0.8\textwidth]{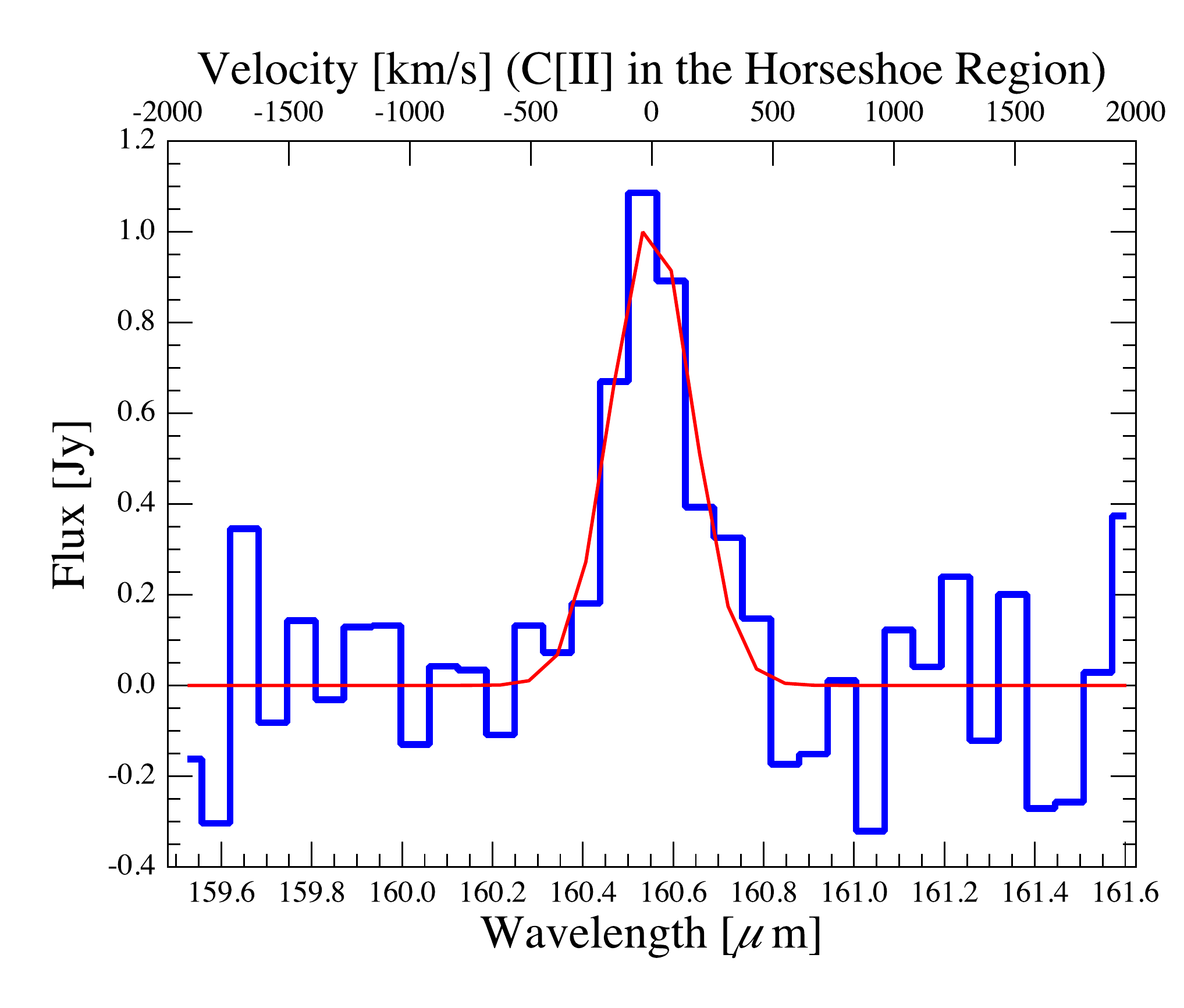}
  \end{minipage}%
  \begin{minipage}{0.49\textwidth}
    \centering
    \includegraphics[width=0.8\textwidth]{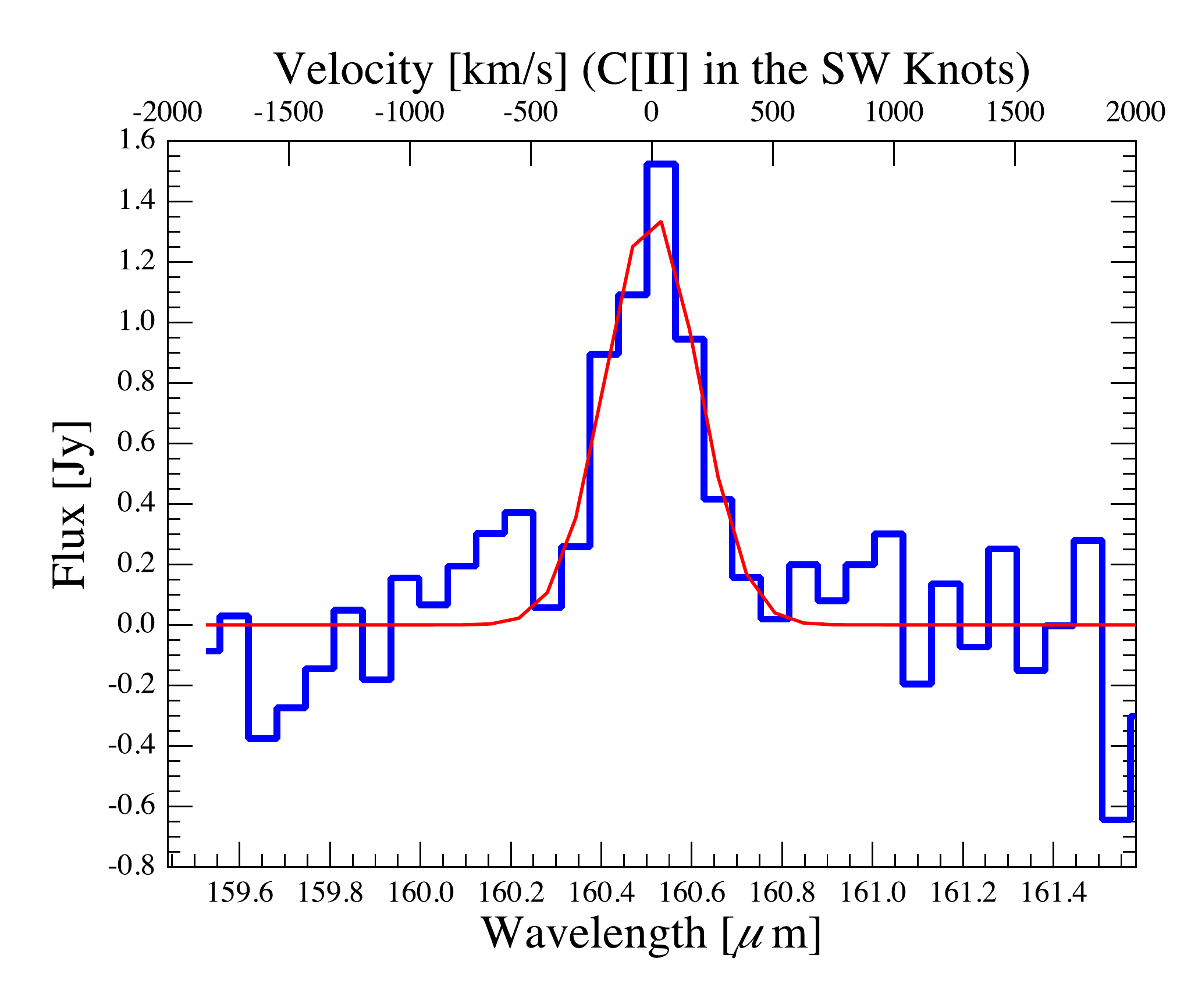}
  \end{minipage}\\
  \begin{minipage}{0.49\textwidth}
    \centering
    \includegraphics[width=0.8\textwidth]{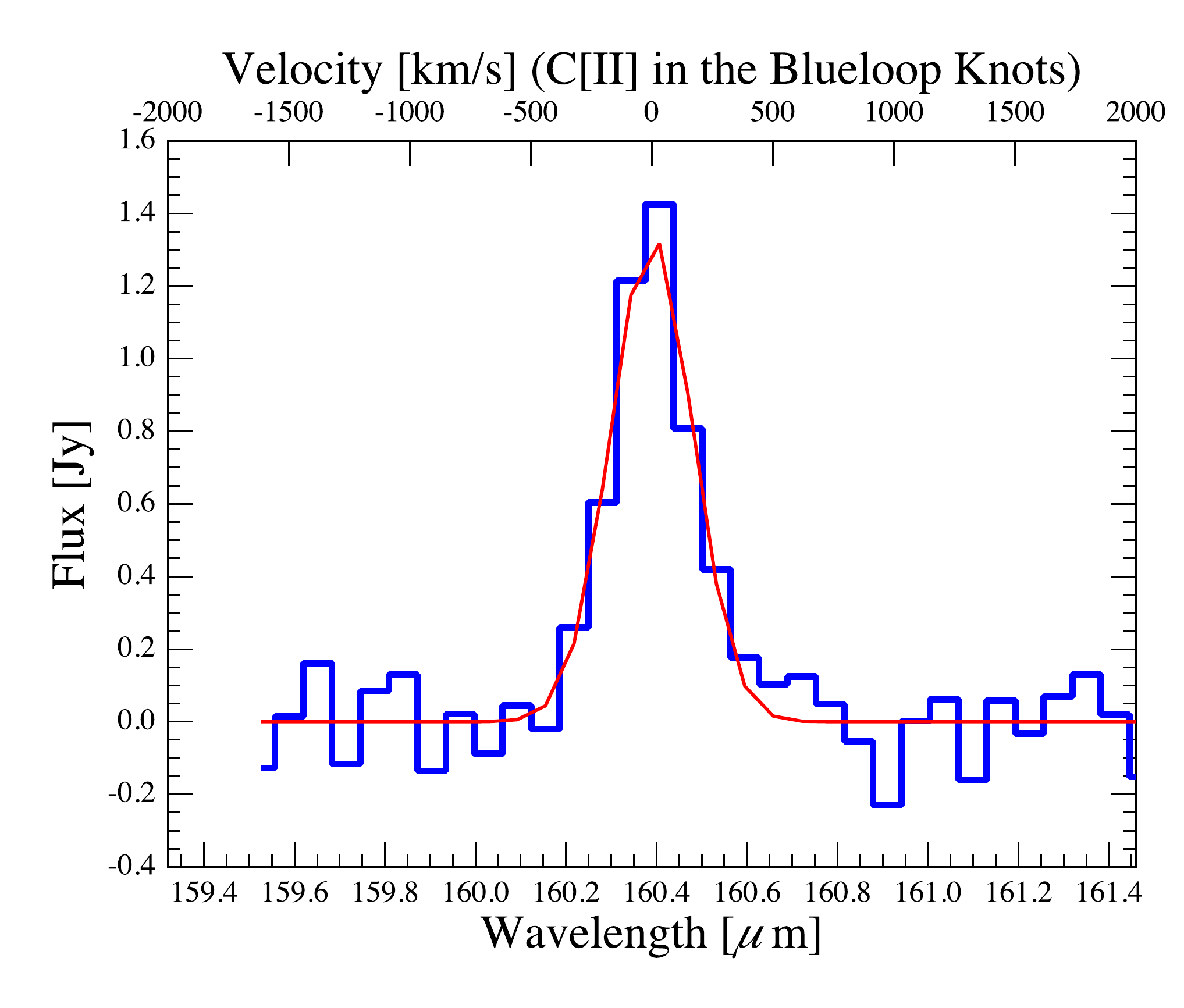}
  \end{minipage}%
  \begin{minipage}{0.49\textwidth}
    \centering
    \includegraphics[width=0.8\textwidth]{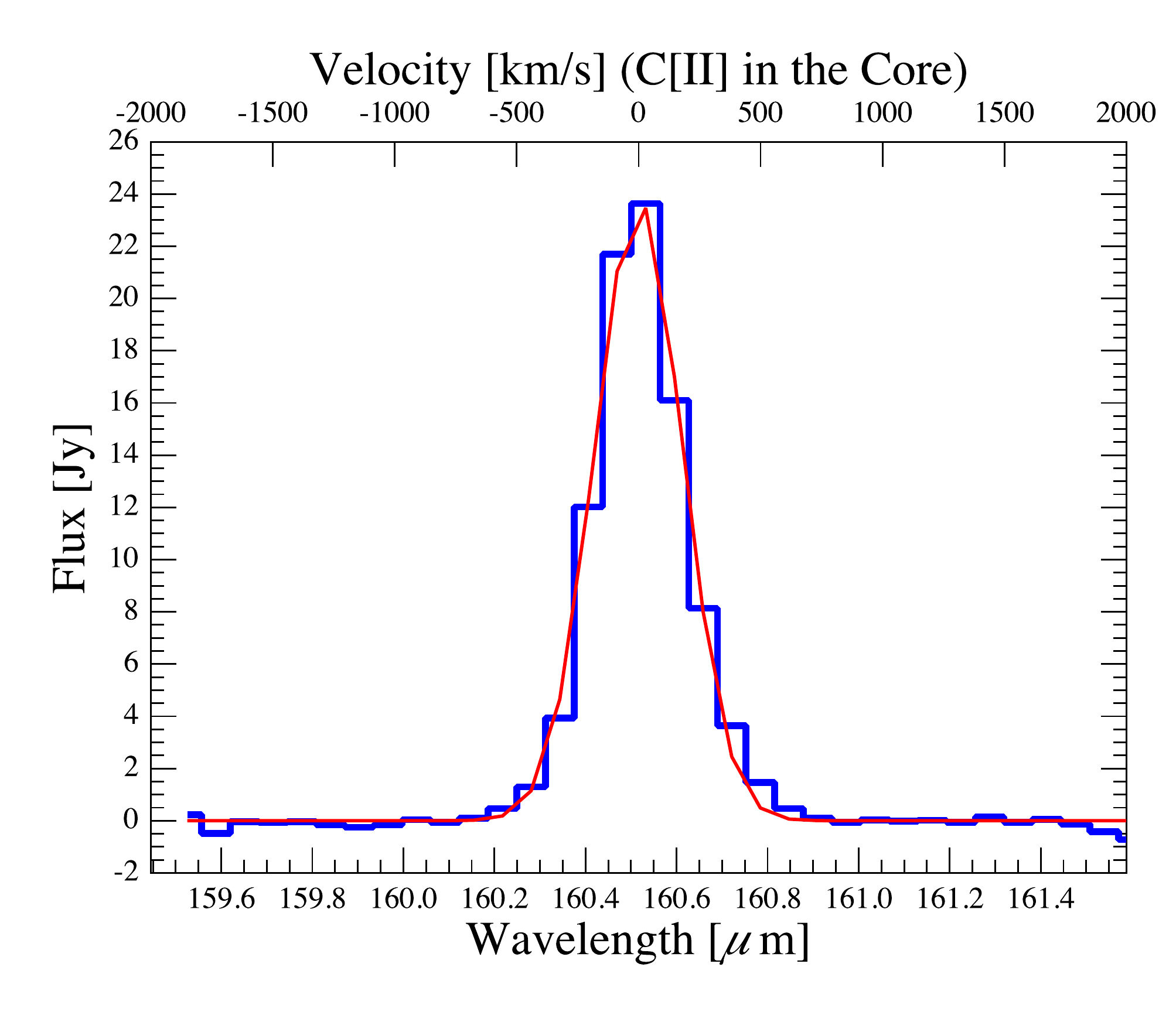}
  \end{minipage}
  \caption{\small The {\cii} emission from the Horseshoe knot in the
    northwest (upper left panel), the SW1/SW2 knots in the southwest
    (upper right panel) and the Blue Loop knots in the southeast
    (lower left panel). These apertures are marked in
    Fig.~\ref{CIIHalpha}. We also show the {\cii} spectrum for the
    core region (lower right panel), marked in the upper left panel of
    Fig.~\ref{ciivelocity}.}
  \label{filaments}
\end{figure*}

\subsection{Kinematics}
\label{kin}

\begin{figure*}
  \begin{minipage}{0.49\textwidth}
    \centering
    \includegraphics[width=\textwidth]{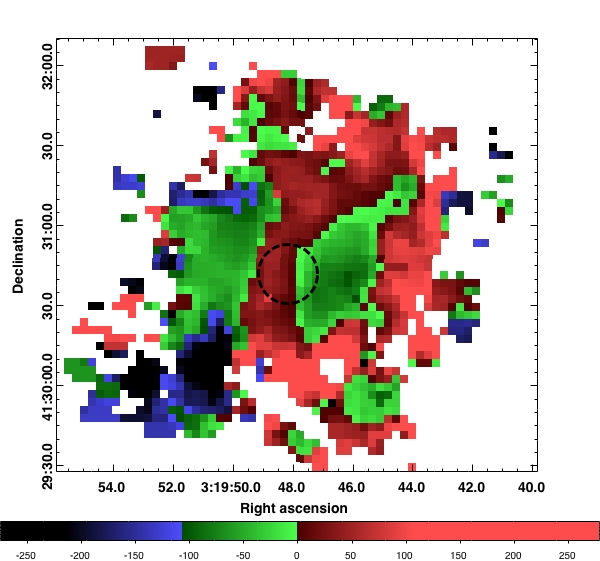}
  \end{minipage}%
  \begin{minipage}{0.49\textwidth}
    \centering
  \vspace*{-0.75cm}
    \includegraphics[width=\textwidth]{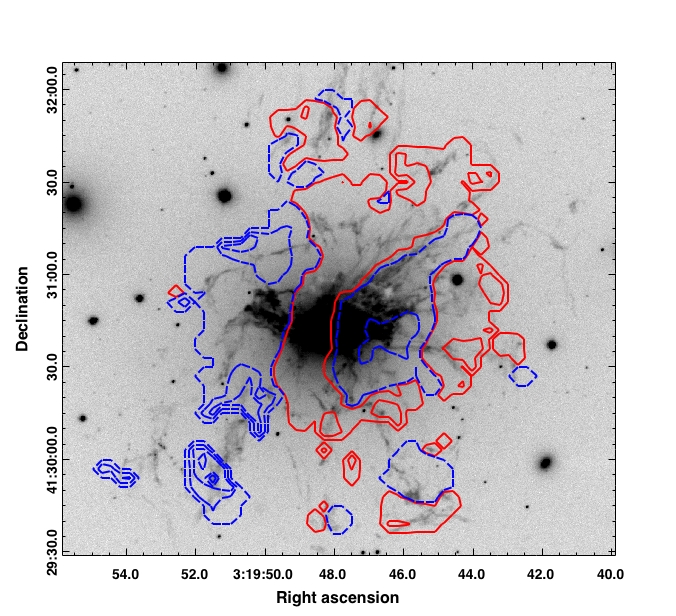}
  \end{minipage}\\
    \centering
    \includegraphics[width=\textwidth]{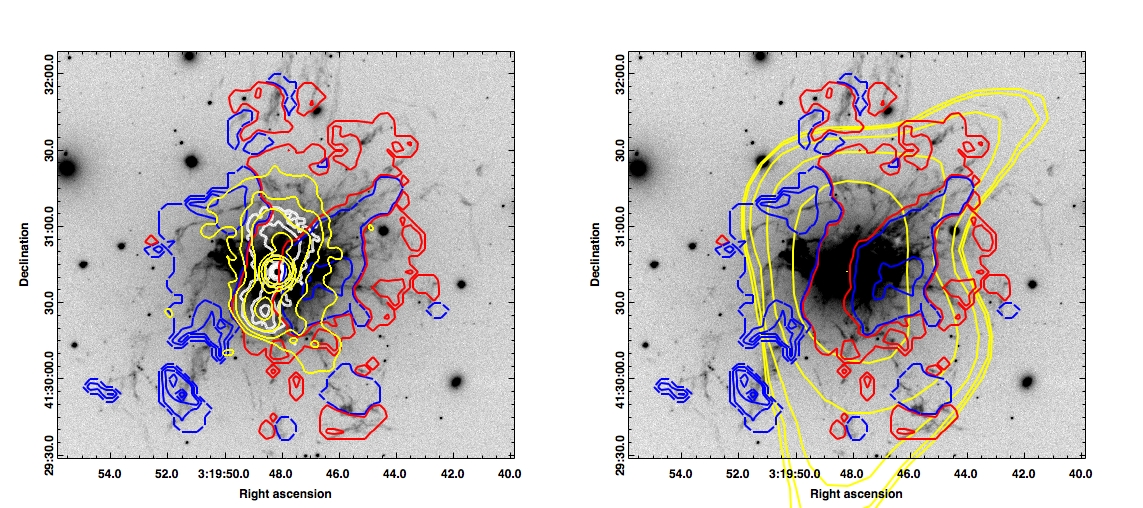}
    \caption{\small {\it Upper Left:} The detailed velocity
      distribution of the {\cii} gas. The colour-bar represents the
      velocities in km~s$^{-1}$. The dashed circle represents gas
      compatible with rotation. {\it Upper Right: } The velocity
      contours superimposed on the {\ha} image. The dashed blue
      contours correspond to blueshifted (negative) velocities and the
      solid red contours correspond to the redshifted (positive)
      velocities. {\it Bottom Left}: 1.4~GHz radio contours at two
      different resolutions, $\pp{2}$~(white) and $\pp{5}$~yellow,
      overlaid on the {\ha} image. {\it Bottom Right}: 74~MHz radio
      contours at $\pp{25}$ resolution~(yellow) overlaid on the {\ha}
      image. }
  \label{ciivelocity}
\end{figure*}

Fig.~\ref{ciivelocity}, the {\cii} velocity distribution shows the
full extent of the complex kinematical structure of the filaments in
{\pers}. The colour scheme is such that the red shaded regions
represent redshifted gas (with positive velocities with respect to the
systemic velocity of the BCG), whereas green and blue shaded regions
represent blueshifted gas (with negative velocities with respect to
the systemic velocity of the BCG). The velocity pattern is likely to
be the combination of inflowing and outflowing gas, along with
projection and small-scale rotation effects (see below).

The dashed circle at the center marks the `core' region (see
Table~\ref{coreratios}), which shows a gradient in the {\cii} line
velocity. \cite{Wilman2005} conducted near-infrared spectroscopy of
the central region in NGC~1275 with the UIST IFU instrument on the
United Kingdom Infrared Telescope. They detected ro-vibrational H$_2$
emission originating from a region $\sim 50~$pc from the nucleus, with
the results indicating a strong velocity gradient in the peak position
of the H$_2$ line. The measured rotation curve in the central IFU slit
placed east-west along the nucleus shows a velocity change from
150~km~s$^{-1}$ east of the nucleus to -100~km~s$^{-1}$ west of the
nucleus. From the sharp decrease in the magnitude of the velocity on
either side of the nucleus, they concluded that the molecular gas is
distributed in a disk-like structure with the rotation axis oriented
north-south. This is also consistent with the double-horn structure
seen in the CO(2-1) spectrum extracted from the same core region
\citep{Salome2011}, which may be interpreted as indirect evidence for
a central rotating disk. The resolution of the {\cii} data is not good
enough to resolve any rotational structure over the scales observed by
\cite{Wilman2005}. The smallest structure in {\cii} that can be
resolved is about 4~kpc. Similarly, the spectral resolution of the
{\cii} data (230~km~s$^{-1}$) is much poorer than that of the CO(2-1)
line observed with the IRAM 30~m telescope on Pico Veleta
(40~km~s$^{-1}$). From Fig.~\ref{ciivelocity}, the rotational pattern
claimed by \cite{Wilman2005} and \cite{Salome2011} is not
surprising. However, it is difficult to say whether the double horn
feature is due to small-scale disk-like rotation or large-scale flows.
If there is a rotating disk present in the center, it involves a
relatively small gas disk of $\sim 5$~kpc radius. The majority of the
gas distribution sampled by {\cii} does not show any regular rotation.

Also interesting is the redshifted ridge of gas passing through the
center of the BCG with a north-south extension. The CO velocity
measurements made by \cite{Salome2011} in regions 4 and 21 marked in
Fig.~1 of their paper provide a confirmation of the presence of this
redshifted gas \citep[also see][]{Lim2008}. This wide vertical
distribution of gas has a cylindrical symmetry and is suggestive of
material being dragged upward and downward.

\begin{figure}
    \centering
    \includegraphics[width=\columnwidth]{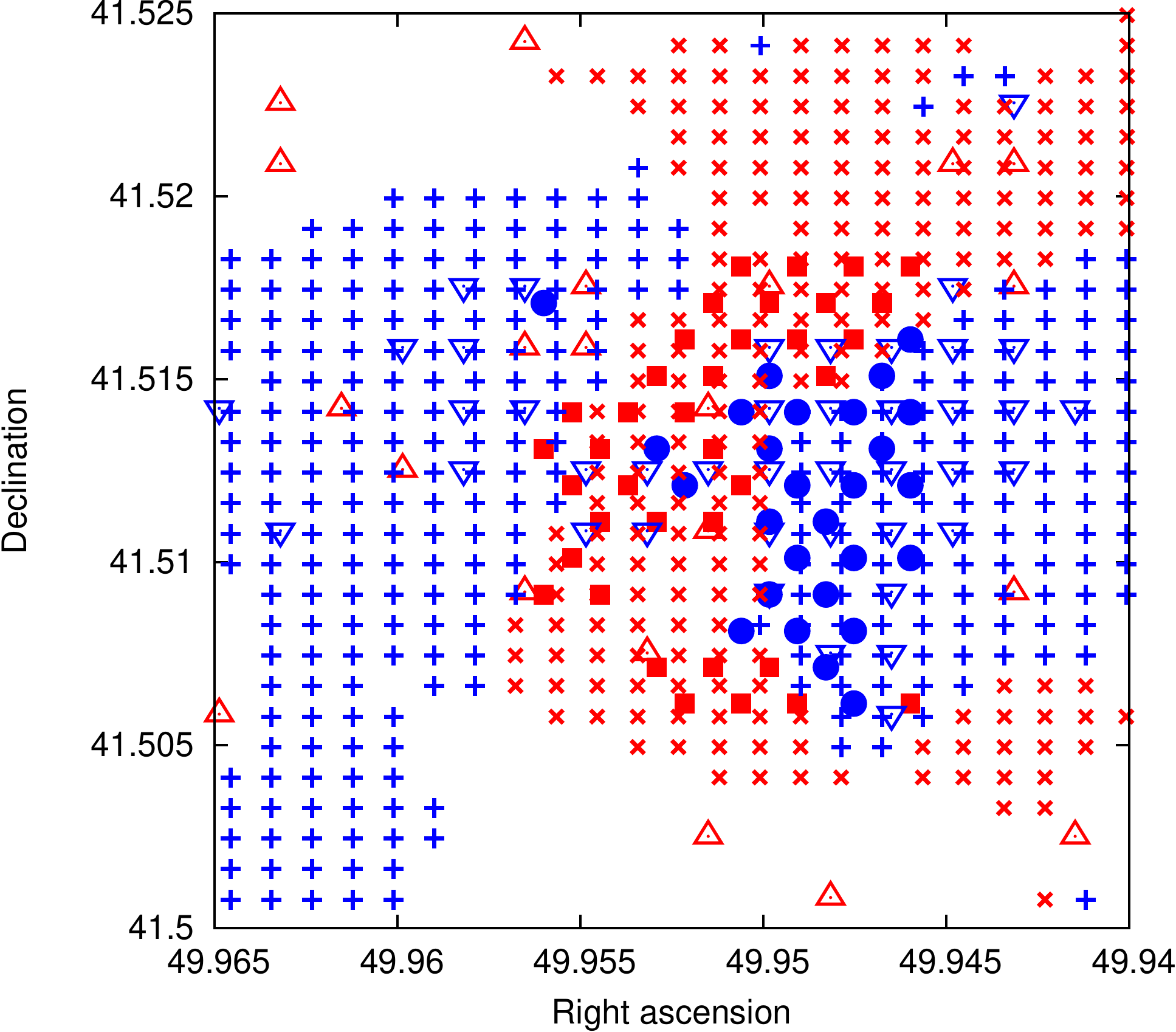}
    \caption{\small The kinematics of the three phases, optical,
      far-infrared and millimeter, represented by {\ha}, {\cii} and
      CO(2-1) line emissions. Show are the redshifted (red symbols)
      and blueshifted (blue symbols) velocities for the three
      phases. The {\cii} velocities are shown as plus (blueshifted)
      and cross (redshifted) symbols, the {\ha} velocities are shown
      as filled circles (blueshifted) and squares (redshifted) and
      CO(2-1) velocities are shown as open down-triangles
      (blueshifted) and up-triangles (redshifted).}
  \label{ciihaco}
\end{figure}

The upper left panel of Fig.~\ref{ciivelocity} shows that there is a
negative velocity region on either side of the center along the
east-west direction. The east region extends all the way along to the
south, where it merges with the Blue Loop knots. The blueshifted
components on both sides of the major axis of the emission are also
visible in the kinematics inferred from the CO spectra
\citep{Salome2006}.  The CO flux and velocity distribution shows a
close association with the optical filaments seen in {\ha}
\cite[e.g.][]{Salome2006}. In Fig.~\ref{ciihaco}, we show the
kinematics of the gas in the inner $\pp{90}$x$\pp{90}$ in three phases
-- optical, FIR and millimeter -- as represented by {\ha} [taken from
\cite{Conselice2001}], {\cii} and CO(2-1) [taken from
\cite{Salome2006}] line emissions. This plot shows a clear overlap in
the redshifted gas, with absolute velocities greater than
5264~km~s$^{-1}$ (red symbols), along the ridge associated with all
three line emissions. Similarly, there is an overlap in the
blueshifted gas, with absolute velocities less than 5264~km~s$^{-1}$
(blue symbols), west of the ridge. Especially remarkable is the
curvature in the ridge seen in both {\ha} (red filled squares) and
{\cii} (red crosses). From this we conclude a kinematical correlation
among {\cii}, {\ha} and CO emissions, which further reinforces the
idea that different emissions have the same origin.

Radio observations of central radio sources often provide useful
insights in understanding the kinematics of cool-core BCGs. Overlaid
on the bottom panels of Fig.~\ref{ciivelocity} are contours of the
radio emission associated with 3C~84 at 1.4~GHz and 74~MHz. The bottom
left panel shows 1.4~GHz radio emission at two different resolutions
-- $\pp{2}$~(white contours; courtesy of VLA/NRAO\footnote{The
  National Radio Astronomy Observatory is a facility of the National
  Science Foundation operated under cooperative agreement by
  Associated Universities, Inc.}) and $\pp{5}$~(yellow; kindly
provided by Greg Taylor). The higher resolution contours reflect a
north-south radio morphology, demonstrating a good alignment with the
redshifted {\cii} contours. The lower resolution contours reflect an
inverted-S shaped morphology, previously noted in several studies
\citep[e.g.][]{Pedlar1990,Boehringer1993,Fabian2000}. Although the
southern jet shows a correlation with the {\cii} emission, the tip of
the northern jet does not. The 74~MHz contours (also kindly provided
by Greg Taylor), shown in yellow in the bottom right panel, indicate a
reversal of the east-west component of the jet direction on both sides
of the core, wherein the northern jet once again overlaps with the
{\cii} emission. \cite{Conselice2001} noted a similar alignment
between the linear extension of the {\ha} filaments to the north and
the low-brightness radio emission seen at low frequencies.

Several studies have presented scenarios wherein radio outbursts are
responsible for the dredge-up of cold, metal-rich gas from the core in
the direction of the buoyantly rising radio plasma
\citep[e.g.][]{Simionescu2008,Simionescu2010,Gitti2011,Revaz2008,Tremblay2012b,Tremblay2012a}.
Evidence for these scenarios is based on the X-ray-derived temperature
and metallicity maps, which show a spatial correlation between radio
emission and cool gas extending away from the core with a metal
content higher than that of the ambient medium. The positive
correlation between the radio emission and the redshifted ridge of
{\cii} gas in {\pers} is reminiscent of cold gas being dredged up by
the radio lobes. On the other hand, this interpretation implies that
both the radio jets (jet and counter-jet) are pointed away from the
line-of-sight. On milliarcsec scale, the radio morphology comprises a
one-sided jet, such that the southern jet is deemed to be approaching,
and the northern jet (counter-jet) receding
\cite[e.g.][]{Pedlar1990,Vermeulen1994,Dhawan1998,Taylor2006}. Both
the radio jets reveal complex kinematics in the plane of the sky. The
southern jet is initially elongated along position angle~(PA) $\sim
235^{\circ}$ (at $\sim 10$~mas) but suddenly bends toward
PA$\sim160^{\circ}$ (at $\sim 20$~mas) and continues though a series
of such bends, at approximately the same angles, out to about an
arcminute \citep{Pedlar1990,Dhawan1998}. The northern jet exhibits a
similar complex structure. On the kiloparsec scale, it is therefore
possible that, similar to the kinks observed in the plane of the sky,
the jets undergo bends along the line-of-sight, such that both the
jets are receding. There are dredge-up interpretations offered for
{\pers}, suggesting that {\ha} gas is being dredged up by the radio
source.  \cite{Fabian2003} recognized two filaments in the north-west,
including the Horseshoe, bent on either side of the north-west ghost
bubble. They showed that the {\ha} emission associated with the
Horseshoe is just behind the bubble and is likely dragged out by
it. Similarly, \cite{Sanders2005} discovered a high-abundance ridge
using Chandra observations, which they hypothesized is formed by
material entrained by a fossil radio bubble.

\begin{figure}
    \centering
    \includegraphics[width=0.4\textwidth]{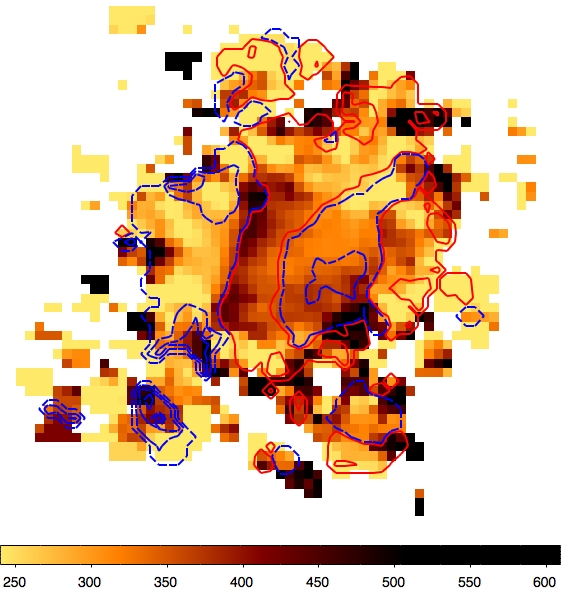}
    \caption{\small The {\cii} linewidth in km~s$^{-1}$. The dashed
      blue contours correspond to blueshifted (negative) {\cii}
      velocities and the solid red contours correspond to the
      redshifted (positive) {\cii} velocities (same as the upper right
      panel of Fig.~\ref{ciivelocity}).}
  \label{ciilinewidth}
\end{figure}

Lastly, the {\cii} velocity structure may be related to the disordered
motion of gas clouds at larger radii that is not affected by the inner
radio structure. Shown in Fig.~\ref{ciilinewidth} is the {\cii}
linewidth in km~s$^{-1}$. An interesting feature of this map is that
one of the regions the linewidth peaks is adjacent to the redshifted
ridge where the line velocity flips from positive to negative on the
eastern side of the galaxy. This is where the CO(3-2) HARP maps shows
double lines implying that there are multiple gas components along the
line of sight (Edge et al., in prep.). However, the linewidths in this
region may also be high due to the superposition of rapidly varying
line-of-sight velocity elements.

\subsection{Spatial variation of {\oi} to {\cii} ratio}
\label{ciioi}

\begin{figure}
    \centering
    \includegraphics[width=0.4\textwidth]{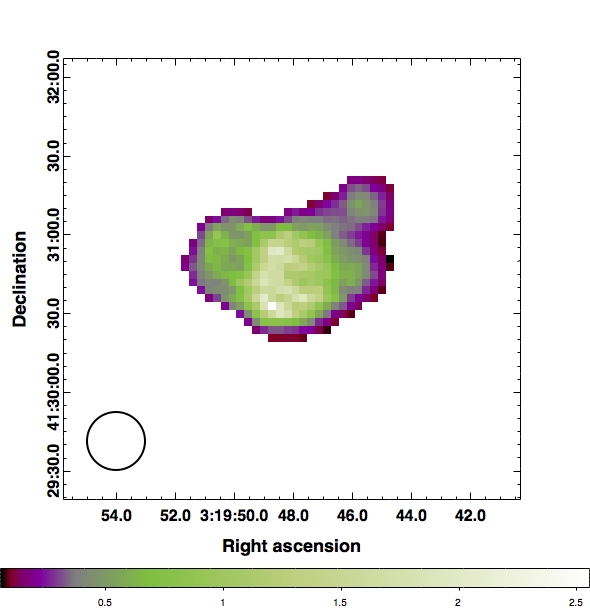}
    \caption{\small {\oi} to {\cii} ratio. {\oi} has been convolved
      with a gaussian such that the resulting full-width at half
      maximum~(FWHM) equals that of the PSF of the PACS spectrometer
      at the wavelength of {\cii}. The circle shown in the bottom left
      corner represents the size of the FWHM. The ratio map shows that
      {\oi} is stronger than {\cii} in the core. The trend reverses at
      $\sim 4$~kpc from the center.}
  \label{oicii}
\end{figure}

Here we briefly discuss the spatial variation of {\oi} to {\cii} ratio
and what may be inferred from it. The resolution of {\cii} line is
slightly poorer than the {\oi} line. For compatibility, we convolved
the {\oi} line map with a gaussian such that the resulting FWHM is
similar to that of the {\cii} line. Fig.~\ref{oicii} is obtained by
dividing the resulting smeared {\oi} map by the {\cii} map (the
threshold for both {\cii} and {\oi} was set to SNR$>$2). This figure
indicates that while {\oi} is stronger than {\cii} in the core, the
opposite is true at larger radii, namely, the {\cii} line emission
becomes stronger at radii larger than $\sim$4~kpc, which corresponds
to the spatial resolution of {\cii}.

The relative strength of {\oi} to {\cii} is an indicator of gas
density, such that a higher {\oi}/{\cii} ratio represents a higher
density gas for the reasons described in section~\ref{best-fit}. Hence
a higher {\oi}/{\cii} ratio in the center implies a relatively denser
gas in the cluster core, as expected.

\subsection{Dust SED}
\label{dustsed}

\begin{table}
  {\small
  \centering
  \caption{\small A compilation of the fitted radio, sub-mm and IR flux-densities for {\pers}. 
    The columns are (1) wavelength (2) instrument (3) year of the observation (4) aperture 
    (available only for this work) and (5) the measured flux-density. Note that the measurements
    correspond to the total flux-densities for the given instrument.}
  \label{compilation}
  \begin{tabular}{| c | c | c | c | c|}
    \hline
    $\lambda$ ($\mu$m)  & Instrument & Year & Aperture  &  Flux (mJy) \\
    \hline\hline
    20     & SPITZER IRS$^{\st d}$        &  2004 & --  &  2410$\pm$241 \\ 
    25     & IRAS$^{\st c}$                   &  1983 & --  &  3539$\pm$176 \\ 
    30     & SPITZER IRS$^{\st d}$        &  2004 & --  &  3820$\pm$382 \\ 
    60     & IRAS$^{\st c}$                   &  1983 & --  &  7146$\pm$286 \\ 
    70     & PACS Herschel$^{\st a}$   &   2011 & $\pp{55}$  &  7405$\pm$741 \\ 
    100   & PACS Herschel$^{\st a}$   &   2010 & $\pp{55}$  &  8541$\pm$854  \\ 
    160   & PACS Herschel$^{\st a}$   &   2010 & $\pp{55}$  &  6979$\pm$1396   \\
    250   & SPIRE Herschel$^{\st a}$   &   2010 & $\pp{60}$  &  3805$\pm$571 \\ 
    350   & SPIRE Herschel$^{\st a}$   &   2010 & $\pp{70}$  &  3095$\pm$464 \\ 
    500   & SPIRE Herschel$^{\st a}$   &   2010 & $\pp{112}$  &  2992$\pm$449 \\
    1153    & IRAM PdB$^{\st b}$        &   2010 & --  &  7220$\pm$1083 \\
    1303    & IRAM PdB$^{\st b}$        &   2010 & --  &  7741$\pm$1161 \\
    1428    & IRAM PdB$^{\st b}$        &   2010 & --  &  7299$\pm$1095 \\
    2710    & IRAM PdB$^{\st b}$        &   2010 & --  &  10000$\pm$1500 \\
    2913    & IRAM PdB$^{\st b}$        &   2010 & --  &  10713$\pm$1607 \\
    3019    & IRAM PdB$^{\st b}$        &   2010 & --  &  10330$\pm$1550 \\
    20675  & UMRAO$^{\st e}$           &   2010 & --  &  23230$\pm$130 \\
    \hline 
    \multicolumn{5}{p{8cm}}{{\bf a} This work. The errorbars correspond to the absolute 
      flux uncertainties: 10\% at PACS BS and BL and 20\% at R, and 15\% at all SPIRE 
      wavelengths; {\bf b} \cite{Trippe2011}; {\bf c} \cite{Moshir1990}; 
      {\bf d} \cite{Weedman2005}; {\bf e} University of Michigan Radio Astronomy Observatory
      data for year 2010 (courtesy of M.~Aller and H.~Aller).} \\
  \end{tabular}
}
\end{table} 

\begin{figure*}
  \begin{minipage}{0.4\textwidth}
    \centering
  \centering
    \includegraphics[width=\textwidth]{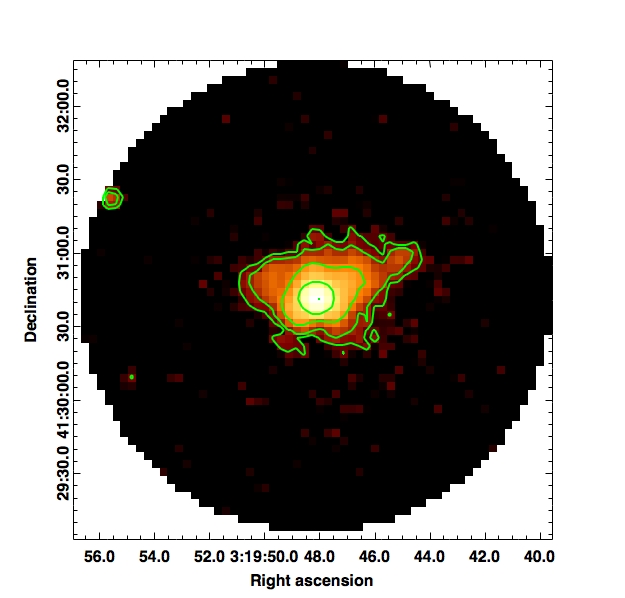}
  \end{minipage}%
  \begin{minipage}{0.4\textwidth}
    \centering
    \includegraphics[width=\textwidth]{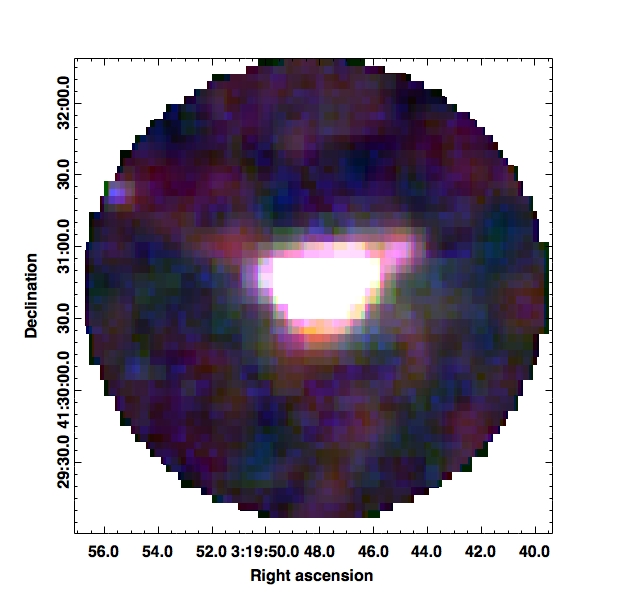}
  \end{minipage}%
  \caption{\small Dust emission. {\it Left:} PACS 100{\mm} image at a
    resolution of $\pp{7}$ showing the extended dust features.  {\it
      Right:} PACS images convolved with a gaussian to yield a FWHM of
    $\pp{12}$. The 70{\mm} image is shown in blue, the 100{\mm} image
    in green and the 160{\mm} image in red.}
  \label{dustemission}
\end{figure*}

We detected emission at all three PACS wavelengths and all three SPIRE
wavelengths. The PACS images are shown in
Fig.~\ref{dustemission}. Dust emission can usually be modelled as a
simple modified black body function; {\pers}, though, poses a
complication. This is because of the strong radio source at the center
of {\pers}, with a large contribution in the sub-millimeter and FIR
range \citep{Irwin2001}. The radio source, additionally, shows a
large-amplitude variability on timescales of decades \citep[see][ and
references therein]{Nagai2012}. Detailed monitoring at 3~mm has shown
an increase in flux from 3.5~Jy in 2002 to 11~Jy in 2010
\citep{Trippe2011}.

\begin{figure}
  \centering
  \includegraphics[height=0.3\textheight]{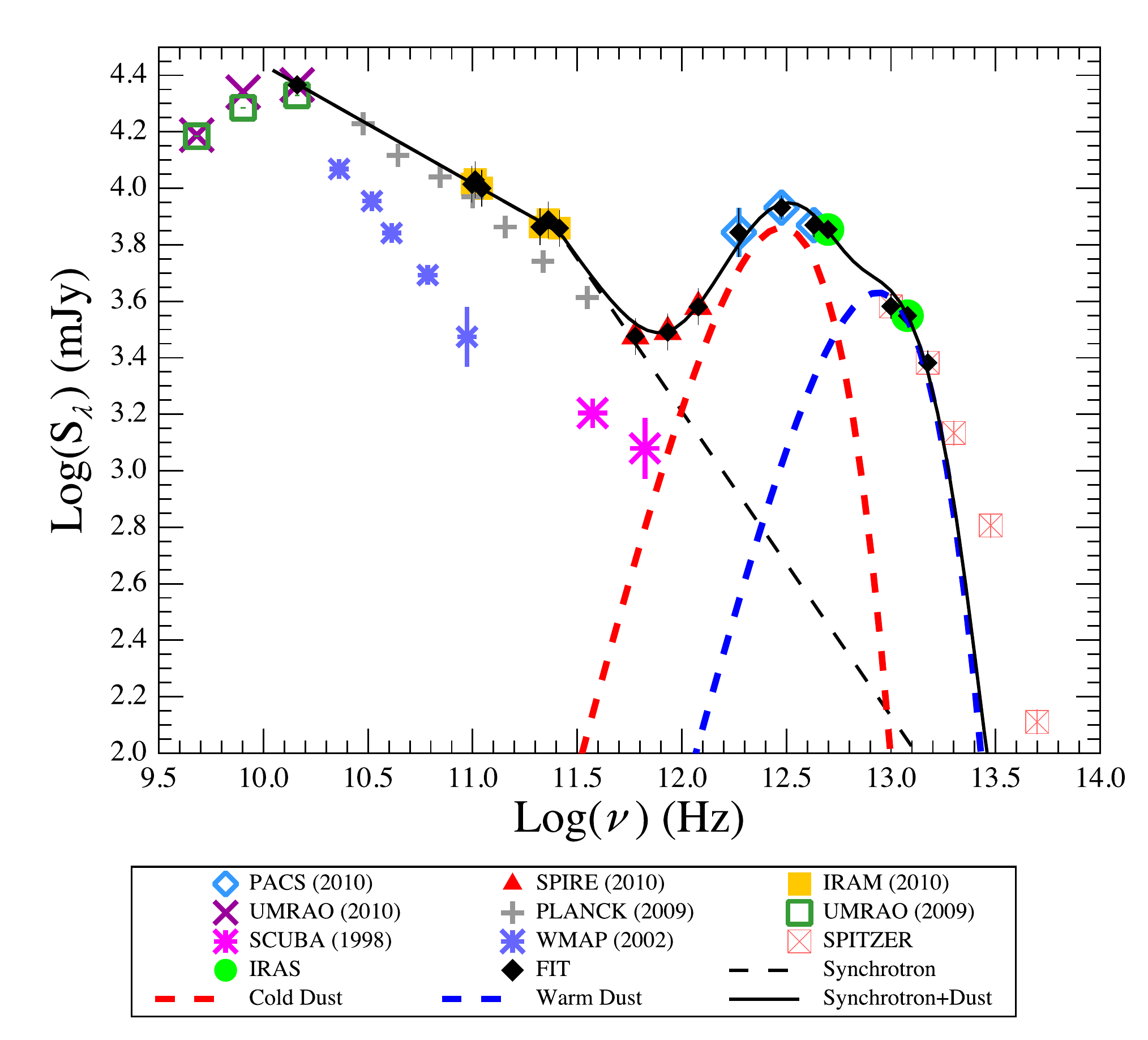}
  \caption{\small The spectral energy distribution for {\pers}. The
    dust emission dominates at FIR wavelengths, whereas the AGN
    synchrotron emission dominates at sub-mm wavelengths. The points
    overlaid with black diamonds were used to find the best-fit
    parameters of the model.}
  \label{SED}
\end{figure}

In order to obtain accurate dust parameters, it is essential to
estimate the fraction of the FIR flux originating from the synchrotron
emission from 3C~84.  In the following, we attempt to simultaneously
fit the dust and AGN emissions. We list in Table~\ref{compilation} and
show in Fig.~\ref{SED} the various flux measurements (black filled
diamonds) that were used to obtain an optimal model. While the dust
contribution to the SED is constrained by the PACS (blue open diamond)
and SPIRE (red filled triangles) data, the synchrotron contribution is
constrained by radio and sub-mm data. The two sub-sets of sub-mm
flux-densities clustered around 1~mm and 3~mm (yellow filled squares)
correspond to the IRAM Plateau de Bure Interferometer
\citep{Trippe2011} observations. These measurements are from a
monitoring program and chosen so as to be closest in time (spanning
days between 2010-08-21 to 2010-08-29) to the SPIRE and PACS
observations. The 5~GHz, 8~GHz and 15~GHz radio measurements (purple
diagonal-plus symbols) correspond to the observations made in 2010
with the 26-m telescope at the University of Michigan Radio Astronomy
Observatory~(UMRAO, courtesy of M.~Aller and H.~Aller). We used only
the 15~GHz data point for the fitting since the spectrum turns over
below this frequency, indicative of synchrotron self-absorption. The
50{\mm} and 25{\mm} measurements (green filled circles) correspond to
the IRAS observations \citep{Moshir1990}. To obtain a robust estimate
of the temperature and mass of the plausible second (warm) dust
component, we also used the 20{\mm} and 30{\mm} continuum data (red
open crossed-squares) determined from the Spitzer IRS observations
\citep{Weedman2005}. We did not fit 6{\mm}, 10{\mm} and 15{\mm} data
due to a possibly increasing contribution from a passively evolving
population of stars at wavelengths $\lesssim 15${\mm}. The AGN
synchrotron contribution to the total emission is expected to be small
at wavelengths shorter than 70{\mm} and so the variation in the flux
density due to the different times of the IRAS/Spitzer observations
from the rest of the fitted data can be neglected.

To model both the components - the dust emission and AGN synchrotron
emission - we used the following fitting function:

\begin{equation}
S_{\nu} = S_{\st{syn}, \nu} + S_{\st{dust}, \nu} \; , \quad \st{where}\\
\label{FIRemission}
\end{equation}
\begin{equation}
 S_{\st{syn}, \nu} = \left\{ \begin{array}{rl}
 S_0 \left( \frac{\nu}{\nu_0} \right)^{\alpha_1}   &\mbox{ if $\nu < \nub$} \\
 S_0 \left( \frac{\nub}{\nu_0} \right)^{\alpha_1-\alpha_2} \left( \frac{\nu}{\nu_0} \right)^{\alpha_2}  & \mbox{ if $\nu \ge \nub$}
      \end{array} \right.
\label{syn}
\end{equation}
\begin{equation}
  S_{\st{dust}, \nu} = \frac{\Omega}{(1+z)^3} \, \left[ B_{\nu}(\td) - B_{\nu}(\tcmb) \right] \, (1- \st{e}^{-\tau_{\nu}(\md)}),
\label{dust}
\end{equation}
Eqn.~\ref{syn} quantifies the synchrotron emission assumed to have a
broken powerlaw form. $S_0$ denotes the normalization at
$\nu_0=$100~GHz (3~mm), $\alpha_1$ and $\alpha_2$ and the (negative)
powerlaw indices on either side of the powerlaw break frequency,
$\nub$, respectively ($\alpha_1$ representing the powerlaw at radio
frequencies and $\alpha_2$ representing the powerlaw at submm/FIR
frequencies). We find that for {\pers} it is not necessary to modify
the synchrotron emission with an exponential term like
\cite{Privon2012} did to model the spectrum of Cygnus~A. The need for
including an exponential term in the case of Cygnus-A arises from the
lack of synchrotron emission beyond the submm regime. {\pers}, on the
other hand, shows evidence of an AGN contribution all the way out to
submm and, possibly, IR wavelengths \citep{Krabbe2000}. Hence, we
preferred a broken powerlaw, such that $|\alpha_2| > |\alpha_1|$, to
an exponential term to represent the slowly decaying synchrotron
emission.

 \begin{table*}
   \centering
   \caption{\small The best-fitting model parameters of the function describing 
     the FIR emission (Eqn~\ref{FIRemission}) for the dust emissivity index, $\beta=1$. 
     For comparison, we also give the parameters for $\beta=1.5$ and $\beta=2$. 
     The uncertainties correspond to the 1-$\sigma$ errorbars.} 
   \label{sed-params}
    \begin{tabular}{| c | c | c | c |}
     \hline
      Parameter   &  $\beta=1.0$ & $\beta=1.5$ &   $\beta=2.0$ \\
      \hline
      \hline
     Cold dust temperature,  $\tdc$                     & (38.0$\pm$2.0)~K                                 & (35.0$\pm$1.6)~K                               & (31.8$\pm$1.3)~K              \\
     Cold dust mass,  $\mdc$                              & $10^{6.96\pm0.12} \ms$                           & $10^{6.91\pm0.12} \ms$                          & $10^{6.89\pm0.12} \ms$        \\
     Warm dust temperature,  $\tdw$                   & (115.5$\pm$8.5)~K                              & (107.0$\pm$7.4)~K                              & (95.7$\pm$6.0)~K            \\ 
     Warm dust mass,  $\mdw$                            & $10^{4.71\pm0.17} \ms$                           & $10^{4.37\pm0.17} \ms$                         & $10^{4.60\pm0.18} \ms$         \\
     Synch. powerlaw norm.,  S$_{0}$                     & (10434$\pm$393)~mJy                       &  (10364$\pm$393)~mJy                        &    (10448$\pm$393)~mJy    \\ 
     Synch. powerlaw index 1,  $\alpha1$             & -0.41$\pm$0.02                                 & -0.42$\pm$0.02                                  & -0.41$\pm$0.02                 \\
     Synch. powerlaw index 2,  $\alpha2$             & -1.08$\pm$0.28                                 & -0.92$\pm$0.24                                  & -0.83$\pm$0.21                 \\
     Synch. powerlaw break frequency,  $\nub$    &  (2.6$\pm$0.5)$\times10^{11}$~Hz     & ( 2.6$\pm$0.6)$\times10^{11}$~Hz      &  (2.5$\pm$0.7)$\times10^{11}$~Hz  \\ 
     \hline 
     \multicolumn{4}{c}{Derived Quantities} \\
     \hline\hline
     Dust luminosity (8{\mm}-1000{\mm}), $\lfir$ &  (1.4$\pm$0.05)$\times10^{11}~\ls$ & (1.3$\pm$0.05)$\times10^{11}~\ls$ & (1.3$\pm$0.05)$\times10^{11}~\ls$   \\
     Star formation rate, SFR   &  24$\pm$1~$\mpy$ &  23$\pm$1~$\mpy$ &  22$\pm$1~$\mpy$ \\
     Total gas-to-dust mass ratio &  4500 to 7800 &  5000 to 8000 &  5300 to 9200  \\
     \hline
   \end{tabular}
 \end{table*}  

Assuming the dust to be in thermal equilibrium, Eqn.~\ref{dust}
quantifies the dust emission, which is a blackbody function modified
by a term that depends on the dust optical depth, $\tau_{\nu}$,
defined as
\begin{equation}
\tau_{\nu} = \kappa_{\nu} \frac{\md}{\Da^2 \Omega}.
\end{equation}
The modification from a standard blackbody function is due to the fact
that for typical dust temperatures, the dust grains are smaller than
the peak wavelength of the Planck function and hence do not radiate as
perfectly as a blackbody. Here, $\md$ is the dust mass, $\td$ is the
dust temperature, $\kappa_{\nu} = \kappa_{\nu_0}(\nu/\nu_0)^{\beta}$
is the dust absorption coefficient and we adopted $\kappa_{\nu_0} =
1$~m$^2$~kg$^{-1}$ at 1200~GHz (250{\mm})
\citep{Hildebrand1983}. $\beta$ is the dust emissivity index, which
based on empirical results likely lies in the range 1 to
2. $B_{\nu}(T)$ is the Planck function at frequency $\nu$, and
temperature $T$.  $B_{\nu}(\tcmb)$ is the contribution from the cosmic
microwave background at $\tcmb = 2.73(1+z)$~K. $\Omega$ is the solid
angle subtended by the source, here assumed to be the total extent of
the FIR emission at 70{\mm}.

The powerlaw indices for the AGN emission, $\alpha_1$ and $\alpha_2$,
were constrained to have negative values, so that the synchrotron
emission decreases with increasing frequency. \cite{Dunne2001} showed
that a two-component dust model with $\beta$ close to 2 better fits
the observed SEDs than a single-dust component.  This is also
consistent with our findings based on the SEDs of cool-core BCGs
investigated so far \citep{Edge2010a,Edge2010b,Mittal2011b}. We
therefore fitted the data with a model comprising a cold and warm dust
component parametrized by temperature and mass ($\tdc$, $\mdc$) and
($\tdw$, $\mdw$), respectively. We used the Levenberg-Marquardt
non-linear least square fitting algorithm from Numerical Recipes to
obtain the best-fit model parameters appearing in
Eqn.~\ref{FIRemission}: $\mdc$, $\tdc$, $\mdw$, $\tdw$, $\beta$,
$S_0$, $\alpha_1$, $\alpha_2$ and $\nub$. We explored a range of
values between 0.5 and 2.5 for the dust emissivity index, $\beta$. The
$\chi^2$-minimization gave a best-fit value $\beta$ which was $< 1$.
Note that there is a strong degeneracy between $\beta$ and the dust
temperature such that all values of $\beta$ explored yield models,
with varying temperatures, that are compatible with the observed
SED. I.e., for any of the explored values of $\beta$ between $0.5$ and
$2.5$, the best-fit $\chi^2$ value was less than $11$; for a
chi-squared distribution with $17-8=9$ degrees of freedom (because we
are not optimizing over $\beta$), this corresponds to a $p$-value
\citep{Gregory2005} greater than $0.10$, i.e., a model consistent with
the data. Based on the work of \cite{Dunne2001}, however, $\beta$ is
expected to lie between 1 and 2, and so we fixed $\beta$ to unity and
found the best-fit values for the other parameters subject to that
choice.

The best-fit model is shown in Fig.~\ref{SED} (black solid line), as
are the AGN (thin-dashed line) and the dust contributions
(thick-dashed lines; red representing the cold dust component and blue
representing the warm dust component). In addition to the fitted
flux-densities, we show the supplementary Spitzer/IRAS and UMRAO data,
along with SCUBA \citep{Irwin2001} and WMAP \citep{Wright2009} data
from around year 2000. The SCUBA (big magenta crosses) and WMAP data
(small indigo crosses) clearly demonstrate the strong variability in
the radio source and, hence, the need for coeval flux-densities for
SED fitting. Similarly, the Planck \cite[gray plus
symbols][]{Planck2011} and UMRAO data (open green squares) from the
year 2009 fall slightly below the IRAM and UMRAO data, both from the
year 2010, respectively.

The best-fit parameter values were used to calculate the total dust
emissivity by integrating the emission between 8{\mm} and
1000{\mm}. Assuming that young ($\le 10^8$~yrs), hot stars dominate
the interstellar radiation field across the UV-optical band, the star
formation rate can be estimated using the Kennicutt relation
\citep{Kennicutt1998}.  In early-type galaxies, including BCGs, the
cooler emissions ($> 100${\mm}) may arise from dust heated by a
passively evolving old stellar population, which warrants caution in
the Kennicutt calibration. The best-fit parameters for $\beta=1$ and
the derived quantities are given in Table~\ref{sed-params} (we also
give the best-fit parameters for $\beta=1.5$ and $\beta=2.0$ for
comparison) and the predicted dust and AGN flux contributions are
given in Table~\ref{contributions}. A total gas-to-dust mass ratio
between 4500 and 7800 was estimated using the molecular gas mass $\sim
4\times10^{10}~\ms$ derived in \cite{Salome2006}. The total gas mass
depends upon the conversion factor used to calculate the atomic plus
molecular mass. We used a factor of 1.36 as given in
\cite{Edge2001}. The total gas-to-duss mass ratio, although high, is
within the range of the derived mass ratios in other cool-core BCGs
\citep{Edge2001}. At the other extreme is NGC~4696, for which a
3-$\sigma$ upper-limit of $\sim 450$ was obtained on the total
gas-to-dust mass ratio \citep{Mittal2011b}.

While the SPIRE and PACS 100{\mm} observations were made close in time
to the IRAM sub-mm observations (2010-08-24 and 2010-09-09), the PACS
70{\mm} observations were made half a year later (2010-03-14), and so
the underlying contribution from the AGN may have varied because of
the variability in the AGN output. According to the best-fit model,
this is not an issue of concern since the AGN contribution to the
total flux at 70{\mm} is small. This was also verified by discarding
the 70{\mm} data point and refitting the data.

 \begin{table}
   \centering
   \caption{\small Dust and AGN contributions based on the best-fitting model 
     given in Table~\ref{sed-params}.} 
   \label{contributions}
    \begin{tabular}{| c | c |c }
     \hline
      Wavelength {\mm}   &  Dust (mJy) & AGN (mJy) \\
      \hline
      \hline
          20 &      2349 $\pm$ 221    &               88 $\pm$  62           \\
          25 &      3384 $\pm$ 163    &              112 $\pm$  74           \\
          30 &      3890 $\pm$ 282    &              137 $\pm$  85           \\
          60 &      6772 $\pm$ 295    &              289 $\pm$ 143           \\
          70 &      7840 $\pm$ 389    &              341 $\pm$ 160           \\
         100 &     8513 $\pm$ 775     &             502 $\pm$ 202            \\
         160 &     5532 $\pm$ 746     &             833 $\pm$ 265            \\
         250 &     2493 $\pm$ 369     &            1349 $\pm$ 323            \\
         350 &     1175 $\pm$ 171     &            1939 $\pm$ 354            \\
         500 &        485 $\pm$  67    &              2850 $\pm$ 363          \\
        1153 &          50 $\pm$   6   &               7023 $\pm$ 714         \\
        1303 &          35 $\pm$   4   &               7389 $\pm$ 398         \\
        1428 &          27 $\pm$   3   &               7673 $\pm$ 400         \\
        2710 &            4.2 $\pm$   0.5 &               10008 $\pm$ 396        \\
        2914 &             3.4 $\pm$   0.4&                10313 $\pm$ 394       \\
        3019 &             3.1 $\pm$   0.3&                10466 $\pm$ 392       \\
       20675 &            0.0 $\pm$   0.0 &               23231 $\pm$ 130        \\
     \hline
   \end{tabular}
 \end{table}  

\begin{figure}
  \centering
  \includegraphics[height=0.3\textheight]{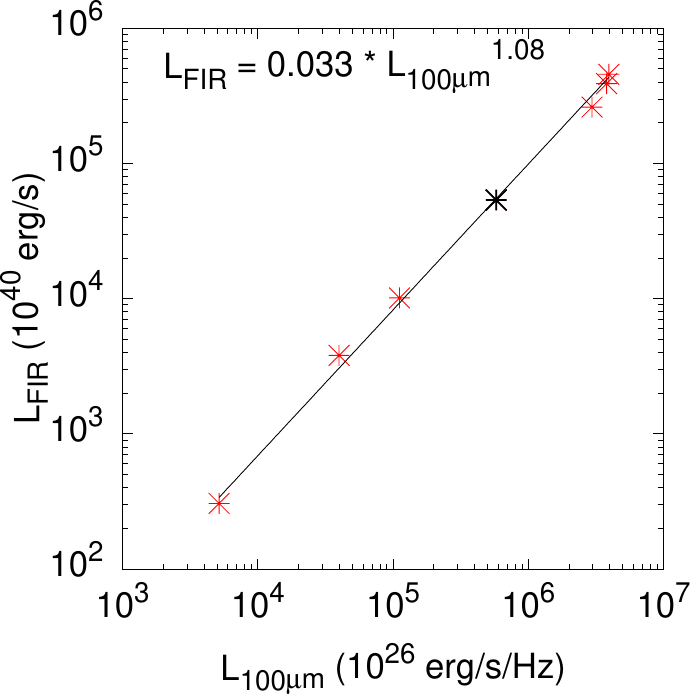}
  \caption{\small The total far-infrared luminosity (8{\mm} to
    1000{\mm}) vs. the 100{\mm} luminosity for seven of the 11 BCGs in
    the Herschel cool-core BCG sample. {\pers} is denoted as a black
    cross.}
  \label{100mu_vs_LFIR}
\end{figure}

 \section{What are the prevailing heating mechanisms in the inner
   4~kpc of {\pers}}
\label{core}

The filamentary nebula extends out to $\sim 50$~kpc from the core. The
excitation mechanisms in the outskirts of the galaxy are very likely
different from those prevailing in the core. This is evident also from
the reversal of the relative strengths of {\cii} and {\oi} line
emission (section~\ref{ciioi}) with cluster-centric distance. The core
of {\pers}, where we expect photoionization from stars and AGN to play
an important role, needs to be modelled separately from the
filaments. To this end, we conducted simulations using the radiative
transfer code {\sc cloudy} \citep{Ferland1998}, the main goal of which
was to determine whether or not an additional form of heating is
required to reproduce some of the observed emission lines emerging
from the core. For example, in the case of {\cen}, the BCG of the
Centaurus cluster, the {\cii}/$\lfir$ and {\ha}/{\cii} ratios clearly
call for another heating component, in addition to photoionization
\citep{Mittal2011b}.

From the point of view of energetics of a cooling plasma, earlier
works, such as
\cite{Johnstone1987,Heckman1989,Voit1994,Jaffe1997,Donahue2000}, have
shown that the observed filamentary emissions are far too luminous to
be just due to the recombination phase of gas cooling from the
ICM. The observed luminosities imply far too much cooling, with mass
deposition rates inconsistent by orders of magnitude with the star
formation rates. While such calculations have been performed for
emissions such as {\ha} and molecular hydrogen lines, none of the
current cooling calculations simulate the FIR emission lines. This is
because the existing cooling calculations (that use codes like {\sc
  cloudy}, {\sc apec}, {\sc mkcflow}) usually stop when the gas
reaches a temperature of $\sim10^4$~K, not low enough to produce the
FIR lines under consideration. Hence it is presently not possible to
estimate the luminosity of the FIR lines from a simple cooling-flow
plasma.

We modelled a composite cloud comprising a photo-dissociation
region~(PDR) adjacent to an ionized (H{\sc ii}) region. For the
equation of state we assumed a constant gas pressure throughout the
cloud. As input, we provided the normalization for an old stellar
population (OSP), {\norm}, the normalization for a young stellar
population (YSP), $G0$ (in Habing units), the starting hydrogen
density, $n$ (in cm$^{-3}$), and the hydrogen column density, {\hcol}
(in cm$^{-2}$). For further details on the role played by these input
parameters, we refer the reader to \cite{Mittal2011b}. The parameters,
along with the range explored, are given in Table~\ref{pdr-params}.
Based on some initial investigations carried out to assess the effect
of an OSP component on the output emissions, we concluded that the
{\cii}/$\lfir$ ratio can be better reconciled if an OSP component is
included. An old generation of stars produces optical and UV photons
that contribute to the heating of dust but not so much to the
photoelectric heating of the gas. In the absence of an OSP component,
the {\cii}/$\lfir$ for a given $G0$ is higher. The results, however,
are not very sensitive to the presence of an OSP, except for low $G0$
($\le 100$~Habing) but a low $G0$ is not consistent with other
observed ratios, as we will show below.

We used the starburst99 stellar synthesis library
\citep{Leitherer1999} to model both the OSP and YSP. The main input
parameters to the starburst99 simulations are the stellar mass and age
(continuous or starburst), and the initial stellar mass function. We
did not explore the range of possible values these input parameters
can take -- that is a subject of another detailed ongoing study
(Mittal~et~al. 2012, in prep.). For the purpose of conducting the {\sc
  cloudy} simulations described here, we fixed the age and mass of the
OSP to $\sim 10$~Gyr and $5\times10^{11}~\ms$, and the age and the
mass of the YSP to $\sim 2$~Myr and $5\times10^{9}~\ms$. The SED
corresponding to a starburst 2~Myr old is indistinguishable from the
one corresponding to a continuous star formation scenario, where the
oldest stars are assumed to be 2~Myr. This is because the emission in
both the cases is dominated by young stars that are about 2~Myr
old. Hence the young population of stars need not have formed at the
same time.

The elemental abundances were initially set to their default ISM
values given in Table~\ref{abundances}. However, X-ray observations
made with Chandra indicate a slight drop in the central metallicity
with the average value around $0.6~\zs$ \citep{Sanders2007}. With this
in mind, we also conducted simulations with a lower metallicity,
which, as shown below, indeed fit the observed ratios better.  We did
not include any polycyclic aromatic hydrocarbon (PAH) grains since the
mid-infrared spectra do not contain any PAH features
\citep{Weedman2005}. Although several BCGs show strong PAH features in
their IR spectra indicative of star formation \citep{Donahue2011},
there are a few such as {\pers} and NGC~4696 \citep{Kaneda2005} that
do not. PAH molecules are small in size and can easily be destroyed
through physical sputtering or thermal evaporation
\citep{Dwek1992,Micelotta2011}.

\begin{table}
  \centering
  {\small
    \caption{\small The default ISM gas phase chemical composition used in 
      {\sc cloudy} simulations. The abundances are given relative to H.}
    \label{abundances}
    \begin{tabular}{| c | c |}
      \hline
      Element & $\log_{10}$ abundances (ISM)\\
      \hline\hline
      He      &     -1.0088     \\
      C        &     -3.8222     \\
      N        &     -4.2010     \\
      O        &     -3.7171     \\
      Ne      &     -4.1319     \\
      Mg     &      -5.1215     \\
      Si       &      -5.7222     \\
      S        &      -4.7113     \\
      Cl       &     -7.2218     \\
      Ar      &      -5.7716     \\
      Fe      &      -6.4218     \\
      \hline
    \end{tabular}
  }
\end{table}

\begin{table*}
  \centering
  \caption{\small A comparison of FIR and optical emission lines in the core region. 
    Given in the last column are the ratios with respect to {\cii}. The {\ha} flux has been 
    corrected for the Galactic extinction. We also give the {\ha} flux after correcting for the 
    internal extinction assuming E(B-V)=0.37 in brackets.} 
  \label{coreratios}
  \begin{tabular}{| c | c | c | c | c | c | c | c|}
    \hline 
   Region      &  RA                    & Dec                    &  Aperture Radius &  Line   &  Velocity & Flux    & Ratio \\
    &                          &                           & (arcsec)     &           &   km~s$^{-1}$ & ($10^{-15}$~erg~s$^{-1}$~cm$^{-2}$) & \\
    \hline\hline
   Radio Core            &03h19m48.16s & +41d30m42.1s  & 11     &    {\cii}  & $11\pm3$   &  665.1 $\pm$ 6.3 &   1.00      \\
    &                          &                           &        &    {\oi}   & $27\pm4$ & 1310.5 $\pm$ 18.4 &    1.97      \\
    &                          &                           &        &    {\oib} & $50\pm9$ & 81.2   $\pm$ 2.1 &    0.12      \\
    &                          &                           &        &    {\nii}  & $-2\pm7$ & 53.7     $\pm$ 1.1 &    0.08        \\
    &                          &                           &        &    {\oiii} & $-26\pm15$ & 59.0   $\pm$ 3.1 &   0.09       \\
    &                          &                           &        &    {\ha}  &         &  1810 (4185) \comment{(4417)}      & 2.7 (6.3) \\ 
    &                          &                           &        &    ${\lfir}$  &         &  $\sim 5.9\times10^{5}$      & $\sim 800$ \\
  CII Core            & 03h19m48.01s & +41d30m44.9s  & 11     &    {\cii}  & $8\pm3$   &  724.8 $\pm$ 7.3 &   1.00      \\
    &                          &                           &        &    {\oi}   & $23\pm4$ & 1320.2 $\pm$ 17.3 &    1.82      \\
    &                          &                           &        &    {\oib} & $47\pm8$ & 84.3   $\pm$ 2.0 &    0.12     \\
    &                          &                           &        &    {\nii}  & $-4\pm8$ & 58.1     $\pm$ 1.2 &    0.08       \\
    &                          &                           &        &    {\oiii} & $-30\pm14$ & 66.1   $\pm$ 3.0 &   0.09       \\
    &                          &                           &        &    {\ha}  &         &  1810 (4185) \comment{(4417)}      & 2.5 (5.8) \\
    &                          &                           &        &    ${\lfir}$  &         &  $\sim 5.9\times10^{5}$      & $\sim 700$ \\
    \hline
  \end{tabular}
\end{table*} 

The absolute and relative strengths of the FIR and {\ha} lines are
very important diagnostics of the various heating contributors in
cool-core BCGs.  These emission lines and their ratios relative to
{\cii} are listed in Table~\ref{coreratios}. We note that there is a
slight offset of $\sim \pp{3}$ between the peak of the {\cii} emission
and the radio core emission (the latter coincides with the peak of
{\ha} emission to within $\sim \pp{0.5}$). This offset is on the order
of the $1\sigma$ pointing inaccuracy of the PACS spectrometer and
unlikely real. However, we estimated the FIR line fluxes for both the
cases - (a) assuming the offset is not real and (b) assuming the
offset is real. These cases are referred to as ``{\cii} core'' and
``radio core'', respectively. The two cases have only marginal
differences in their FIR line fluxes and are compared only for the
purpose of illustrating the level of uncertainty in the estimated line
parameters. In the following, we used the ``{\cii} core'' as the
nominal case and used the other set of line fluxes to derive
uncertainties on the line ratios.

The total FIR flux associated with the core can not be directly
estimated from fitting the SED due to insufficient resolution
available beyond $\sim 160${\mm}. We instead used the 100{\mm} flux as
a proxy for the total luminosity. Shown in Fig.~\ref{100mu_vs_LFIR} is
the total FIR luminosity in the range 8{\mm} to 1000{\mm} vs the
100{\mm} luminosity for seven of the 11 BCGs in the Herschel cool-core
BCG sample, including {\pers}. We considered only those here because
their FIR luminosity can be determined with least uncertainty.  For
{\pers} we set $\lfir$ equal to the total dust luminosity. This
relation is being investigated for the whole sample separately (Oonk
et al. 2012, in prep.). There is a clear correlation between the total
FIR luminosity and the 100{\mm} luminosity, such that $\lfir \propto
\lhundred^{1.09}$. The trend is also observed in the flux-flux plane
(not shown) and so the correlation is not spuriously induced due to
the common dependence of the two quantities on redshift. Using this
correlation and the 100{\mm} core flux, the total FIR flux for the
core may be estimated. The contribution from the AGN synchrotron
emission at 100{\mm} is negligible ($<6\%$, see
Table~\ref{contributions}) and may be ignored.

The measured 100{\mm} flux within the core aperture of $\pp{11}$
radius is (5608$\pm7$)~mJy. Even though some of the flux may fall
beyond the chosen aperture, necessitating an aperture correction, the
factor is very small for the given radius. We determined the required
aperture correction factor by assuming that the FIR emission
correlates with the {\ha} emission. Assuming the WIYN {\ha} image to
represent the true model for the FIR surface-brightness distribution,
we convolved the {\ha} image with the PACS 100{\mm} PSF and estimated
the aperture correction from the ratio of the flux from the smoothed
to the unsmoothed image. The ratio is close to unity and therefore we
did not apply any aperture correction. The measured 100{\mm} flux
yields a FIR (8{\mm} to 1000{\mm}) flux that implies a $\lfir$/{\cii}
ratio $\sim$ 700 to 800.

\subsection{Best-fit energy model for the core}
\label{best-fit}

\begin{figure*}
  \begin{minipage}{0.45\textwidth}
    \centering
  \centering
    \includegraphics[width=0.85\textwidth]{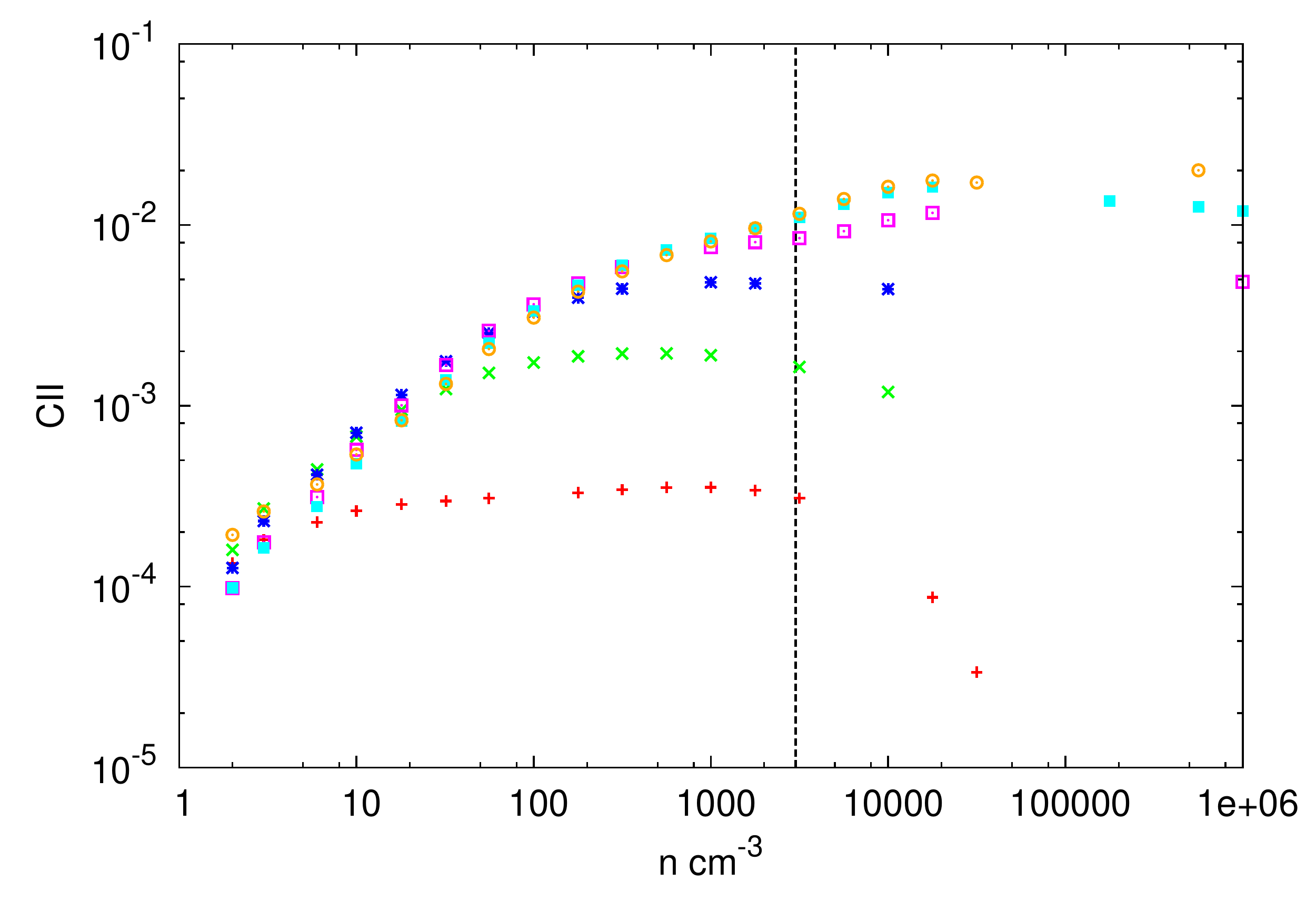}
  \end{minipage}\hfill
  \begin{minipage}{0.45\textwidth}
    \centering
    \includegraphics[width=0.85\textwidth]{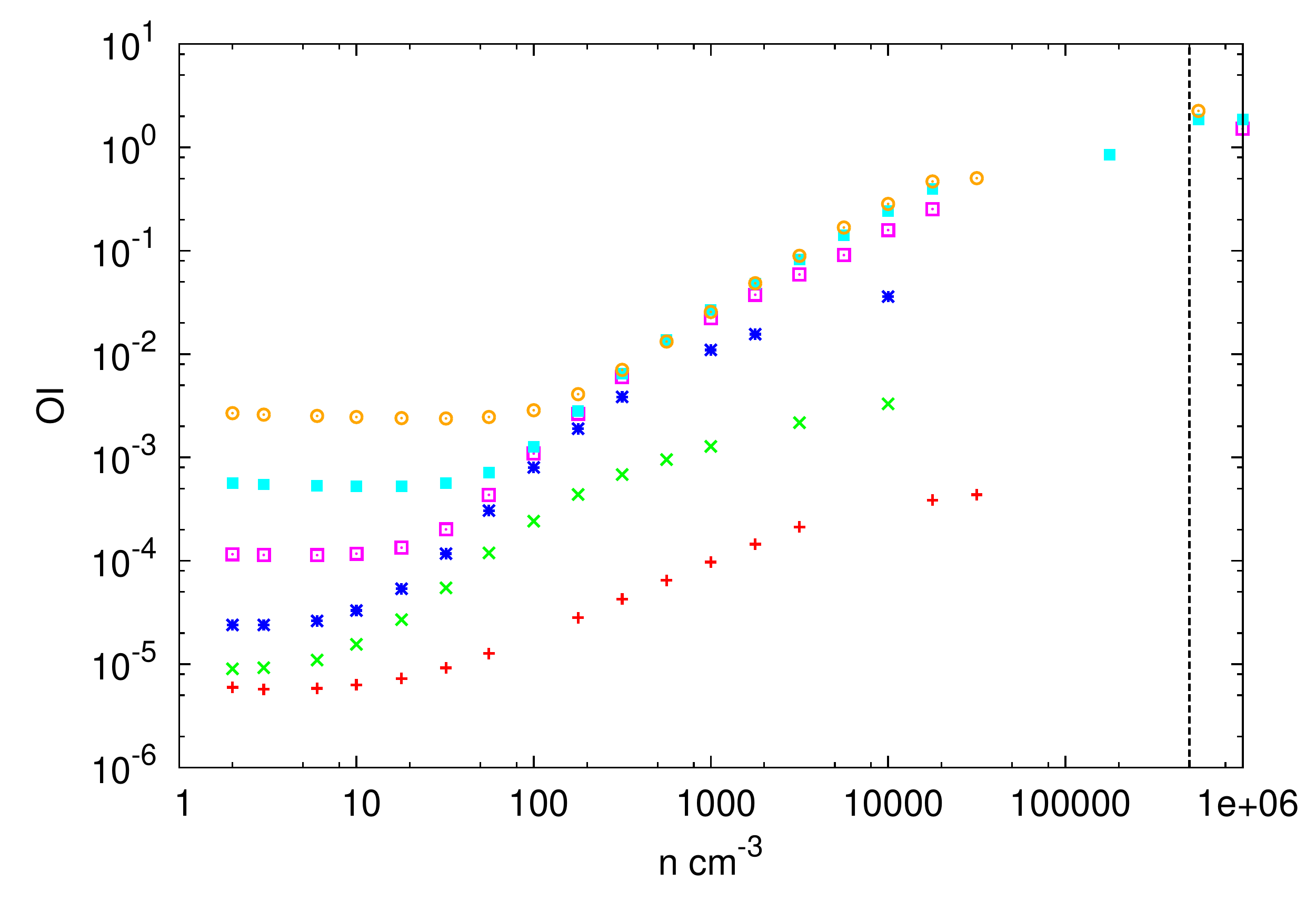}
  \end{minipage}\\
  \begin{minipage}{0.45\textwidth}
    \centering
    \includegraphics[width=0.85\textwidth]{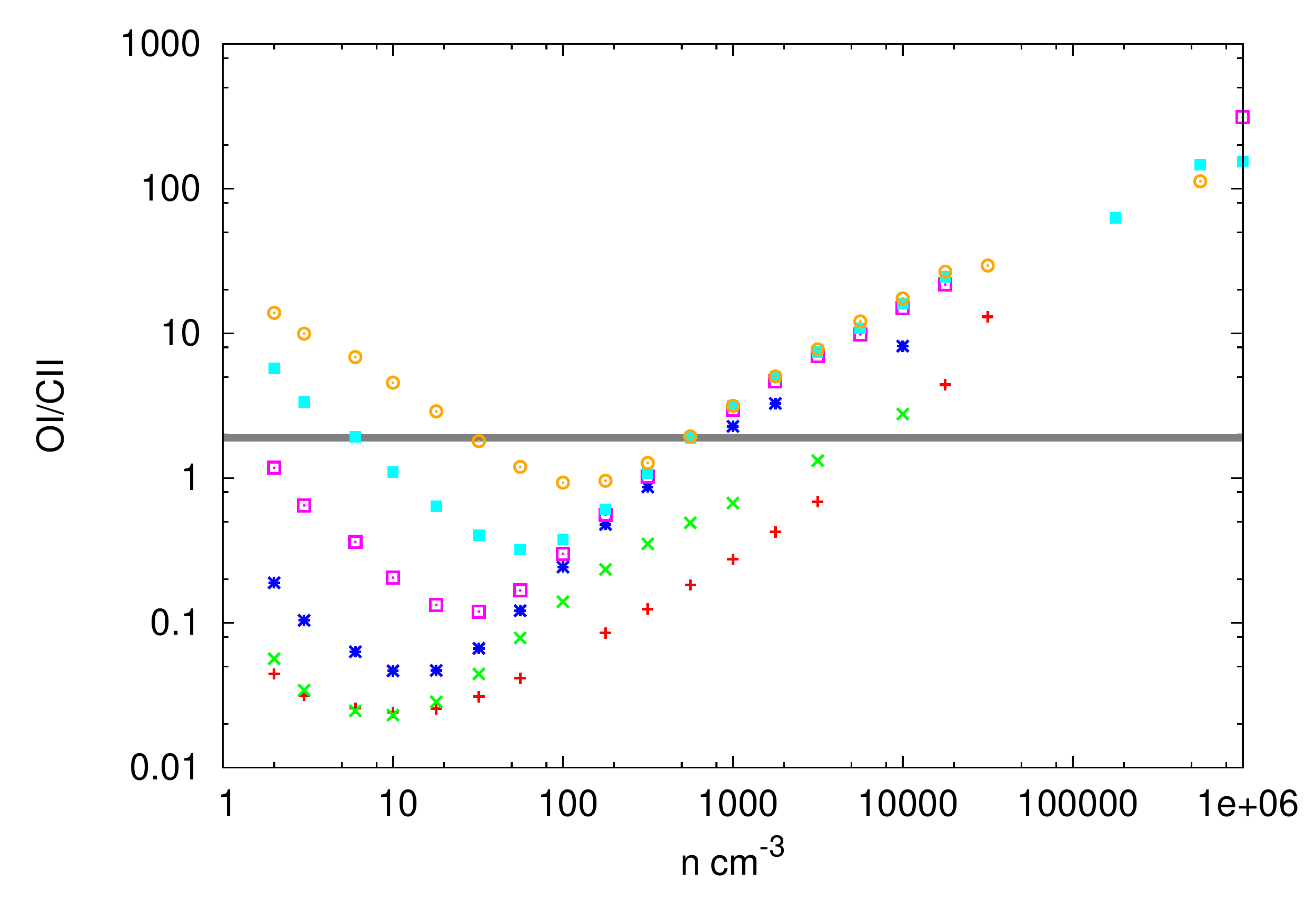}
  \end{minipage}%
  \begin{minipage}{0.15\textwidth}
    \centering
    \includegraphics[width=\textwidth]{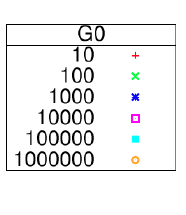}\vspace*{1.8cm}
  \end{minipage}
  \caption{\small {\oi}/{\cii} sensitivity to the gas density. The
    simulated {\cii} intensity in erg~s$^{-1}$~cm$^{-2}$ (upper left
    panel), {\oi} intensity in erg~s$^{-1}$~cm$^{-2}$ (upper right
    panel) and {\oi} to {\cii} ratio (lower panel) as a function of
    gas density, $n$, and the normalization of the young stellar
    population (the photon parameter), $G0$. The column density has
    been fixed to {\hcol}$=10^{23}~$cm$^{-2}$. The two vertical dashed
    lines represent the critical densities,
    $\ncr=3\times10^3$~cm$^{-3}$ for {\cii} and
    $\ncr=3\times10^3$~cm$^{-3}$ for {\oi}. The horizontal solid line
    corresponds to the observed ratio of {\oi} to {\cii} for the inner
    4~kpc region.}
  \label{oicii_cloudy}
\end{figure*}

The Herschel FIR coolants serve as strong constraints for evaluating
the physical parameters of the ISM in the BCG. Fig.~\ref{oicii_cloudy}
shows {\oi}/{\cii} versus $n$ for different values of the photon
parameter, $G0$. This plot can be understood in terms of the critical
densities for {\cii} and {\oi}.  This is the density at which the
probabilities of collisional de-excitation and radiative de-excitation
are equal \citep{Osterbrock2006}. Above the critical density the level
populations may be considered to be in local thermal equilibrium as
the level populations become dominated by collisions.  The critical
density is lowered by a factor that is roughly the optical depth of
the line due to photon trapping when the line is optically thick.
Fig.~\ref{oicii_cloudy} shows that, for $n < \ncr$, the intensity of
an optically thin line increases linearly with $n$ and for $n>\ncr$
the intensity becomes independent on $n$. This result is fairly
insensitive to details such as the cloud energy source, as
Fig.~\ref{oicii_cloudy} shows.  While $\ncr$ for {\cii} is $\sim 3
\times 10^3$~cm$^{-3}$, it is substantially higher for {\oi} $\sim 5
\times10^5$~cm$^{-3}$. Therefore, at a fixed column density and photon
parameter, $G0$, for high densities $3 \times 10^3 $~cm$^{-3}< n < 5
\times 10^5$~cm$^{-3}$ the ratio, {\oi}/{\cii}, continues to increase
with density. This makes {\oi}/{\cii} a sensitive probe of density if
the lines are optically thin. The individual trends displayed by the
{\oi} and {\cii} lines in Fig.~\ref{oicii_cloudy} are complicated by
the fact that for a given column density these lines become optically
thick for certain combinations of $n$ and $G0$.

\begin{table*}
  \centering
  \caption{\small Input parameters for the {\sc cloudy} simulations.} 
  \label{pdr-params}
   \begin{tabular}{| c | c | c | c |}
    \hline
     Parameter   &  Symbol  & Input Range & Likely Values\\
     \hline
     \hline\rul
    Total Hydrogen Density (cm$^{-3}$)        &  $n$           & 10 to 10$^6$ & 500 to 700 \\\rul
    FUV Intensity Field  (Habing$^*$)                   &  $G0$         & 1 to 10$^6$ & 700 to 900  \\\rul
    Hydrogen Column Density (cm$^{-2}$)& {\hcol} & 10$^{19}$ to 10$^{26}$ & $>10^{23}$ \\
    \hline
    Metallicity  & $Z$ & --- & 0.6 \\
    Nitrogen abundance  (relative to $Z$) & {\nitrogen} & --- & 2 \\
    Normalization for the OSP$^*$ (10$^{-16}$~erg~s$^{-1}$~cm$^{-2}$~Hz$^{-1}$)&  {\norm} & --- & 52.54 \\
    \hline 
    \multicolumn{4}{c}{$^{\st a}$~1~Habing $=1.6\times10^{-3}$~erg~s$^{-1}$~cm$^{-2}$, OSP: Old Stellar Population } \\
  \end{tabular}
\end{table*} 

Similarly, Fig.~\ref{n2o3} shows {\nii} and {\oiii} versus $n$ for
different $G0$ and for two different column densities. {\cii} is
produced in both ionized~(for e.g. H{\sc ii} regions) and neutral
media~(for e.g. PDRs), however, {\nii} and {\oiii} are produced only
in ionized region due to their higher ionization potentials. This has
the effect that {\nii} and {\oiii} are produced profusely at the
surface of the cloud facing the ionization source and reduce with the
depth into the cloud. This explains the drop in both {\nii} and
{\oiii} intensities relative to {\cii} with {\hcol}. This makes
{\nii}/{\cii} and {\oiii}/{\cii} a sensitive probe of the cloud depth
or hydrogen column density, {\hcol}.

Of the {\hcol} range investigated, the optimal cloud depth required to
reproduce the observed ratios is $10^{23}$~cm$^{-2}$. Unless $G0$ is
high ($>10000$~Habing), a higher {\hcol} does not affect the predicted
ratios, implying that the cloud becomes radiation bounded. In
Fig.~\ref{best-PDR}, we present the modelled ratios as function of $n$
and $G0$ for {\hcol}$=10^{23}$~cm$^{-2}$, assuming stellar
photoionization only. The different sets of curves correspond to the
observed lower- and upper-limits of {\cii}/$\lfir$ (solid red),
{\oi}/{\cii} (dotted blue), {\ha}/{\cii} (dashed green), {\nii}/{\cii}
(dash-dotted orange), {\oib/{\cii} (dash-dot-dotted pink) and
  {\oiii}/{\cii} (long-dashed gray). Note that the observed
  {\ha}/{\cii} ratio has a wide range due to the uncertainties
  mentioned in section~\ref{ha}. The FIR ratios, in contrast, are much
  better constrained. Consequently, our strategy was to first try and
  reproduce the FIR line and continuum ratios and then use
  {\ha}/{\cii} to check for consistency against the best-fit model.

  The left panel of Fig.~\ref{best-PDR} assumes ISM abundances, as
  adopted in \cite{Ferland2008}. While {\oi}/{\cii}, {\oib}/{\cii} and
  {\oiii}/{\cii} converge at $n \sim 1000$~cm$^{-3}$ and $G0 \sim
  1000$~Habing, the predicted {\nii}/{\cii} for this combination of
  $n$ and $G0$ is less than the observed ratio. The middle panel of
  Fig.~\ref{best-PDR} assumes the nitrogen abundance is twice the ISM
  value, yielding a much better convergence between the FIR
  ratios. Curiously, the best-fitting ($n$,$G0$) for {\cen} in the
  Centaurus galaxy cluster can also be reconciled with the observed
  {\nii}/{\cii} only if the nitrogen abundance is increased by a
  factor of two over the assumed overall metallicity
  \citep{Mittal2011b}. In the case of {\cen}, the need for an
  overabundance in nitrogen is confirmed by independent observations
  in the optical and X-ray bands. An overabundance of nitrogen could
  imply an abundance of intermediate-mass asymptotic giant
  branch~(AGB) stars with pre-main sequence mass between $3~\ms$ and
  $8~\ms$, which may convert dredged-up carbon in their outer shells
  into nitrogen \citep{Wood1983,Iben1975}. We are conducting such
  detailed radiative transfer modelling of all the cool-core BCGs in
  the Herschel sample and it will be interesting to see if this is a
  generic feature of galaxies at the center of cooling-flows.

  As mentioned in section~\ref{core}, X-ray observations suggest a
  lower metallicity in the core of Perseus ($\le \pp{20} \sim
  7$~kpc). Hence, in the right panel of Fig.~\ref{best-PDR}, we show
  the predicted ratios for a metallicity equal to $0.6$ times the ISM
  value. The FIR ratios coincide nicely at $n \sim 650$~cm$^{-3}$ and
  $G0 \sim 800$~Habing. We can now derive constraints on the physical
  scale of the cloud by comparing the {\cii} flux emerging from the
  modelled cloud and the observed {\cii} flux. For the range of $n$
  and $G0$ given in Table~\ref{pdr-params}, the emergent {\cii} flux
  is in the range $3.2\times10^{-3}$~erg~s$^{-1}$~cm$^{-2}$ to
  $3.7\times10^{-3}$~erg~s$^{-1}$~cm$^{-2}$. The size of the cloud
  that produces {\cii} flux at the distance of Earth between
  $665.1\times10^{-15}$~erg~s$^{-1}$~cm$^{-2}$ and
  $724.8\times10^{-15}$~erg~s$^{-1}$~cm$^{-2}$ is $\gtrsim 1$~kpc,
  yielding a volume filling factor of $1.7\times10^{-2}$.

\begin{figure*}
  \begin{minipage}{0.49\textwidth}
    \centering
  \centering
    \includegraphics[width=0.8\textwidth]{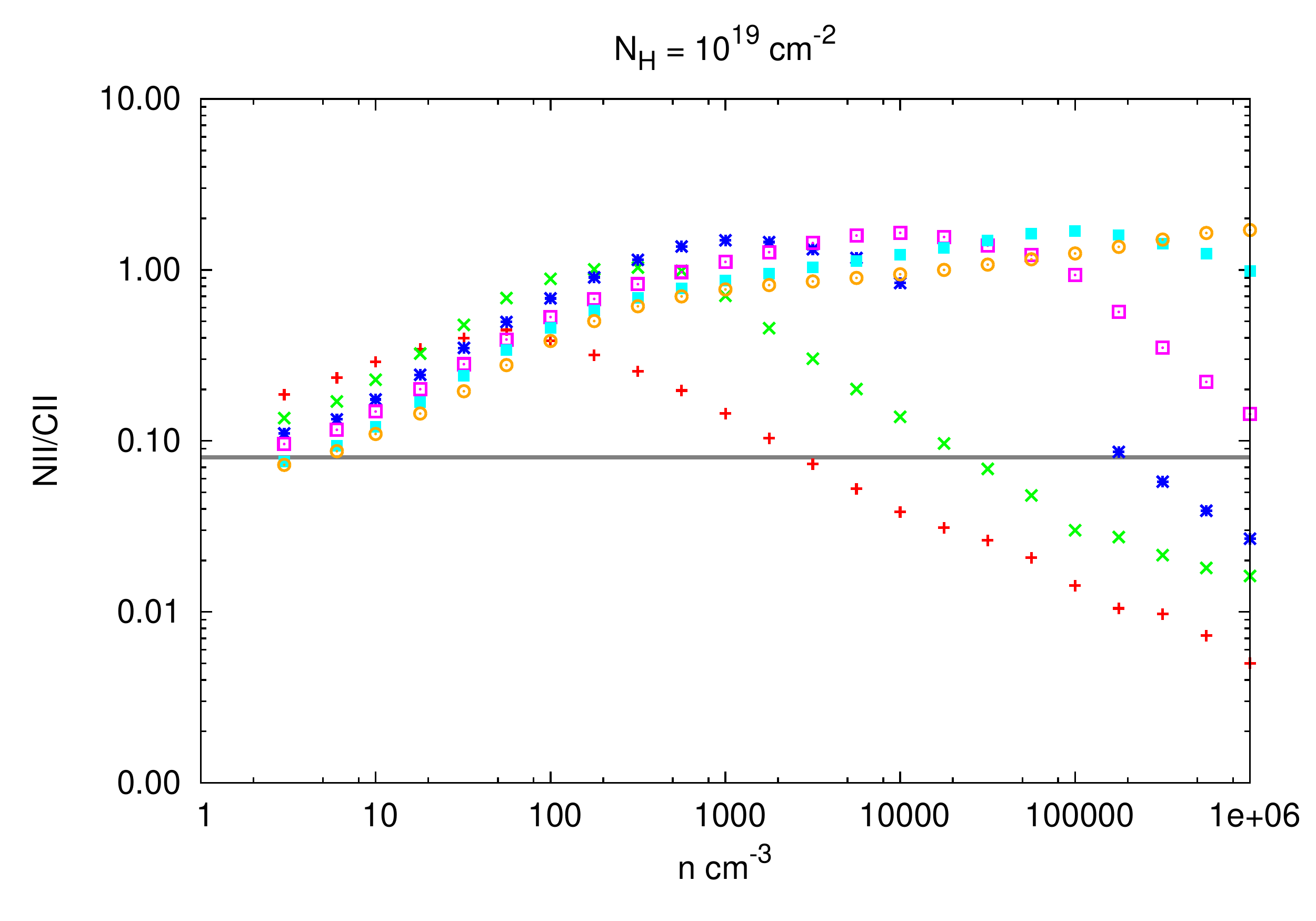}
  \end{minipage}\hfill
  \begin{minipage}{0.49\textwidth}
    \centering
    \includegraphics[width=0.8\textwidth]{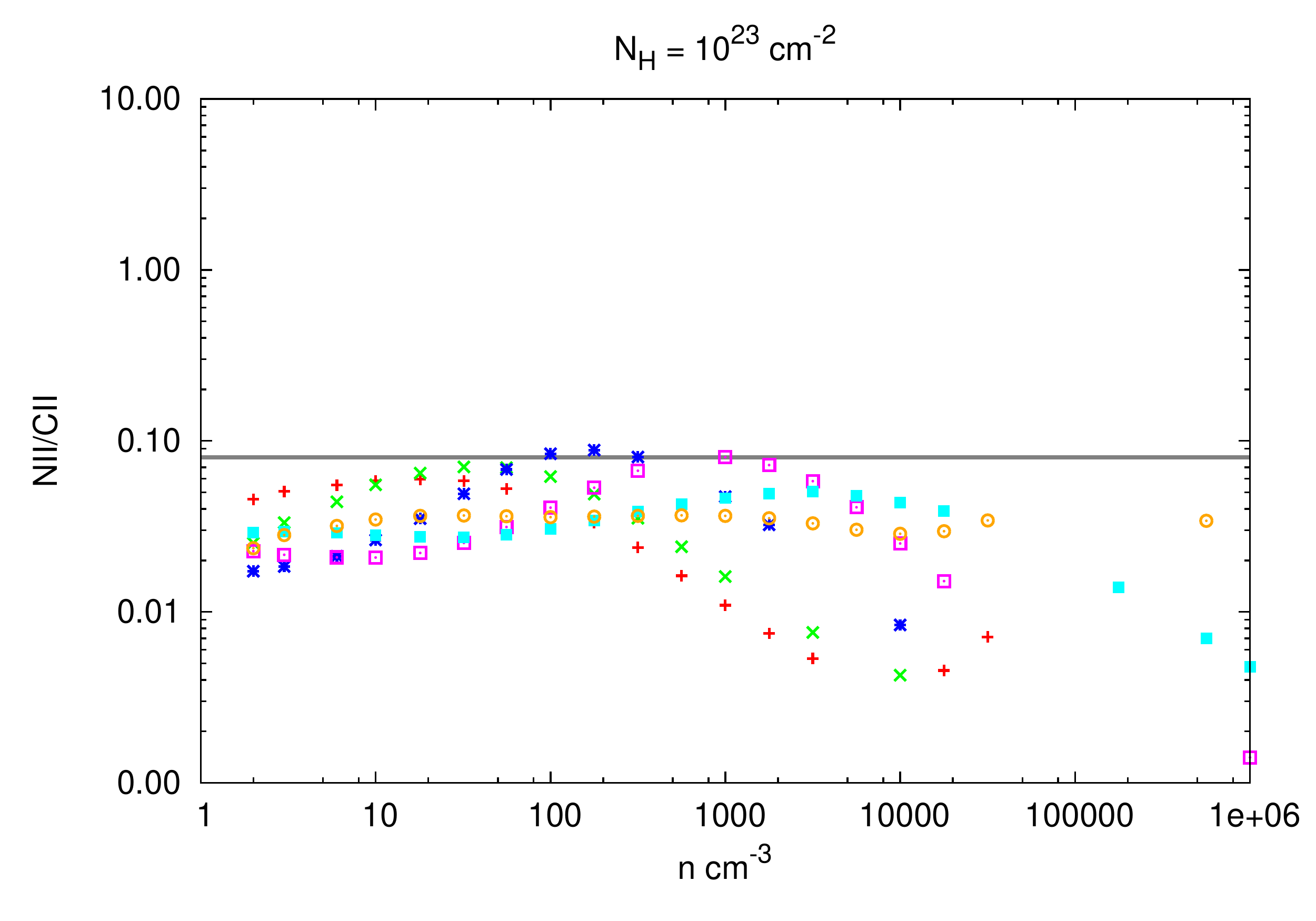}
  \end{minipage}\\
  \begin{minipage}{0.49\textwidth}
    \centering
    \includegraphics[width=0.8\textwidth]{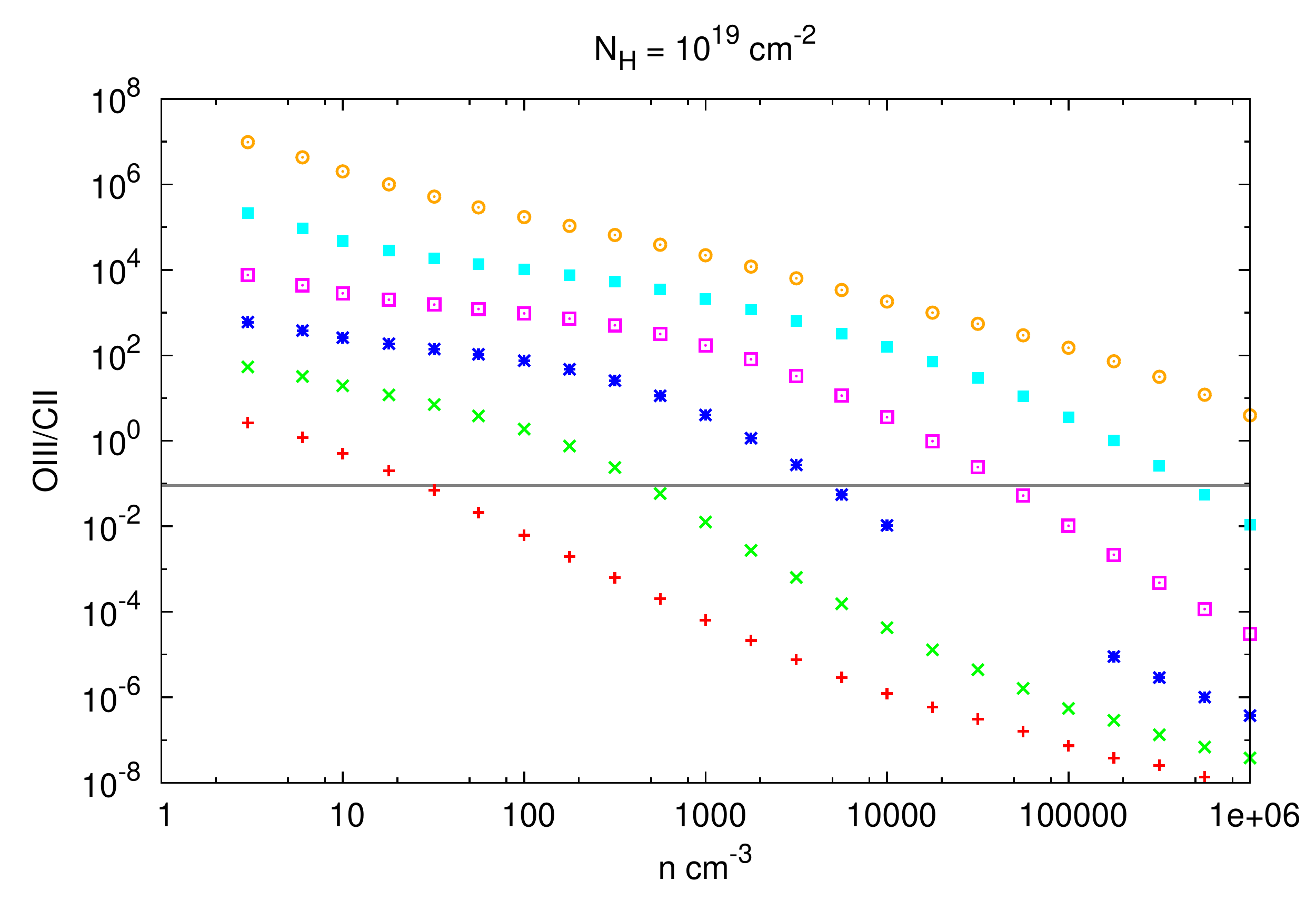}
  \end{minipage}
  \begin{minipage}{0.49\textwidth}
    \centering
    \includegraphics[width=0.8\textwidth]{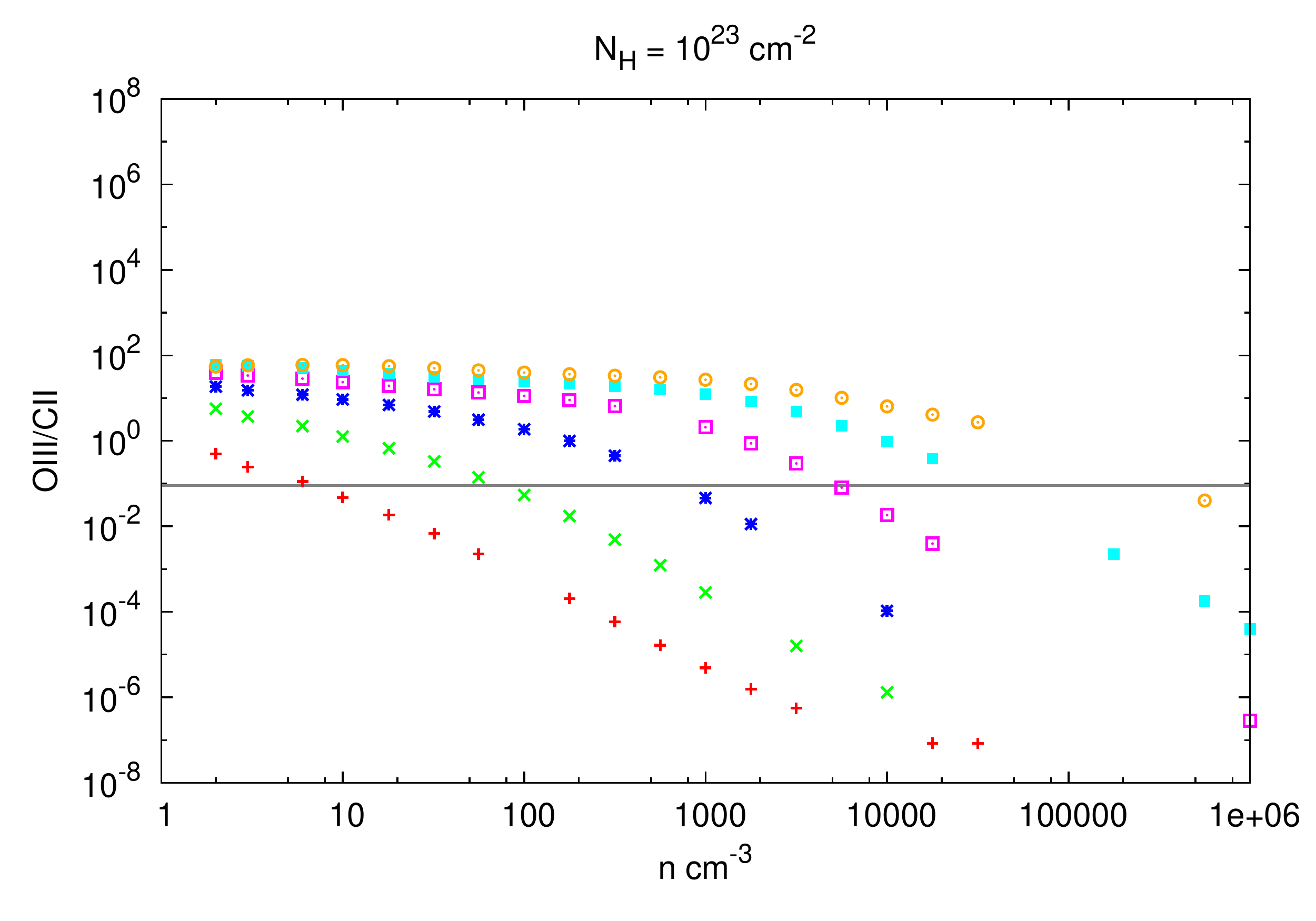}
  \end{minipage}
  \caption{\small {\nii}/{\cii} (upper row) and {\oiii}/{\cii} (lower
    row) versus density, $n$. Shown are the ratios for two different
    column densities, \hcol$=10^{19}~$cm$^{-2}$ (left) and
    \hcol$=10^{23}~$cm$^{-2}$ (right). {\nii} and {\oiii} are produced
    in abundance at the surface of the cloud facing the ionization
    source and deplete with the depth of the cloud and, hence,
    sensitive probes of {\hcol}. The different points have the same
    meaning as in Fig.~\ref{oicii_cloudy}. The horizontal lines
    represent the observed ratios in the inner 4~kpc region.}
  \label{n2o3}
\end{figure*}

\begin{figure*}
  \begin{minipage}{0.33\textwidth}
    \centering
  \centering
    \includegraphics[width=\textwidth]{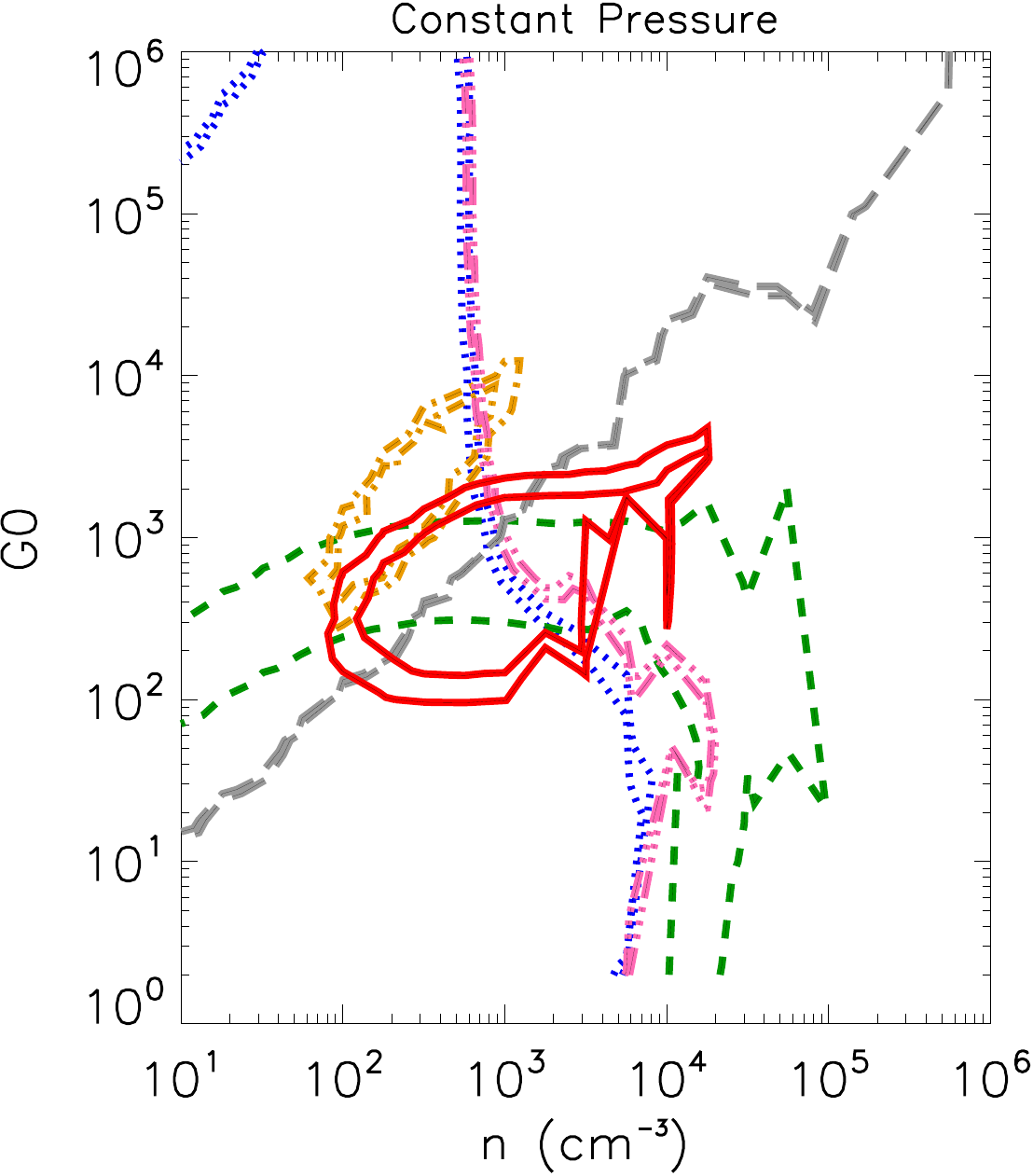}
  \end{minipage}%
  \begin{minipage}{0.33\textwidth}
    \centering
    \includegraphics[width=\textwidth]{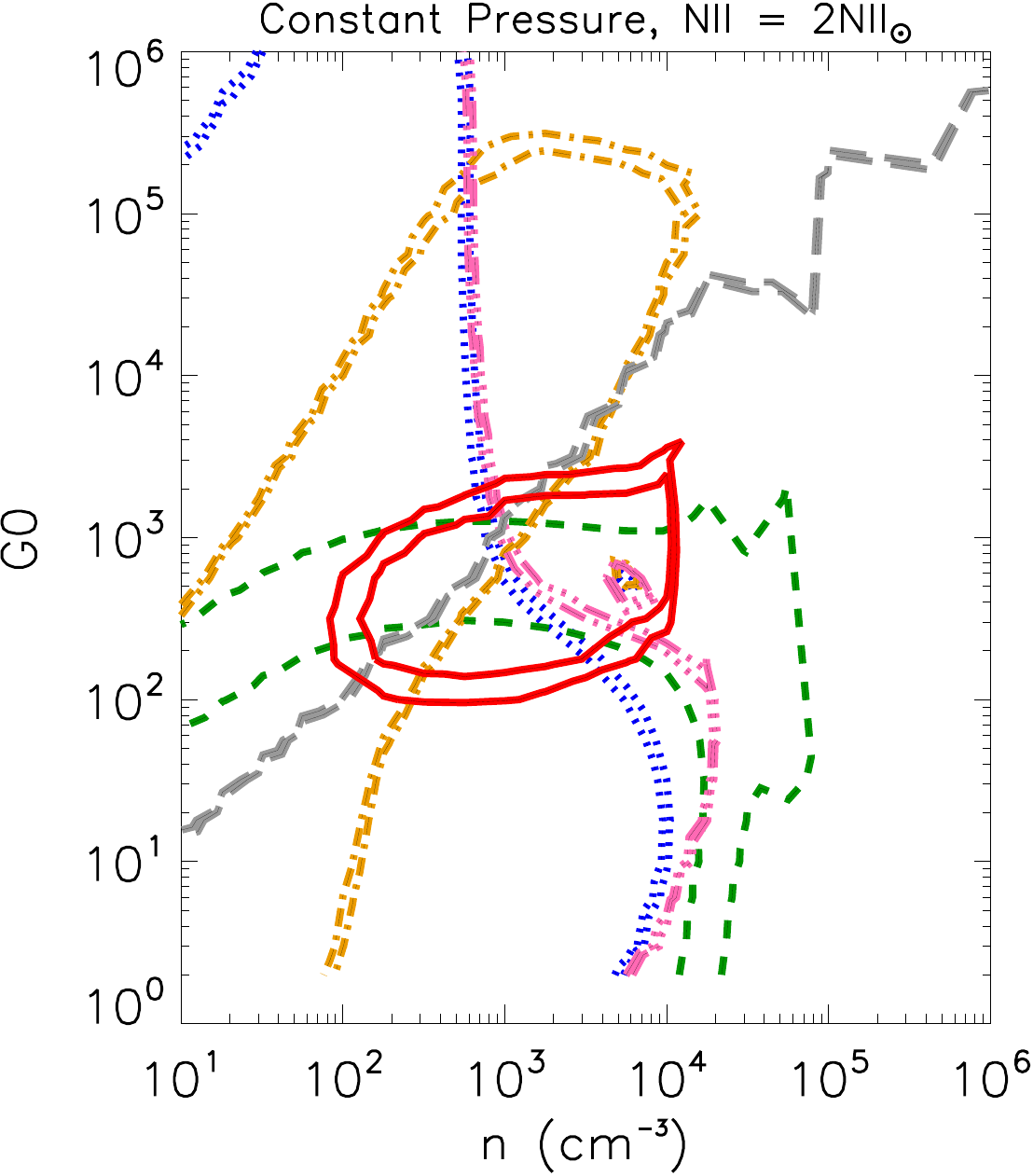}
  \end{minipage}%
  \begin{minipage}{0.33\textwidth}
    \centering
    \includegraphics[width=\textwidth]{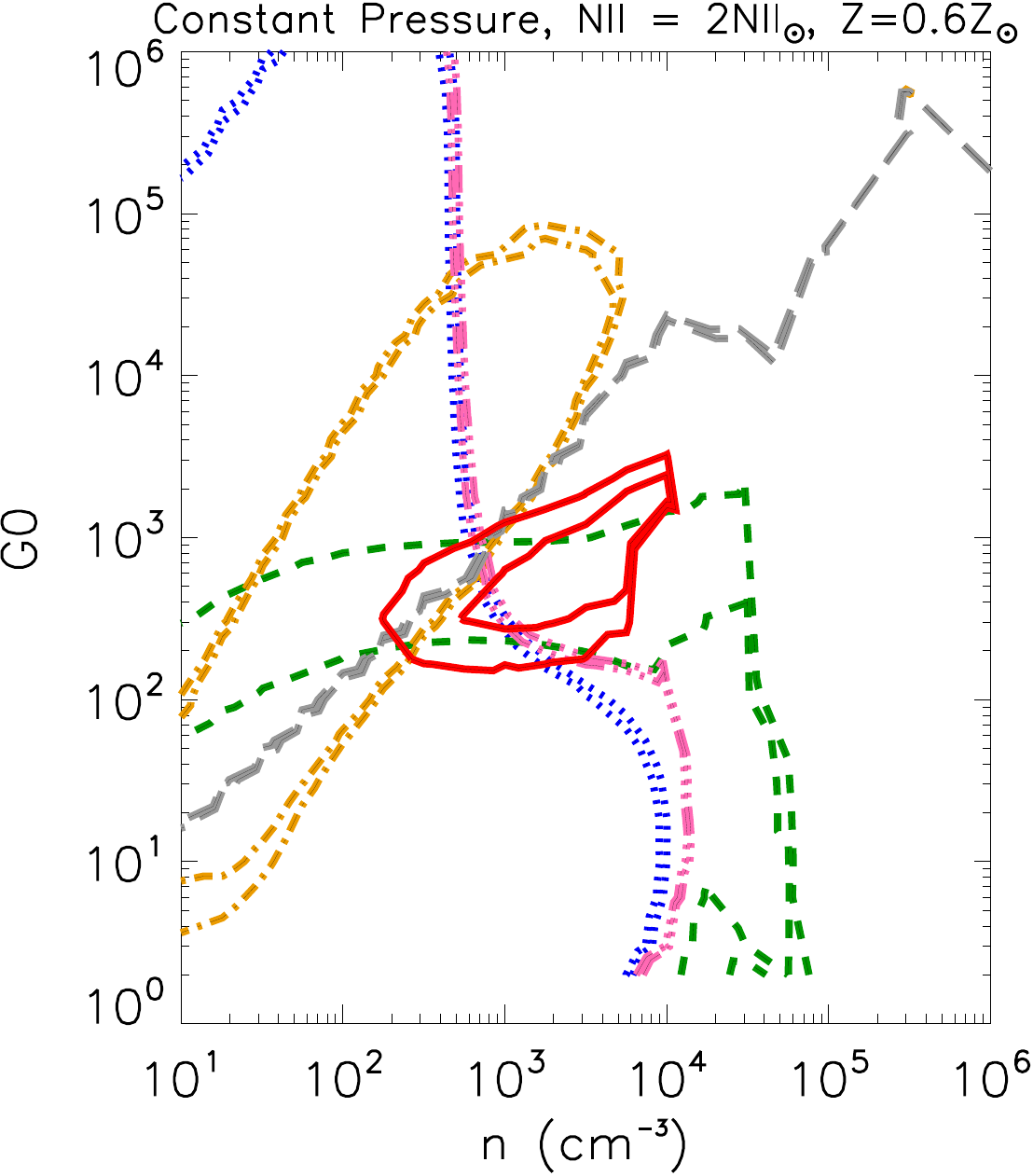}
  \end{minipage}
  \caption{\small Simulations based on a photoionization model
    containing only stars. Shown are contours corresponding to the
    lower and upper observed limits (given in square brackets below)
    for $\lfir$/{\cii} (solid red) $\rightarrow$ [700:800];
    {\oi}/{\cii} (dotted blue) $\rightarrow$ [1.79:2.00]; {\ha}/{\cii}
    (dashed green) $\rightarrow$ [2.5:6.3]; {\nii}/{\cii}
    (dashed-dotted orange) $\rightarrow$ [0.078:0.082]; {\oib}/{\cii}
    (dash-dot-dotted pink) $\rightarrow$ [0.113:0.125] and
    {\oiii}/{\cii} (long-dashed gray) $\rightarrow$ [0.083:0.095]. For
    this simple model, the regions allowed by the data, between the
    paired lines, would be expected to overlap at a single Habing flux
    ($G0$) and cloud density ($n$).  {\it Left}: The element
    abundances are fixed to their default ISM values. {\it Middle}:
    Nitrogen overabundance is fixed to two. {\it Right}: The element
    abundances are fixed to 0.6 times the ISM value and nitrogen
    overabundance to two. The best-measured ratio, $\lfir$/{\cii},
    overlaps for this model.}
  \label{best-PDR}
\end{figure*}

The main conclusion from modelling the ISM associated with the core of
{\pers} is that a stellar-like ionizing source with
$G0\sim1000$~Habing is capable of reproducing the observed FIR
emission and {\ha} emission. No additional form of heating is required
to explain these observations.

\subsection{Consistency with optical flux ratios}
\label{optfluxratios}

According to previous studies
\citep{Johnstone1988,Donahue1991,Johnstone2007,Ferland2009}, a pure
stellar or a pure active nucleus origin as the source of the optical
filaments has difficulty explaining certain optical flux ratios. This
may be true for the filaments but we suspect that stellar
photoionization becomes increasingly important with decreasing
distance to the core of the galaxy. Although fitting line and
continuum ratios measured in different bands across the
electromagnetic spectrum (submm, IR, optical, UV) simultaneously is
beyond the scope of this paper, we compared a subset of the optical
ratios measured by \cite{Johnstone1988} to our best-fit model
predictions. This comparison is shown in Table~\ref{optical}. We give
a range for the observed ratios since the ratios display a strong
radial gradient. `Model~I' refers to the best-fit energy model
described above. While the predicted values of the {\niiopt}/{\ha} and
{\siiopt}/{\ha} ratios are consistent with the observed values,
{\oiiopt}/{\hb} is lower by a factor of a few and {\oiiiopt}/{\hb} and
{\oiopt}/{\niiopt}+{\ha} are lower by about an order of magnitude than
the observed values. We devised another model, referred to as
`Model~II', obtained by slightly modifying the best-fit model by
decreasing the age of the input young stellar population by an order
of magnitude. Using Model~II we are able to reproduce some of the
optical ratios, specifically, the ratio {\oiii}/{\hb}. There is also
an increase in the ratio, {\oiopt}/{\niiopt}+{\ha}, by a factor of
two. The modified model does not impact the best-fit model parameters
constrained by the relative strengths of FIR and {\ha} lines
(Table~\ref{coreratios}). This is due to the fact that lowering the
age of the oldest stars in the YSP alters the SED only below
1000$~\AA$, such that {\oiiiopt} and {\oiopt} are affected. FIR lines,
on the other hand, are mainly produced by gas heated by lower-energy
UV and optical photons.

From this exercise we conclude that photoionization from stars is the
predominant source of energy in the cold gas clouds in the core of
{\pers}.  It may be that in order to obtain consistency with the
complete set of observed flux ratios at different wavebands, a
composite model that includes a stellar component and one or more
additional forms of heating is required. Particle-heating models are
usually very hard to distinguish from slow shocks
\citep[e.g.][]{Ferland2008,Farage2010,Rich2011}. Even though the work
of \cite{Ferland2009} showed that ionizing particles are the likely
heating agents in the filaments of {\pers}, the particle heating model
was only marginally more successful than heating by dissipative MHD
wave energy. The core of {\pers} displays large-scale motions, which
may very well entail shocks \citep[although strong shocks have been
ruled out by X-ray data,][]{Fabian2003,Sanders2007}. Similarly, due to
the proximity to the AGN, particle heating (cosmic rays) may be
significant as well. Hence additional heating required to explain the
FIR and optical line ratios in entirety may either manifest itself in
the form of weak shocks or particle-heating or both.


\begin{table}
  \centering
  {\small
    \caption{\small The optical emission line ratios in the central 4~kpc region of {\pers} 
      \citep{Johnstone1988}. The columns are (1) the species, (2) the measured ratios 
      (3) the predicted ratios based on the best-fit model (4) the predicted ratios 
      based on a slightly modified model. The observed ratios are from the observations
      off to the south-east side of the nucleus to avoid the high-velocity system.}
    \label{optical}
    \begin{tabular}{| c | c | c| c|}
      \hline
      Ratio & Value & Model I & Model II \\
      \hline\hline
      {\oiiopt}/{\hb}                 &     3.5 -- 8                 &  1-2               &   1-2  \\
      {\oiiiopt}/{\hb}                &     1.2  --  1.8         &  0.1-0.2         &  0.7-2   \\             
      {\oiopt}/{\ha}+{\niiopt}   &     0.06 --   0.14     &  0.004-0.007 &  0.008-0.01   \\                  
      {\niiopt}/{\ha}                 &     0.8 - 1.4             &  0.6-0.9        &  0.6-0.9   \\             
      {\siiopt}/{\ha}                 &     0.2 -- 0.7           &  0.6-0.8        &  0.6-0.8    \\            
      \hline
    \end{tabular}
  }
\end{table}

\subsection{Star formation}
\label{fuv}

Far-ultraviolet~(FUV) emission is a direct indicator of morphological
and spatial extent of recent star-formation sites. In Fig.~\ref{FUV},
we show the FUV emission associated with the BCG of Perseus. This
image was created using two datasets from the HST archive -- (1) Space
Telescope Imaging Spectrograph~(STIS) FUV-MAMA/F25SRF2 data (proposal
ID: 8107, PI: C.~O'Dea) and (2) Advance Camera of Surveys Solar Blind
Channel~(ACS/SBC) FUV-MAMA/F140LP data (proposal ID: 11207, PI:
R.~O'Connell). The first dataset consisted of a single pointing only
(centered on the core). The second dataset, however, consisted of 8
different pointings, designed to map the bright UV filaments extending
NW and SE from the core. We obtained the single flat-fielded,
dark-subtracted exposures from the HST archive for both the datasets
and combined them using the {\sc iraf} task, {\sc multidrizzle}.

The filaments NW of the core have the same spiral morphology as the
HVS. This spatial correlation seems to suggest that the star clusters
in the filaments are associated with the foreground galaxy
\citep[e.g.][]{Keel2001}. There is also evidence to the contrary,
suggesting that the star clusters belong to the LVS
\citep[e.g.][]{Goudfrooij1995,Brodie1998}.

For the present study, we will focus only on the FUV emission
originating from the core region (see Fig.~\ref{FUVzoom}), which, due
to proximity and symmetry, is expected to be associated with the
LVS. The integrated flux-density within the core is
$(5.3\pm0.1)\times10^{-15}$~erg~s$^{-1}$~cm$^{-2}~\AA^{-1}$ at
1456~$\AA$; however the FUV emission close to the center is likely to
be associated with the AGN. For the purpose of quantifying FUV
emission from stars, we determined the FUV emission in the annulus
(0.5~kpc -- 3.8~kpc) as
$(3.6\pm0.1)\times10^{-15}$~erg~s$^{-1}$~cm$^{-2}~\AA^{-1}$, and
correcting for the Galactic extinction results in
$(1.2\pm0.3)\times10^{-14}$~erg~s$^{-1}$~cm$^{-2}~\AA^{-1}$. There is
a further correction factor that needs to be applied in order to
account for the dust internal to the ISM of {\pers} (see
section~\ref{extinction}). Assuming an internal reddening of
$E(B-V)=0.37$, we obtain a FUV flux-density of
$(1.9\pm0.5)\times10^{-13}$~erg~s$^{-1}$~cm$^{-2}~\AA^{-1}$. This can
be directly compared to the flux density expected from a synthetic
spectrum of a young stellar population, such as the one used as input
for the {\sc cloudy} simulations. The expected flux density was
determined by convolving the redshifted synthetic spectrum with the
bandpass of the FUVMAMA/F25SRF2 filter on STIS using the {\sc iraf}
tool {\sc synphot}. A synthetic spectrum corresponding to an
instantaneous starburst containing $5\times10^9~\ms$ and 2~Myr in age
predicts a flux density of
$1.7\times10^{-11}$~erg~s$^{-1}$~cm$^{-2}~\AA^{-1}$. Holding the age
constant and scaling down the mass so that the flux density agrees
with the extinction-corrected observed flux density implies a SFR of
$27~\mpy$. This is similar to the SFR $\sim24~\mpy$ derived in
section~\ref{dustsed} (Table~\ref{sed-params}) from the FIR
measurements. This is also in very good agreement with the SFR $\sim
25~\mpy$ derived by \cite{Norgaard1990} from the IUE~(International
Ultraviolet Explorer) spectroscopic data.

Since the star formation rate traced via FIR data is expected to be
most accurate (due to negligible extinction at FIR wavelengths), and
the FUV rate matches it, the average internal reddening over the inner
4~kpc radius region is probably close to that suggested by the
observed Galactic-extinction corrected Balmer decrement (4.07) in a
region $\pp{18}$ SW of the nucleus (section~\ref{extinction}). On the
other hand, the SFR derived above using FUV data is for the core
region, whereas the SFR derived in Table~\ref{sed-params} is for the
whole galaxy. For example, \cite{Canning2010} obtained an optical star
formation rate of about $20~\mpy$ over the Blue Loop region, which is
outside the core region. Although there is hardly any FIR emission
detected in the Blue Loop region, it is possible that it contributes
to the total SFR derived from FIR measurements to some extent. This
could imply that the average internal reddening over the core is lower
than assumed, making the FUV-derived star formation rate lower as
well. Hence $E(B-V)\sim 0.37$ is an upper-limit. Note, the above
calculation is based on two assumptions: (1) the dust in {\pers}
follows the same extinction law as the Milky Way and (2) the Balmer
line ratios are close to the case-B recombination values.

It is interesting that if we use the dust attenuation derived from the
Balmer decrement and apply it at smaller (UV) wavelengths, the FIR-
and UV-derived SFRs are consistent with each other. This is in
contrast to studies that have indicated that the reddening values
diverge between nebular line emission and UV continuum by a factor of
two \citep[e.g.][]{Calzetti1997,Buat2002}, such that nebular lines,
like {\ha} and {\hb}, are more attenuated than the stellar continuum
emission. This discrepancy is usually attributed to an uneven
distribution of dust in front of stars and ionized gas, with the
latter being more closely associated with the dust than the former.
\cite{Calzetti1997} suggests using the relation $E(B-V)_{\st{star}} =
0.44 \times E(B-V)_{\st{gas}}$.  However, \cite{Garn2010} argue that
the while the continuum at longer wavelengths (optical) may be probing
older stellar population residing in less dusty environments, and
hence may not be co-spatial with the ionized gas, the continuum at
shorter wavelengths (UV) should be more closely tied to the young
stars producing ionized gas. Hence the correction factor (0.44) for
FUV emission may be higher than suggested by the above relation.

Using the relation suggested by \cite{Calzetti1997}, \cite{Buat2002}
derived $A_{\st{UV}(\lambda=2000)} = 1.6 \times A_{\st
  H\alpha}$. Instead of applying the attenuation at FUV wavelengths by
extrapolating the reddening value calculated from the Balmer
decrement, if we apply the above calibration we obtain a SFR of
$6~\mpy$. This is a factor of four lower than the FIR-derived
SFR. This implies that in the case of {\pers}, the dust covering
factor for the ionized gas resulting in the Balmer lines and the FUV
emission must be rather homogeneous, such that the attenuation tracked
by the ionized gas is closely related to that of the FUV emission.

\begin{figure}
  \centering
  \vspace{-0.5cm}
  \includegraphics[width=0.5\textwidth]{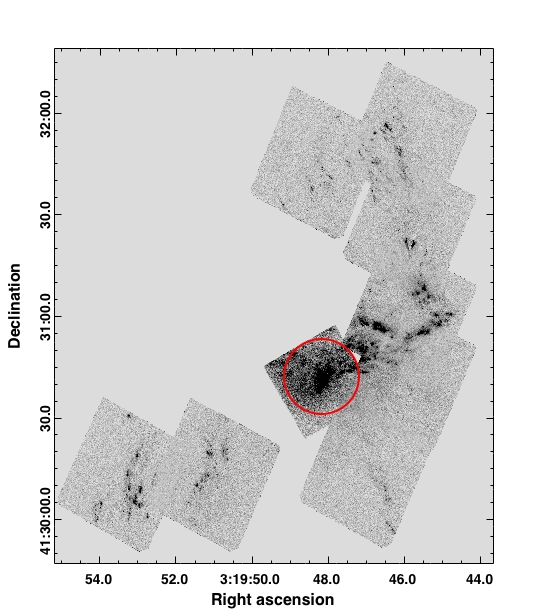}
  \caption{\small Combined HST SBC/F140LP and STIS/F25SRF2 FUV image
    of {\pers}. The red circle indicates the central 4~kpc core
    region.}
  \label{FUV}
\end{figure}

\begin{figure}
  \centering
  \includegraphics[width=0.5\textwidth]{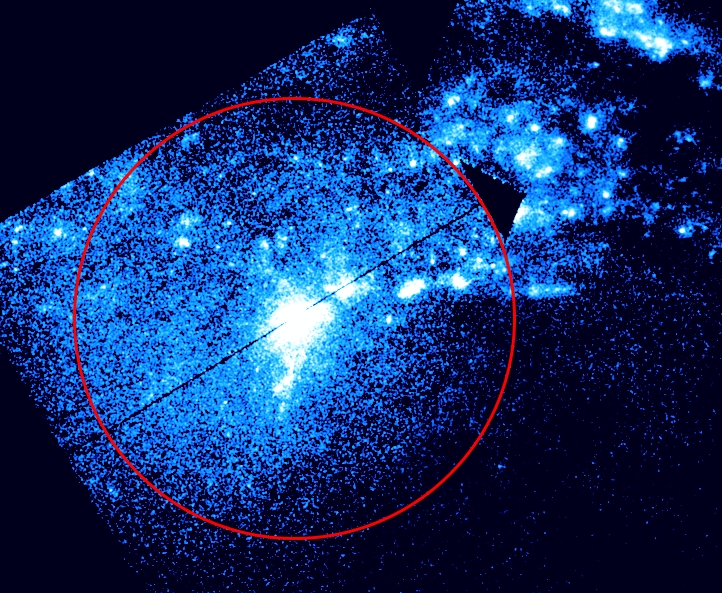}
  \caption{\small A zoom-in FUV image of the center of the galaxy, the
    red circle indicates the central 4~kpc core region.}
  \label{FUVzoom}
\end{figure}

\section{Discussion of the outer filaments}
\label{disc}

\begin{table*}
  \centering
  \caption{\small A comparison of {\cii}, {\oi} and {\ha} line fluxes in the Horseshoe 
    knot, the SW1/SW2 knots and the Blue Loop knots. Given in the last column are 
    the ratios with respect to {\cii}. {\oi} is not detected in any of the 
    extended filaments --  we give the 3$\sigma$ upper-limits.} 
  \label{filamentratios}
  \begin{tabular}{| c | c | c | c | c | c | c | c|}
    \hline 
   Region      &  RA                    & Dec                    &  Aperture Radius &  Line   &  Velocity & Flux    & Ratio \\
    &                          &                           & (arcsec)     &           &   km~s$^{-1}$ & ($10^{-15}$~erg~s$^{-1}$~cm$^{-2}$) & \\
    \hline\hline
    Horseshoe knot  & 03h19m45.15s & +41d31m33.1s  & 11   &    {\cii}   &  79.5 & 26.78 $\pm$ 0.26 & 1.00 \\
    &                          &                           &        &    {\oi}  &   & $<44$  & $<1.64$ \\
    &                          &                           &        &    {\ha}  &   & 36 (88) & 1.34 (3.27) \\
    Southwest knots  & 03h19m45.48s & +41d29m49.4s & 11 &  {\cii}   &  3.3 & 40.67 $\pm$ 3.55 & 1.00  \\
(Southern Filament)  &                          &                           &        &    {\oi}   &  & $<34$ & $<0.84$ \\
    &                          &                           &        &    {\ha}  &   & 87 (211) & 2.14 (5.19) \\
    Blue Loop knots & 03h19m50.72s & +41d30m07.2s & 11 &  {\cii}   &  -224.5 & 35.19 $\pm$ 1.87 & 1.00  \\
(Southeast Filament)  &                          &                           &        &    {\oi}   &  &  $<30$ & $<0.85$ \\
                            &           &                           &        &    {\ha}  &  &  73  (177) & 2.07 (5.03) \\
    \hline
  \end{tabular}
\end{table*}

\citet[][ F09 hereafter]{Ferland2009} conducted a thorough analysis of
the Horseshoe region in {\pers}. They used the infrared and optical
line intensities to distinguish between two types of heating: extra
heating, as would be produced by passing shocks, and heating by
energetic particles such as cosmic rays.  Extra heating refers to
heating by dissipative MHD waves, and it is assumed that this kind of
heating increases the thermal energy of the gas only. Hence the local
gas kinetic temperature dictates the collisional processes that are
energetically possible. Energetic particle heating refers essentially
to ionizing particles, and, on the other hand, is not only capable of
depositing thermal energy via elastic collisions with the particles of
the ISM but can also ionize a neutral medium via creation of a
population of suprathermal secondary electrons. The physics of
molecular gas exposed to such heating sources is described by
\cite{Ferland2011a} and \cite{Ferland2011b}.

The model in F09 assumes that the gas in the filaments is in pressure
balance with the surrounding X-ray-emitting intracluster gas. Based on
the X-ray measurement of the electron pressure of 0.128~keV~cm$^{-3}$
of the hot gas surrounding the horseshoe by \cite{Sanders2007}, F09
assumed a constant pressure of $nT = 10^{6.5}$~cm$^{-3}$~K. The
simulations conducted in F09 are further based on the assumption that
the filaments are composed of cloudlets with a range of density, $n$,
and temperature, $T$, but which have this single pressure, $nT$. This
assumption is motivated by the fact ionic, atomic, and molecular
emissions are all observed in the filaments. This implies that
different phases of gas occupy a telescope beam, even at HST
resolution\comment{(see Fig.~\ref{CP_spectrumA})}. In other words,
there is observational evidence that both dense molecular and diffuse
ionized emission arise from spatially coincident regions.  Because of
the constant pressure assumption, low density cloudlets have high
temperature and produce emission from ionized gas, while dense clouds
are cold and account for the molecular component.

The formalism adopted in F09 uses a cumulative filling factor $f(n)$,
which is a powerlaw in density, as the weighting function to co-add
clouds of different densities. This factor describes the fractional
volume filled with gas with density $n$ or lower. The spectra for
various emission lines, in particular the infrared H$_2$ and optical H
{\sc i} emission lines, are determined using {\sc cloudy} using a
range of electron densities, temperatures, and non-radiative heating
rates. The emission for a given line is integrated over the ensemble
of clouds and then compared to observations.

\cite{Ferland2009} found that both the forms of non-radiative heating,
extra heating and energetic particle heating, match the optical and
infrared observations to within a factor of two for the majority of
the lines. There are a few discriminant lines, such as the optical
emission lines He {\sc i}~$\lambda~5876~\AA$, [Ne {\sc
  iii}]~$\lambda~3869~\AA$, and the infrared emission line [Ne {\sc
  ii}]~$\lambda~12.81${\mm}, which show a few orders of magnitude
difference and indicate that ionizing particles are responsible for
heating and ionizing the gas. \citet{Fabian2011} argue that the
surrounding hot intracluster medium is the source of the ionizing
particles, rather than true cosmic rays as would be found in the
adjoining radio lobes.

All of the predicted lines used to compare with observations were
optically thin, so their intensity relative to similar forbidden lines
has no dependence on cloud column density. The F09 model also made
predictions for the far-infrared emission lines, such as those
observed by {\herschel}. In Table~\ref{filamentratios}, we list the
detected Herschel emissions in the three regions of the extended
filaments $-$ the Horseshoe region, the southwest~(SW) knots. We also
give the 3-$\sigma$ upper-limits for the non-detections. These lines
may be used to distinguish between the two heating scenarios. In
particular, the predicted {\oi}/{\cii} ratio is $\sim 3$ for the extra
heating and $\sim 21$ for the energetic particle heating.  The
observed {\oi}/{\cii} ratios, on the other hand, have an upper-limit
of 1.64 in the Horseshoe knot and 0.85 in both the SW and Blue Loop
knots.  This discrepancy was first pointed out by \citet{Mittal2011b}
in a different context.

There are two plausible explanations for why the F09 model fails to
reproduce the Herschel observations. The first reason is tied to the
critical densities, $\ncr$, of {\cii} and {\oi} gas (see
Fig.~\ref{oicii_cloudy}). Since the F09 model assumes a constant
pressure of $nT = 10^{6.5}$~cm$^{-3}$~K, the FIR lines which are
produced at low temperatures ($\sim 100$~K to a few hundred K )
correspond to high densities ($3 \times 10^4 $~cm$^{-3}> n > 5 \times
10^3$~cm$^{-3}$). Such high densities lead to high {\oi}/{\cii} ratios
due to the reasons given in section~\ref{best-fit}.  The model could
be brought into agreement with the observations by postulating a large
component of low density gas, which would strongly emit the Herschel
lines with the observed ratio.  F09 note, in section 7, point 7, that
large reservoirs of cold gas could be present yet not detected with
the selection of lines they had available.

The second likely cause for the incompatibility of the F09 model with
the {\oi}/{\cii} ratio is that the F09 model assumes that the emission
lines are optically thin. If the gas has high enough column density,
the Herschel lines become optically thick. This is normally the case
in galactic PDRs \citep{Tielens1985}. The line luminosity is no longer
determined simply by the product of the line emissivity and the volume
of the cloud but rather the geometry of the emitting cloud, especially
its column density, will affect the line intensities. F09 note that
the observed surface brightness of the {\ha} line gives the
line-of-sight thickness of the cloud ${\st d}l \sim 0.3$~pc. This
implies a hydrogen column density of $N(H)=10^{22.5}$~cm$^{-2}$.  If
the emission forms in a single cloud with this column density, the
Herschel lines would be strongly affected but there would be little
impact on the NIR and optical lines, other than the effects of
internal reddening.

Clearly much remains to be learned about the geometry of the filaments
at sub-HST resolution scales. Both scenarios outlined here are
consistent with what is known from available observations.

\section{Conclusions}
\label{conclusions}

We have presented far-infrared Herschel observations of the center of
the Perseus galaxy cluster. The brightest cluster galaxy, {\pers}, is
surrounded by filaments, previously imaged extensively in {\ha} and CO
emission. In this work, we report the presence of coolants, such as
{\cii}, {\oi}, {\nii}, {\oib} and {\oiii}, in addition to {\ha} and
CO. All the Herschel lines except {\oiii} are spatially extended, with
{\cii} extending up to 25~kpc from the nucleus of the galaxy, and
cospatial with {\ha} and CO. Furthermore, {\cii} shows a similar
velocity distribution to CO, and the latter has been shown in previous
studies to display a close association with the {\ha} kinematics. The
spatial and kinematical correlation among {\cii}, {\ha} and CO gives
us confidence that the different components of the gas may be modelled
with a common heating model.

The velocity structure inferred from {\cii} observations reveals
blueshifted components on either side of the nucleus and a ridge of
redshifted gas passing through the center in the north-south
direction. A combination of disk rotation on $\sim 10$ kpc scales and
more disordered motion of a few distinct clouds at larger radii is one
plausible scenario consistent with {\cii} (Herschel) and CO data. It
may also be that outflows from the radio source at the center are
dredging up cold material away from the line of sight, resulting in a
ridge of redshifted gas.

We have detected continuum emission at all PACS and SPIRE wavelengths
ranging from 70{\mm} to 500{\mm}. The Herschel continuum data combined
with a suite of existing radio, submm and IR data, have allowed us to
perform a spectral energy distribution fitting using a model
comprising two modified-blackbody dust components, dominant at IR
wavelengths, and synchrotron AGN emission, dominant at radio and submm
wavelengths. The total infrared luminosity (8{\mm} to 1000{\mm}) is
$(1.5\pm0.05)10^{11}~\ls$, making {\pers} a luminous infrared galaxy
(LIRG). The infrared-inferred star formation rate is about
24~$\mpy$. This is comparable to the SFR estimated in a core region
($\sim 4~$kpc in radius) using the HST-FUV observations, and also {\it
  XMM-Newton} RGS observations, which suggest a residual cooling rate
of $20~\mpy$.

We have investigated in detail the source of the emissions emerging
from a core region $4~$kpc in radius. This investigation was done by
carrying out radiative transfer simulations assuming a photoionization
model consisting of only old and young stellar populations as heating
agents. We find that the Herschel and {\ha} emissions can be
reproduced by such a model, yielding hydrogen density between
500~cm$^{-3}$ and 700~cm$^{-3}$ and FUV intensity field between
700~Habings to 900~Habings. Optical flux ratios indicate that a second
heating component may be needed; however, stellar photoionization
seems to be the dominant mechanism.

We have also detected {\cii} in three previously well-studied regions
of the filaments: the Horseshoe, Blue Loop and southwest knots. The
observed upper-limits of the {\oi}/{\cii} ratio are 1~dex smaller than
predicted by the best-fit model of \cite{Ferland2009}, wherein two
sources of ionization have been considered individually: particle
heating and extra-heating (the latter increases the thermal energy of
the gas). The {\cii} line has a excitation potential of 91~K and a
critical density much lower than the lines considered in that study.
This suggests that the lines are optically thick, as is typical of
galactic PDRs, and implies that there is a large reservoir of cold
atomic gas. This has not been included in previous inventories of the
filament mass and may represent a significant component. It seems
likely that the model used by \cite{Ferland2009} needs to be augmented
in order to reproduce the observed strengths of the various emissions
in the filaments of {\pers}. It can also be that a composite model,
like the one used for the study of Centaurus \citep{Mittal2011b},
consisting of more than one heating agent, is required to achieve
compatibility with observations.


\section*{Acknowledgments}

We thank Christopher Conselice for providing the WIYN {\ha} image and
Greg Taylor for the radio images. We thank the referee sincerely for a
very useful feedback. This work is based (in part) on observations
made with Herschel, a European Space Agency Cornerstone Mission with
significant participation by NASA. Support for this work was provided
by NASA through an award issued by JPL/Caltech. We thank M.~Aller for
providing us with the UMRAO data. This research has made use of data
from the University of Michigan Radio Astronomy Observatory which has
been supported by the University of Michigan and by a series of grants
from the National Science Foundation, most recently AST-0607523. This
research has made use of the NASA/IPAC Extragalactic Database (NED)
which is operated by the Jet Propulsion Laboratory, California
Institute of Technology, under contract with the National Aeronautics
and Space Administration.  This work is supported in part by the
Radcliffe Institute for Advanced Study at Harvard University. GJF
acknowledges support by NSF (0908877; 1108928; and 1109061), NASA
(10-ATP10-0053, 10-ADAP10-0073, and NNX12AH73G), JPL (RSA No 1430426),
and STScI (HST-AR-12125.01, GO-12560, and HST-GO-12309). STSDAS is a
product of the Space Telescope Science Institute, which is operated by
AURA for NASA. HRR thanks the Canadian Space Agency Space Science
Enhancement Program for support.


{\small
\bibliographystyle{mn2e}
\bibliography{ref}}

\end{document}

%% file: makros.tex

\newcommand{\st}[1]{\mathrm{#1}} 
\newcommand{\pow}[2]{$\st{#1}^{#2}$}
\newcommand{\grad}{\hspace{-0.15em}\r{}}
\newcommand{\lm}{\lambda}
\newcommand{\average}[1]{\left\langle #1 \right\rangle}
\newcommand{\pp}[1]{#1^{\prime\prime}}
\newcommand{\p}[1]{#1^{\prime}}

\newcommand{\oi}{[O{\sc i}]}
\newcommand{\cii}{[C{\sc ii}]}
\newcommand{\nii}{[N{\sc ii}]}
\newcommand{\oiii}{[O{\sc iii}]}
\newcommand{\si}{[Si{\sc i}]}
\newcommand{\oib}{[O{\sc ib}]}
\newcommand{\mm}{~$\mu$m}
\newcommand{\ha}{H$\alpha$}
\newcommand{\hb}{H$\beta$}
\newcommand{\niiopt}{\nii$\lambda6583$}
\newcommand{\niioptii}{\nii$\lambda6583,6548$}
\newcommand{\oiopt}{\oi$\lambda6300$}
\newcommand{\oiiopt}{[O{\sc ii}]$\lambda\lambda3727,3729$}
\newcommand{\oiiiopt}{[O{\sc iii}]$\lambda5007$}
\newcommand{\siioptdoub}{[S{\sc ii}]$\lambda\lambda6716,6731$}
\newcommand{\siiopt}{[S{\sc ii}]$\lambda6731$}
\newcommand{\civ}{[C{\sc iv}]$\lambda1549$}
\newcommand{\md}{M_{\st{d}}}
\newcommand{\td}{T_{\st{d}}}
\newcommand{\mdc}{M_{\st{d,c}}}
\newcommand{\tdc}{T_{\st{d,c}}}
\newcommand{\mdw}{M_{\st{d,w}}}
\newcommand{\tdw}{T_{\st{d,w}}}
\newcommand{\mg}{M_{\st{g}}}
\newcommand{\tcmb}{T_{\st{cmb}}}
\newcommand{\ncr}{n_{\st{cr}}}
\newcommand{\hextra}{H_{\st{extra}}}
\newcommand{\hcol}{N$_{\st H}$}
\newcommand{\pdv}{$P{\st d}V$}
\newcommand{\nitrogen}{$Z_{\odot}$(N)}

\newcommand{\h}{~h_{71}~}
\newcommand{\hinv}[1]{~h_{71}^{#1}~}
\newcommand{\eq}[1]{(\ref{eq-#1})}
\newcommand{\wrt}{with respect to\ }
\newcommand{\mms}{\frac{M_{\odot}}{M}}
\newcommand{\mpy}{\ms~\st{yr}^{-1}}
\newcommand{\ct}{t_{\st{cool}}}
\newcommand{\rc}{r_{\st{cool}}}
\newcommand{\lc}{L_{\st{cool}}}
\newcommand{\tvir}{T_{\st{vir}}}
\newcommand{\rvir}{R_{\st{500}}}
\newcommand{\mvir}{M_{\st{500}}}
\newcommand{\mbh}{M_{\st{BH}}}
\newcommand{\ms}{M_{\odot}}
\newcommand{\ls}{L_{\odot}}
\newcommand{\zs}{Z_{\odot}}
\newcommand{\nub}{\nu_\st{break}}
\newcommand{\lxb}{L_{\st {Xb}}}
\newcommand{\lfir}{L_{\st{FIR}}}
\newcommand{\lhundred}{L_{\st{100\mu m}}}
\newcommand{\lfirtot}{L_{\st{FIR,tot}}}
\newcommand{\lx}{L_{\st {X}}}
\newcommand{\lr}{L_{\st {R}}}
\newcommand{\lbcg}{L_{\st{BCG}}}
\newcommand{\lt}{L_{\st{X}}{\st -}\tvir}
\newcommand{\Dl}{D_{\st{L}}}
\newcommand{\Da}{D_{\st{A}}}
\newcommand{\mdr}{\dot{M}_{\st{classical}}}
\newcommand{\smdr}{\dot{M}_{\st{spec}}}
\newcommand{\ohb}{[\st{O_{~III}}]~\lambda 5007/{\st H}\beta}
\newcommand{\norm}{$\eta_{\st{OSP}}$}

\newcommand{\chandra}{\textit{Chandra}}
\newcommand{\vla}{\textit{VLA}}
\newcommand{\gmrt}{\textit{GMRT}}
\newcommand{\atca}{\textit{ATCA}}
\newcommand{\XMM}{\textit{XMM-Newton}}
\newcommand{\einstein}{\textit{Einstein}}
\newcommand{\asca}{\textit{ASCA}}
\newcommand{\rosat}{\textit{ROSAT}}
\newcommand{\herschel}{\textit{Herschel}}
\newcommand{\iras}{\textit{IRAS}}
\newcommand{\spitzer}{\textit{Spitzer}}
\newcommand{\hiflux}{\textit{HIFLUGCS}}

%% file: journals.tex
%
\def\aj{AJ}%
\def\araa{ARA\&A}%
\def\apj{ApJ}%
\def\apjl{ApJ}%
\def\apjs{ApJS}%
\def\ao{Appl.~Opt.}%
\def\apss{Ap\&SS}%
\def\aap{A\&A}%
\def\aapr{A\&A~Rev.}%
\def\aaps{A\&AS}%
\def\azh{AZh}%
\def\baas{BAAS}%
\def\jrasc{JRASC}%
\def\memras{MmRAS}%
\def\mnras{MNRAS}%
\def\pra{Phys.~Rev.~A}%
\def\prb{Phys.~Rev.~B}%
\def\prc{Phys.~Rev.~C}%
\def\prd{Phys.~Rev.~D}%
\def\pre{Phys.~Rev.~E}%
\def\prl{Phys.~Rev.~Lett.}%
\def\pasp{PASP}%
\def\pasj{PASJ}%
\def\qjras{QJRAS}%
\def\skytel{S\&T}%
\def\solphys{Sol.~Phys.}%
\def\sovast{Soviet~Ast.}%
\def\ssr{Space~Sci.~Rev.}%
\def\zap{ZAp}%
\def\nat{Nature}%
\def\iaucirc{IAU~Circ.}%
\def\aplett{Astrophys.~Lett.}%
\def\apspr{Astrophys.~Space~Phys.~Res.}%
\def\bain{Bull.~Astron.~Inst.~Netherlands}%
\def\fcp{Fund.~Cosmic~Phys.}%
\def\gca{Geochim.~Cosmochim.~Acta}%
\def\grl{Geophys.~Res.~Lett.}%
\def\jcp{J.~Chem.~Phys.}%
\def\jgr{J.~Geophys.~Res.}%
\def\jqsrt{J.~Quant.~Spec.~Radiat.~Transf.}%
\def\memsai{Mem.~Soc.~Astron.~Italiana}%
\def\nphysa{Nucl.~Phys.~A}%
\def\physrep{Phys.~Rep.}%
\def\physscr{Phys.~Scr}%
\def\planss{Planet.~Space~Sci.}%
\def\procspie{Proc.~SPIE}%